\documentclass[%
 reprint,
 superscriptaddress,
 nofootinbib,
 nobibnotes,
 amsmath,amssymb,
 aps,
 prx,
]{revtex4-2}

\usepackage{graphicx}% Include figure files
\usepackage{dcolumn}% Align table columns on decimal point
\usepackage{bm}% bold math
\usepackage[colorlinks,linkcolor=blue,urlcolor=blue, citecolor=blue]{hyperref}
\usepackage{lipsum}

\usepackage{dsfont}
\usepackage{physics}
\usepackage{enumitem}
\usepackage{tcolorbox}
\usepackage{needspace}

\usepackage{amsthm} 
\usepackage{thmtools} 
\usepackage{thm-restate}

\usepackage{soul}

\declaretheorem{corollary,lemma,proposition,definition}

\usepackage{amsmath, amssymb, amsfonts, amsthm}

\usepackage[normalem]{ulem}

\newcommand{\E}{{\rm E}}
\newcommand{\Var}{{\rm Var}}

\usepackage[page]{appendix}

\renewcommand{\appendixtocname}{List of appendices}

\makeatletter
\let\oldappendix\appendices

\g@addto@macro\tableofcontents{%
  \let\tf@toc@orig\tf@toc
}

\renewcommand{\appendices}{%
  \setcounter{section}{0}%
  \renewcommand{\thesection}{Appendix~\Alph{section}}%
  \renewcommand{\thesubsection}{\arabic{subsection}}%
  \setcounter{subsection}{0}%
  \let\tf@toc\tf@app
  \addtocontents{app}{\protect\setcounter{tocdepth}{2}}
  \immediate\write\@auxout{%
    \string\let\string\tf@toc\string\tf@app
  }
  \oldappendix
}%

\g@addto@macro\endappendices{%
  \let\tf@toc\tf@toc@orig
  \immediate\write\@auxout{%
    \string\let\string\tf@toc\string\tf@toc@orig
  }%
}

\renewcommand\tableofcontents{%
  \@starttoc{toc}%
}

\newcommand{\listofappendices}{%
  \begingroup
  \newcommand{\contentsname}{\appendixtocname}
  \let\@oldstarttoc\@starttoc
  \def\@starttoc##1{\@oldstarttoc{app}}
  \tableofcontents
  \endgroup
}
\makeatother

\begin{document}

\preprint{APS/123-QED}

\title{Exponential gain in clock precision using quantum correlated ticks}

\author{Florian Meier}
\email[]{florianmeier256@gmail.com}
\affiliation{Atominstitut, Technische Universit{\"a}t Wien, 1020 Vienna, Austria}

\author{Yuri Minoguchi}
\affiliation{Atominstitut, Technische Universit{\"a}t Wien, 1020 Vienna, Austria}
\affiliation{Wolfgang-Pauli Institut, Universit{\"a}t Wien, Oskar-Morgenstern-Platz 1, 1090 Vienna, Austria}

\author{Gianmichele Blasi}
\affiliation{Department of Applied Physics, University of Geneva, 1211 Geneva, Switzerland}

\author{Géraldine Haack}
\affiliation{Department of Applied Physics, University of Geneva, 1211 Geneva, Switzerland}

\author{Marcus Huber}
\email{marcus.huber@tuwien.ac.at} 
\affiliation{Atominstitut, Technische Universit{\"a}t Wien, 1020 Vienna, Austria}
\affiliation{Institute for Quantum Optics and Quantum Information - IQOQI Vienna, Austrian Academy of Sciences, Boltzmanngasse 3, 1090 Vienna, Austria}

\date{\today}

\begin{abstract}
Creating precise timing devices at ultra-short time scales is not just an important technological challenge, but confronts us with foundational questions about timekeeping's ultimate precision limits. Research on clocks has either focused on long-term stability using an oscillator stabilized by a level transition, limiting precision at short timescales, or on making individual stochastic ticks as precise as possible. Here, we prove the viability of a conceptually different avenue: the autonomous self-correction of consecutive ticks by quantum correlations. This provides a new paradigm that integrates the advantages and insights from quantum transport theory to operate clocks at ultra-short timescales. We fully solve a model of coupled quantum systems and show how the emergent Pauli exclusion principle correlates the clock at the quantum level yielding an exponential advantage in precision. We furthermore demonstrate through simulations with realistic imperfections that this remarkable gain in precision remains stable providing a roadmap for implementation with contemporary quantum technologies.
\end{abstract}

\maketitle

\section{Introduction}
Any recordable and thus irreversible event can be used as a clock.
From sand grains in an hourglass to electronic readout of laser oscillations in an atomic clock, physical implementations vary widely.
Yet, all those clocks define \emph{ticks} through one of two fundamental designs:
\emph{Stochastic clocks} use (a collection of) irreversible events as ticks, such as an accumulation of sand grains or on much grander scales the radioactive decay of unstable elements.
\emph{Oscillator clocks} are based on oscillating processes that are ubiquitous in our Universe, starting from the earth's rotation in sundials to modern atomic clocks, whose oscillator frequency is continuously stabilized on an atomic transition frequency.

State of the art atomic and optical clocks achieve remarkable long-term precision by stabilizing an oscillator via feedback on a sharp level transition~\cite{Essen1955,Ludlow2015,Riehle2015,Bothwell2022}, with even further advances promised by spin-squeezing~\cite{Yang2025SqueezingExp,Meiser2008SqueezingThy} and  nuclear transitions~\cite{Tiedau2024}.
Given their focus on stability and standardization of the tick frequency, the readout---based on sampling the oscillator---has played a secondary role.
This is because performance at long averaging times is dominated by oscillator stability rather than the sampling uncertainty, where quantum and thermal fluctuations fundamentally limit precision~\cite{Milburn2020,Meier2023,Prech2025,Silva2023,Pietzonka2024}.
By design, the operational timescale of oscillator-based clocks is tied to the oscillator frequency, and at finer temporal resolution, stochasticity in the readout becomes prohibitive.
Moreover, resolving each clock tick requires a large overhead of irreversible sampling events, each producing entropy or heat.

\begin{figure*}[ht]
    \centering
    \includegraphics[width=\linewidth]{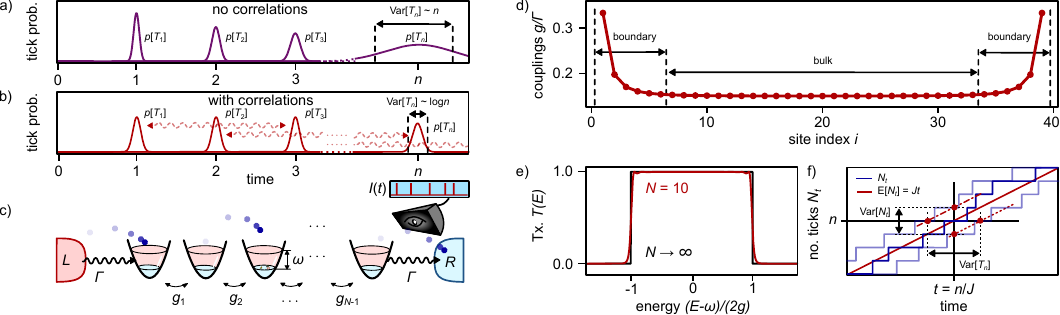}
    \caption{
    Figures of merit and setup. 
    a)~Stochastic events define clock `ticks' and for a clock to be precise the tick distribution should be narrow.
    For uncorrelated ticks, errors add up (see a), whereas correlations can suppress error accumulation (see b).
    c)~An exemplary model based on a spin chain connected to biased leads where the current leaving on the right defines the ticks.
    d)~The couplings between the sites are carefully chosen to exponentially suppress error accumulation.
    It turns out that the optimal couplings are constant in the chain's bulk but increase symmetrically towards the end to match the boundary dissipation.
    e)~The transmission function shows the probability with which an excitation traverses the chain, which approaches a boxcar as the number of sites goes to infinity.
    f)~Random realizations of the tick number trajectories are sketched to illustrate the linear relation between the uncertainty in the tick number and the uncertainty in the tick time.
    }
    \label{fig:Fig1_Setup}
\end{figure*}

Stochastic clocks, where every sampling event contributes directly to the time signal, however, operate at fast timescales and close to thermodynamic optimality.
While these considerations are only of foundational concern thus far, low-dissipation clocks could revolutionize direct quantum control techniques~\cite{MarinGuzman2024,Wadhia2025}, especially in applications where ultra-fast timing is required.
For autonomous and fast operations, stochastic clocks~\cite{US4676661A,Vyas1988,Brandes2008} have thus attracted renewed attention~\cite{Schwarzhans2021,Woods2019,Dost2023,Meier2025a,Zeppetzauer2025}, as modern research focuses on achieving high precision by confining these (nearly) independent stochastic events to narrow time-windows.
This strategy works well at short timescales, but over longer durations, errors accumulate: for uncorrelated events, the overall variance grows linearly in the number of ticks (Fig.~\ref{fig:Fig1_Setup}a).
Optimizing single-tick precision can reduce the slope of this growth, but it cannot eliminate the inherently linear scaling in the number of events.

In this work, we propose a fundamentally different, quantum clock design that radically departs from the independence of stochastic events.
In such a \textit{quantum correlated clock}, we envision correlations between irreversible events that self-stabilize the clock at the most fundamental level, dramatically improving the error scaling (see Fig.~\ref{fig:Fig1_Setup}b).
Because this approach does not rely on an oscillator or active feedback that would impose an additional timescale constraint, the mechanism we propose can operate at unprecedented timescales, while remaining close to thermodynamic optimality.  

To demonstrate this concept, we propose an explicit construction of such a quantum correlated clock using a chain of coupled qubits (Fig.~\ref{fig:Fig1_Setup}c).
A system that is readily realizable on current experimental platforms such as superconducting qubits~\cite{Zhang2023,CastilloMoreno2025}, atom arrays~\cite{Chen2023RydbergExp,Whitlock2017XYRydThy,Duan2003XYThy}, or quantum dots~\cite{Kandel2021,Qiao2021}.
By applying a strong chemical potential bias across the chain, excitations travel from left to right and define ticks as they hop off the chain.
Careful design of the couplings between the sites (Fig.~\ref{fig:Fig1_Setup}d) maximizes the excitation transmission probability (Fig.~\ref{fig:Fig1_Setup}e) and gives rise to a highly correlated temporal tick structure.

Timing uncertainties in this structure are minimized by correlations arising from an emergent Pauli exclusion principle, which prevents excitations from clustering as they propagate along the chain.
This manifests as an exponentially improved scaling of the tick variance,
\begin{align}
\label{eq:VarTn}
    \Var[T_n] \sim \log n,
\end{align}
in comparison to uncorrelated ticks which scale linearly, with $T_n$ the time between any $n$ ticks.

\section{Quantum correlated clock model}

The physical system to demonstrate the scaling in Eq.~\eqref{eq:VarTn} can be written as a chain of $N$ energy-degenerate two-level systems (qubits) as illustrated in Fig.~\ref{fig:Fig1_Setup}c. 
It is characterized by a XY-type Hamiltonian ($\hbar \equiv 1$):
\begin{align}
\label{eq:H}
    H = \sum_{i=1}^N \frac{\omega}{2}\sigma_i^z - \sum_{i=1}^{N-1} g_i \left(\sigma_i^- \sigma_{i+1}^+ + \sigma_{i+1}^- \sigma_i^+\right),
\end{align}
with $\omega$ being the on-site energy, $g_i$ the coupling between neighboring sites $i$ and $i+1$.
We assume single-excitation tunneling events with $\sigma_i^\pm$ the raising and lowering operators for creating and destroying a single excitation on site $i$.
This chain is then coupled to two leads, left ($L$) and right ($R$), with interaction strength $\Gamma$ assumed to be symmetric to allow for a closed form solution.

We consider the transport of excitations across the chain using jump operators $J_L=\sqrt{\Gamma}\sigma_1^+$ for injecting excitations from the left and $J_R=\sqrt{\Gamma}\sigma_N^-$ for dissipating them on the right.
In the limit $\omega\gg\Gamma\gg g_i$, the evolution of the density operator $\rho$ of the chain is correctly described using a local Markov Lindblad master equation~\cite{Hofer2017,Breuer2007} 
\begin{align}
\label{eq:dotRho}
   \dot{\rho} = \mathcal L \rho = -i[H,\rho] + \mathcal D_{J_L}\rho + \mathcal D_{J_R}\rho\,,
\end{align}
with commutator $[A,B]=AB-BA$, anticommutator $\{A,B\}=AB+BA$, and $\mathcal D_{J_\alpha}\rho = J_\alpha\rho J_\alpha^\dagger - 1/2 \{J_\alpha^\dagger J_\alpha,\rho\}$ the dissipator for jump operator $J_\alpha$.

Each tick manifests as a sharp peak in the current $I(t)$ flowing from the chain to the right bath, marking the transport of a single excitation.
We can obtain this current using the master equation~\eqref{eq:dotRho} for the two-terminal transport setup (Appendix~\ref{SM:Stochastic master equation and current}). 
To connect the tick distribution to transport properties, we introduce $N_t=\int_0^t \dd t' \,I(t')$, which counts the total number of ticks (or equivalently, transported excitations) up to time $t$ (see Fig.~\ref{fig:Fig1_Setup}f).
In the stationary limit, the mean current  $J = \lim_{t\rightarrow\infty}\E[I(t)]$ remains constant, establishing the clock's resolution: its average tick rate.
The variance of $N_t$, expressed via current fluctuations, is analytically tractable---a key feature for our subsequent analysis:
\begin{align}
\label{eq:VarNt_classical_main}
    \Var[N_t] = \int_0^t \dd t_1\int_0^t \dd t_2  \,\E[[I(t_2)I(t_1)]]\,,
\end{align}
where $\E[[AB]]=\E[AB]-\E[A]\E[B]$ denotes the second cumulant.

We establish the clock's optimal design by analyzing current fluctuations in the limit of infinite time using the diffusion constant $D=\lim_{t\rightarrow\infty}\partial_t \Var[N_t]$.
Optimizing long-term precision thus amounts to minimizing the Fano factor $D/J$~\cite{Silva2023,Meier2025a,Prech2025}.
Curiously, even as the Fano factor vanishes, suggesting a perfectly regular clock, single ticks are stochastic: the waiting-time distribution between consecutive ticks, $p[T_1]$, retains a finite and universal variance (see Fig.~\ref{fig:Fig2_Results}a), here described by the Wigner--Dyson distribution~\cite{Albert2012,Dasenbrook2015}.
This apparent randomness at the level of consecutive ticks is fully compatible with zero current noise: initially, ticks are uncorrelated, but by coarse-graining over longer times, correlations progressively suppress fluctuations that would arise in an independent sequence. 

We capture this coarse-grained behavior through $T_n$, the waiting time for $n$ ticks.
It is crucial to establish a one-to-one correspondence between the distribution of the number of ticks $N_t$, and the distribution of waiting times $T_n$.
For a constant current, the expected time for $n$ ticks is $\E[T_n]=n/J$.
The variance $\Var[T_n]$, which serves as our primary figure of merit for clock precision in~\eqref{eq:VarTn}, can also be obtained from $\Var[N_t]$.
As illustrated in Fig.~\ref{fig:Fig1_Setup}f, fluctuations in time for the $n$th tick are directly related to the fluctuations in the number of ticks that occur in an interval of length $t=n/J$.
This connects our central object of interest, the waiting time variance, to the analytically tractable current fluctuations:
\begin{align}  
\label{eq:VarTn_sim_VarNt}
    \Var[T_n] \sim J^{-2}\Var[N_{t=n/J}],  
\end{align}
a relation valid for processes where higher order correlations are negligible (Appendix~\ref{SM:calculating_waiting_time_variance}).

\section{Model optimization}
To minimize relative current fluctuations $D/J$, we analyze this model performing a Jordan--Wigner transformation which maps qubits to fermions and retains the quadratic nature of $H$ (details in Appendix~\ref{SM:free_fermion_model}).
Within a Landauer--B\"uttiker approach valid for quadratic Hamiltonians~\cite{Landauer1957,Buettiker1986,Blasi2024} (Appendices~\ref{SM:noise_current_LB} and~\ref{SM:calculating current and noise}), it was shown that the optimal transmission function $T(E)$---the transmission probability of a mode of energy $E$---is a boxcar function~\cite{Whitney2014,Brandner2025,Timpanaro2025,Blasi2025,Khandelwal2025}.
Considering a finite-length chain of $N$ sites, the transmission function can only approximate the ideal, discontinuous boxcar profile (see Fig.~\ref{fig:Fig1_Setup}e).

Therefore, we adjust the coupling strengths $g_i$ to minimize $D/J$ for fixed $N$.
A closed-form set of couplings $g_i = f_N(i)$, expressed as products of trigonometric functions, has been derived in Ref.~\cite{Brandner2025}.
In the limit of infinitely long chains, this closed-form design yields a transmission that approaches the ideal boxcar, but the relative noise still decreases only linearly with chain length, $D/J \sim N^{-1}$.
In contrast, here we determine the $g_i$ numerically by matching them to the boundary coupling $\Gamma$ to achieve maximal transmission and minimal noise.
For the optimized couplings, we obtain the following remarkable scaling improvement:
\begin{align}
\label{eq:D/J_scaling}
D/J \sim N^{-1.86}\,.
\end{align}

\begin{figure}
    \centering
    \includegraphics[width=\linewidth]{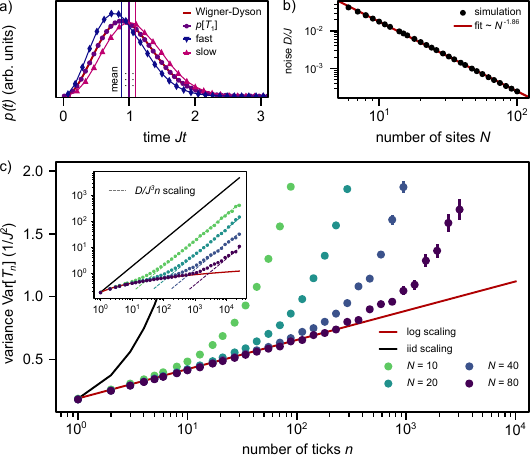}
    \caption{Current and waiting time fluctuations. 
    a)~For $N=20$ sites histogram data is shown for the waiting time probability $p[T_1]$ between subsequent ticks.
    The circles show the unconditional distribution, triangles (diamonds) show the distribution conditioned on the previous tick being faster (slower) than the expectation value.
    Vertical lines indicate the averages, and the solid red curve a Wigner--Dyson fit to $p[T_1]$.
    b)~The relative current noise $D/J\sim N^{-1.86}$ in log-log scale as it decays with increasing chain length $N$.
    c)~A Monte Carlo approximations for the variance of $T_n$, the time of the $n$th tick, is shown as a function of $n$, for different chain lengths $N$ in semi-log scale.
    The inset shows a log-log scale with the same data.
    For comparison, theoretical prediction and the diffusive scaling are shown.} 
    \label{fig:Fig2_Results}
\end{figure}

The noise vanishes almost quadratically faster in $N$ than with the couplings known in closed form (see Fig.~\ref{fig:Fig2_Results}b). 
The numerical solution corresponds to constant couplings in the bulk of the chain, while they increase towards the boundaries as shown in Fig.~\ref{fig:Fig1_Setup}d.
This boundary-matching of the couplings is known as \textit{apodization}~\cite{Sumetsky2003,Chak2006}, and the length of the apodized region is asymptotically independent of the chain length.

\section{Exponential precision gain}
Using the optimized couplings which minimize $D/J$, we numerically simulate the clock's ticks using stochastic Monte Carlo simulations of the chain (details Appendix~\ref{SM:cov_matrix_evolution}).
Already for finite $N$, we recover the log-scaling of $\Var[T_n]$, as shown in Fig.~\ref{fig:Fig2_Results}c.
To analytically derive the log-scaling, we go to the limit of extremely long chains, $N\rightarrow\infty$, where the noise vanishes $D/J\rightarrow 0$ due to apodized boundary.

Because the width of this region stays constant even as $N$ grows, the bulk Hamiltonian dominates in that limit (see Fig.~\ref{fig:Fig1_Setup}d), and is thus well approximated by a constant-coupling $g$, nearest-neighbor interaction with boundary conditions chosen periodically for convenience.
The real space operators $c_j$ can then be mapped to their momentum space counterpart $e_{\kappa}$ using the Fourier transform $c_j = N^{-1/2} \sum_{\kappa \in [-\pi,\pi)} e^{i\kappa j} e_\kappa$, with discretized lattice momentum, $\kappa_m=2\pi m/N-\pi$.
In the bulk, the momentum operators diagonalize the Hamiltonian with dispersion relation $E(\kappa)=-2g \cos(\kappa)$.

Due to the perfect transmission, the momentum space occupation $p(\kappa)=\mathrm{tr}[e_\kappa^\dagger e_\kappa\rho_{\rm ss}]$ of the stationary state $\rho_{\rm ss}$ vanishes for negative momenta, i.e., reflected modes.
Together with a symmetry argument, is is possible to say even more about the stationary state.
Since the evolution equation~\eqref{eq:dotRho} is symmetric under simultaneously flipping left and right (parity) and exchanging excitations with holes (time reversal), $\rho_{\rm ss}$ shares the same symmetry.
In momentum space, this means $p(\kappa)=1-p(-\kappa)$, but since $p(\kappa<0)=0$, this implies that $p(\kappa>0)=1$.
In the limit of $N\rightarrow\infty$, the stationary state represented in momentum space is thus given by a \textit{shifted Fermi-sea}, $\vert{\rm FS}^+\rangle = \prod_{\kappa\in [0,\pi)} e^\dagger_\kappa \vert 0 \rangle$, where $\ket{0}$ denotes the fermionic vacuum.

The shifted Fermi-sea allows us to calculate the average current exactly.
Current conservation in the steady-state regime enforces that the outgoing current is equal to the bulk current at large times: $J =  \langle j_{\ell}\rangle_{\rm ss} $, with the bulk current operator $j_{\ell}=-ig_{\ell}(c_{\ell}^\dagger c_{\ell+1} - c_{\ell+1}^\dagger c_{\ell})$ and $\ell$ far from the boundaries.
Inserting $\ket{\rm FS^+}$ gives the exact result:
\begin{align} 
J = \frac{2g}{\pi}.
\end{align}

Next, we determine the current-current correlations of the ticks.
While this quantity is still difficult to compute analytically in full generality, the perfect transmission provides a useful simplification:
fluctuations of the local current $j_{\ell}(t)$ are proportional to those of the tick current $I(t)$, up to a small constant offset originating from the delay between crossing site $\ell$ and exiting the chain.
Invoking standard quantum optics methods~\cite{Wiseman2009}, we compute the correlation function: $\E[I(t)I(0)]\sim \tr[j_{\ell}e^{\mathcal{L}t}(j_{\ell}\rho_{\rm ss})]$ with steady state $\rho_{\rm ss}$.
Deep in the bulk, lead effects are negligible, so we approximate dissipative evolution by $\mathcal{L}\rho \approx -i[H,\rho]$ and the steady-state using the shifted Fermi-sea.
With this, we obtain $\E[I(t)I(0)] \sim \langle j_{\ell}(t)j_{\ell}(0)\rangle_{\mathrm{FS}^+}$ and substituting into Eq.~\eqref{eq:VarNt_classical_main}, the resulting integral evaluates in closed form in terms of Bessel functions and sine integrals.
To leading order in $t$, the expressions simplify and yield (Appendix~\ref{SM:analytical_derivation_variance}):
\begin{align}
\label{eq:VarNt_main}
    \Var[N_t] \sim \log(Jt).
\end{align}
With the relation~\eqref{eq:VarTn_sim_VarNt} we recover the main result, Eq.~\eqref{eq:VarTn}.

The logarithmic scaling in Eq.~\eqref{eq:VarNt_main} is reminiscent of the logarithmic scaling of the level spacing variance in random matrix theory due to spectral rigidity~\cite{Mehta1967,French1978}.
For fermion transport in a continuum one dimensional medium without lattice effects, tick times follow analytically exactly the universal Sine-kernel distribution of eigenvalues in the spectral bulk of matrices sampled from a unitary ensemble with Dyson index $\beta=2$ (additional details in Appendix~\ref{SM:Connection_RMT}).
In fermionic systems, such random matrix behavior is ubiquitous due to the determinantal state structure coming from the Pauli exclusion principle known as Slater determinant~\cite{Blanter2000,Dean2019,Dixmerias2025}.

\section{Robustness \& experimental prospects}

The analytically derived logarithmic variance scaling rests on the idealized limit of $N \rightarrow \infty$ and perfect transmission. 
In realistic settings, imperfections like thermal effects, moderate disorder or finite size inevitably introduce noise leading to non-zero diffusion $D$.
Still, the clock’s exponential precision advantage proves remarkably robust and persists up to a well-defined cross-over time $t^*\sim 1/D$ (or equivalently tick number $n^* = J/D$ ticks), beyond which ordinary diffusive noise takes over. 
In finite chains with our optimized couplings and where $D/J$ follows Eq.~\eqref{eq:D/J_scaling}, this implies $t^* \sim N^{1.86}/J$, indicating that exponential precision persists for thousands of ticks even in systems of moderate size. 
Numerical simulations confirm this (Fig.~\ref{fig:Fig2_Results}c): 
while a ten-site chain already shows clear logarithmic behavior, an eighty-site chain maintains it for almost $10^4$ ticks before entering the diffusive regime.

Circumventing the spin chain, a direct physical realization could be achieved with a fermionic chain coupled to electronic reservoirs, at a finite inverse temperature $\beta$.
When the chain’s on-site energies lie deep within the transport window set by the chemical potentials $\mu_{L(R)}$ of the left (right) reservoir, $\mu_L \gg \omega \gg \mu_R$, and the leads are near zero temperature, excitations flow unidirectionally as described by Eq.~\eqref{eq:dotRho}. 
Thermodynamically, this regime corresponds to a large entropy dissipation per transported excitation, $\Sigma = \beta(\mu_L - \omega)$, and the associated cross-over time grows exponentially, $t^* \sim J^{-1}\Sigma e^{\Sigma}$ (Appendix~\ref{SM:robustness_realistic_implementation}).
In contrast, in the equilibrium or small-bias regime described in~\cite{Levitov1996,Schoenhammer2007}, the cross-over time scales only as the Planckian time $t^* \sim \beta$.
Our setup therefore extends the logarithmic variance regime to times that are exponentially larger than this Planckian limit by operating in a strongly driven, high-entropy-dissipation regime.
Finally, static disorder in on-site energies $\omega$ or couplings $g_i$ leads to Anderson localization~\cite{Anderson1958}, yet even here the clock remains robust: the cross-over time increases quadratically with vanishing disorder, $t^* \sim J^{-1}/\varepsilon^2$, with $\varepsilon$ the relative disorder strength. 
An extended discussion including theoretical derivation is provided in Appendix~\ref{SM:robustness_realistic_implementation}.

\section{Outlook}
Looking at the broader picture, our work introduces a fundamentally new route to precision in timekeeping.
Instead of making each tick more sharply peaked in time, stability emerges collectively through correlations between ticks. 
In existing quantum-clock architectures, implementing memory-based or feedback-like stabilization has so far been considered infeasible without external control loops and the associated thermodynamic overhead.
That such self-correcting behavior can instead arise from intrinsic quantum correlations alone represents a conceptual shift, revealing that genuine autonomy can sustain the stabilization mechanisms traditionally attributed to engineered feedback

\begin{center}
\vspace{3.25ex plus 1ex minus 0.2ex}
\textbf{\small ACKNOWLEDGMENTS}
\vspace{1.5ex plus 0.2ex}
\end{center}

The authors thank Mark~T.~Mitchison, Bill~Munro, Kae~Nemoto, Keiji~Saito, Tan~Van~Vu, Kjeld~Beeks and Kai~Brandner for discussions.
The computational results presented were obtained using the \href{https://clip.science}{CLIP cluster}.
FM and MH acknowledge funding from the European Research Council Consolidator grant ‘Cocoquest’ 101043705.
GB and GH acknowledge support from the National Center of Competence and Research SwissMAP.
YH acknowledges funding from FWF grant ISOQUANT I6949.
This project is co-funded by the European Union, Quantum Flagship project ASPECTS, Grant Agreement No.\ 101080167 (FM and MH). 
Views and opinions expressed are however those of the authors only and do not necessarily reflect those of the European Union, REA or UKRI. Neither the European Union nor UKRI can be held responsible for them.

\paragraph*{Author Contributions.}
FM conceived the project with inputs from GB, GH and MH.
FM developed the theory and performed the numerical simulations with guidance of YM.
GB and GH gave theoretical inputs on quantum transport.
GB derived the connection between master equation and Landauer--Büttiker formalism.
FM wrote the initial draft of the manuscript.
All authors contributed to the discussions and finalization of the manuscript.
MH supervised the project.

\paragraph*{Competing interests.}
The authors declare that they have a pending patent related to this work (Patent application no.~A55513/2025).

\begin{center}
\vspace{3.25ex plus 1ex minus 0.2ex}
\textbf{\small DATA AVAILABILITY}
\vspace{1.5ex plus 0.2ex}
\end{center}

All simulation data is publicly available on GitHub at \url{https://github.com/FromageDeGruyere/fermion-clock}.

\appendix
\begin{center}
\vspace{3.25ex plus 1ex minus 0.2ex}
\textbf{\small APPENDICES}
\vspace{1.5ex plus 0.2ex}
\end{center}

\setcounter{secnumdepth}{3}
\listofappendices

\begin{appendices}

\section{\label{SM:Stochastic master equation and current}Stochastic evolution equation and current}
In this appendix we remind the reader of the connection between Lindblad master equation~\eqref{eq:dotRho} and measured stochastic current $I(t)$ giving rise to the ticks of the clock.

The master equation evolution from the main text can be unraveled into pure-state stochastic trajectories of when which jump occurred.
While the unraveling of a master equation is generally non-unique, we here work with the photocurrent or unraveling where ticks can be naturally defined: counting variables $N_{R(L)}(t)$ count the number of jumps on the right (left) through discrete increments $\dd N_{R(L)}(t)=0,1$.
With the time increment $\dd t$, the currents are defined by,
\begin{align}
\label{eq:I_def_Stochastic}
    I_{R(L)}(t) = \frac{\dd N_{R(L)}(t)}{\dd t}.
\end{align}
In the main text, we identified $I(t)\equiv I_R(t)$, and mathematically, currents are distributions,
\begin{align}
    I_{R(L)}(t) = \sum_{i\geq 1} \delta(t-t_i^{R(L)}),
\end{align}
which is a sum of Dirac-$\delta$ peaks at the times $t_i^{R(L)}$ when jumps occurs.
Thus, the counting variables are of the form
\begin{align}
    N_{R(L)}(t) = \sum_{i\geq 1}\Theta(t-t_i^{R(L)}),
\end{align}
with the Heaviside $\Theta$ and which is like sketched in Fig.~\ref{fig:Fig1_Setup}f.
The total number of ticks is then recovered as $N_t \equiv N_R(t)$.
The discrete increments satisfy the relations (1) $\dd N_{R(L)}(t)^2 = \dd N_{R(L)}(t)$, which says that the increments can only be $0$ or $+1$, (2) $\dd N_R(t) \,\dd N_L(t) = 0$ stating that no two jumps occur simultaneously, and (3) the equation
\begin{align}
    \E[\dd N_{R(L)}(t)] = \dd t \, \tr[ J_{R(L)} \ketbra{\psi_t}{\psi_t} J_{R(L)}^\dagger ],
\end{align}
setting the expectation value of the increment with respect to the state $\ket{\psi_t}$ at time $t$.

The state $\ket{\psi^c_t}$ conditioned on such a trajectory defined by $I_{R(L)}(t)$ then follows the stochastic Schr\"odinger equation (SSE)~\cite{Wiseman2009},
\begin{align}
    \!\dd \ket{\psi_t}_c &= \dd t \left(\!-iH - \frac{1}{2}\hspace{-0.25cm} \sum_{x=\{R,L\}}\hspace{-0.25cm}\left(J_x^\dagger J_x - \langle \psi_t^c | J_x^\dagger J_x |\psi_t \rangle_c \right)\!\right)\! \ket{\psi_t}_c \nonumber \\
    &\quad + \sum_{x\in\{R,L\}} \hspace{-0.25cm} \dd N_x(t) \, \left(\frac{J_x}{\langle \psi_t^c | J_x^\dagger J_x |\psi_t^c\rangle} - 1\right)\ket{\psi_t}_c.
    \label{eq:dPsi_SSE}
\end{align}
The first line of Eq.~\eqref{eq:dPsi_SSE} is the evolution conditioned on no jump occurring whereas the second line accounts for the jump processes.
The SSE can be readily expanded to the finite-temperature case by adding the backwards jumps discussed in the Methods.
For the generalization to mixed states using density matrices, we refer the reader to pertinent literature~\cite{Wiseman2009}.

Averaging over all stochastic trajectories, denoted by the expectation value $\E[\,\cdot\,]$, the state $\rho(t)=e^{\mathcal L t}\ketbra{\psi_0}{\psi_0}$ from the master equation is recovered, 
\begin{equation}
    \rho(t) = \E[\ketbra{\psi_t}{\psi_t}_c].
\end{equation}

\section{\label{SM:calculating_waiting_time_variance}Relating number variance to tick time variance}

In this appendix we derive how the variance in $T_n$, the time between $n+1$ successive ticks, is related to the variance of the total number of ticks $N_t$ up to time $t$. 
The essence of the derivation is sketched in Fig.~\ref{fig:Fig1_Setup}f.

In the limit of large numbers of ticks $n\gg 1$ (or alternatively long times $t\gg 1$), we would expect that the variation in the tick times 
\begin{align}
    \delta T_n = T_n - \E[T_n],
\end{align}
is proportional that of of the tick number
\begin{align}
    \delta N_{t} = N_t - \E[N_t],    
\end{align}
up to possibly some constant correction and for $t = \E[T_n] = n/J = n\pi / (2g)$.
Fixing the units, we expect that
\begin{align}
\label{eq:deltaT_n_heuristic}
    \delta T_n = -\frac{1}{J}\delta N_{\E[T_n]} + \mathcal O(1),
\end{align}
where the negative sign comes from if $\delta T_n > 0$, there are less than the average ticks in the interval until $\E[T_n]$, that is $\delta N_{\E[T_n]}<0$.
From this heuristic, and together with $\Var[N_t]\sim \log(t)$, we can then conclude that (using $J=2g/\pi$),
\begin{align}
    \Var[T_n] \sim \log(n),
\end{align}
to leading order.

In the following, we make Eq.~\eqref{eq:deltaT_n_heuristic} more rigorous by showing on which assumptions this asymptotic equality depends and how the constant order correction can be derived.
For the derivation we rely on random matrix literature~\cite{French1978,Brody1981} where such a relationship is derived to compare level spacing variance to the number variance.

For the sake of finiteness, we consider a related setting where the clock starts at time $t=0$ and we wish to look at much later times.
To that end, we label the tick times (random variables) by
\begin{align}
    0\leq t_0\leq t_1 \leq t_2 \leq \dots,
\end{align}
and we may add a finite cut-off after $M$ ticks such that there at most $M$ ticks.
It is important to not confuse the random variable $t_k$ with $T_n$, where the latter is the average time between any $n+1$ ticks in the stationary state.
That is, $T_n = t_{k+n}-t_k$ for some sufficiently large $k$ to ensure stationarity, in which case the precise choice of $k$ (as long as it is sufficiently large) is irrelevant.
Further, also the time fluctuations $\delta t_k = t_k - \overline t_k$, where $\overline t_k$ is the expected time of the $k$th tick can be defined.

To obtain the tick number from the tick times, the tick current $I(t)$ can be explicitly written as
\begin{align}
    I^{\rm d}(t) 
    &= \sum_{k=0}^M \delta(t - t_k) \\
    &= \sum_{k=0}^M \delta(t - \overline t_k - \delta t_k).
\end{align}
Note that we added a superscript `d' to highlight that the ticks are discrete, as we later work with a continuous approximation.
Integration yields the number of ticks,
\begin{align}
    \int_0^\tau \dd t I^{\rm d}(t) \equiv N_{[0,t]}^{\rm d},
\end{align}
The fluctuation of the number of ticks an interval of length $t$ (as considered in the previous appendices) in the stationary regime is then given by
\begin{align}
    N_{[\tau,\tau+t]}^{\rm d} &= N_{[0,\tau+t]}^{\rm d} - N_{[0,\tau]}^{\rm d},
\end{align}
for sufficiently large $\tau$.

To relate the fluctuations of the tick number to that of the tick time, we work with a smoothened version of the tick number analogous to~\cite{French1978}.
The reason is that while $N_{[0,\tau]}^{\rm d}$ can only change by in unit increments, the tick times $t_k$ are continuous numbers.

One way to go about this is to generalize the fluctuations $\delta t_k$ to a continuous family $\delta t (x)$ where $x\in \mathbb R_+$ and for $x=k\in \mathbb N$,
\begin{align}
    \delta t(k) \equiv \delta t_k.
\end{align}
Moreover, since $\overline t_k = k/J$, we further have $\overline t(x) = x/J$.
Then, we can define the continuous tick current
\begin{align}
    I^{\rm c}(\tau) &= \int_0^M \dd x' \delta(\tau-\overline  t(x') -\delta t(x')) \\
    &=1 - J\left(\partial_x \delta t\right)(\tau / J) + \mathcal O(J^2\delta t^2),\label{eq:rho^c(x)_linear}
\end{align}
where for the second equality we used the Dirac-$\delta$ identity that for a sufficiently well-behaved function $f(x)$ with a unique root $f(x^*)=0$ stating $\int \dd x \delta(f(x)) = 1/|f'(x')|$.
Furthermore, higher order correlations in $\delta t$ are discarded, so Eq.~\eqref{eq:rho^c(x)_linear} can be thought of as a linearized identity as a function of the fluctuations.
By integration from $0$ to $\tau$ (and setting $x_\tau=J\tau$), we find
\begin{align}
    \delta t(x_\tau) &= - \frac{1}{J}\left(N_{[0,\tau]}^{\rm c} - \overline{N_{[0,\tau]}^{\rm c}}\right) \\
    &= - \delta N_{[0,\tau]}^{\rm c},
\end{align}
and going to the integer values $x_k = k$, we can conclude
\begin{align}
    \delta t_k = -\frac{1}{J}\delta N_{[0,k/J]}^{\rm c}.
\end{align}
Since the continuous and discrete number $N^{\rm c / d}_{[0,\tau]}$ only differ by some constant of order $\mathcal O(1)$, this concludes the justification of the relationship in Eq.~\eqref{eq:deltaT_n_heuristic}.

\section{\label{SM:free_fermion_model}Free fermion formulation}
In this section, we remind the reader how the the spin chain Hamiltonian and the jump operators are mapped to free fermions via the Jordan--Wigner transformation (Appendix~\ref{SM:jordan_wigner_transformation}).
Then, we argue how the clock's state is described in terms of a Gaussian fermionic state (Appendix~\ref{SM:gaussian_fermionic_state}) which will be crucial below.

\subsection{\label{SM:jordan_wigner_transformation}Jordan--Wigner transformation}
The model described in the main text and formulated in terms of the spin-$\frac{1}{2}$ Pauli operators which satisfy the algebra $[\sigma_i^+,\sigma^-_j] = \delta_{ij} \sigma_j^z$.
Using the Jordan--Wigner transformation as detailed in~\cite{Lieb1961}, it is possible to transform our model into an equivalent one using (Dirac) fermions.
At the center of this mapping is the identity
\begin{equation}\label{eq:JWS}
    c_j = \underbrace{\exp\left(-i\pi \sum_{i=1}^{j-1} \sigma^+_i \sigma^-_j \right)}_{ =\mathrm{JWS}} \sigma_j^-.
\end{equation}
The operator $\mathrm{JWS}$ known as \textit{Jordan--Wigner string} is crucial for the nonlocal nature of $c_j$.
From the spin-$\frac{1}{2}$ algebra it follows that operator $c_j$ indeed behave like a Dirac fermion with anticommutation relations.
\begin{align}
    \{c_i,c_j^\dagger\}=\delta_{ij},\quad \{c_i^\dagger,c_j^\dagger\}=0.
\end{align}
Note that here the fermion creation operators $c_j^\dagger$ are defined through hermitian conjugation.
In order to map the spin-$\frac{1}{2}$ Hamiltonian in Eq.~\eqref{eq:H} to a model of free fermions we invert the identity in Eq.~\eqref{eq:JWS} which yields 
\begin{align}
\label{eq:c_j_trafo}
    \sigma_j^- = \underbrace{\exp\left(+i\pi \sum_{i=0}^{j-1}c_i^\dagger c_i\right)}_{=:\Omega_j^\dagger} c_j .
\end{align}
Note that here we defined the Jordan--Wigner string in terms of fermions.
Inserting this in Eq.~\eqref{eq:H} we obtain
\begin{align}
\label{eq:H_fermionic}
    H = \underbrace{\sum_{i=1}^N \omega c_i^\dagger c_i}_{=H_0} - \underbrace{\sum_{i=1}^{N-1} g_i \left(c_i^\dagger c_{i+1} + c_{i+1}^\dagger c_i\right)}_{=H_1}.
\end{align}
The diagonal terms $\omega c_i^\dagger c_i$ can be dropped by going to the interaction picture and noting that the additional time-dependent phases cancel out in the quadratic expressions $c_j^\dagger c_i$ because all sites have the same energy $\omega$.
For simplicity, we thus describe our model using $H_1$ which, with a slight abuse of notation, we will from now on refer to using the symbol $H$.

Let us now consider the jump operators as defined in the main text.
In order to also model the effect of a thermal bath we consider both gain and loss processes.
On the left terminal we get
\begin{align}
\label{eq:J_L}
    J_{L,\rm in} = \sqrt{\Gamma f_L} c_1^\dagger,\quad J_{L,\rm out} = \sqrt{\Gamma (1-f_L)} c_1,
\end{align}
since $\sigma_1^+=c_1^\dagger$ and $\sigma_1^- = c_1$, and with the fermi function $f_L$ of the left lead.
We further note that the phase from the interaction picture acts as a global phase on the state and can thus be safely ignored.
As for the jump operators on the right end of the chain, we pick up a non-trivial operator from the Jordan--Wigner transformation,
\begin{align}
\label{eq:J_R}
    J_{R,\rm out} = \sqrt{\Gamma (1-f_R)} \Omega_N^\dagger c_N,\quad J_{R,\rm in} = \sqrt{\Gamma f_R} \Omega_N c_N^\dagger.
\end{align}
In the limit of a fully biased transport, the above expressions simplify and we have $J_L^{\rm in}=\sqrt{\Gamma}c_1^\dagger$ and $J_L^{\rm out}=0$, as well as $J_R^{\rm out} = \sqrt{\Gamma} \Omega_N^\dagger c_N$ and $J_R^{\rm in}=0$.
For the symmetric setup we consider, the Fermi-functions can be written as
\begin{align}
    f_L = 1-f_R = \frac{1}{1+e^{-\Sigma}},
\end{align}
as well as
\begin{align}
    1-f_L = f_R = \frac{1}{1+e^\Sigma}.
\end{align}
Here, due to symmetry, the entropy factor is equal for left and right $\Sigma_L = \Sigma_R \equiv \Sigma$ where $\Sigma_L = \beta(\mu_L-\omega)$ and $\Sigma_R=\beta(\omega-\mu_R)$.
Further, we can fix units, without loss of generality, such that
\begin{align}
    \Gamma\equiv 1,
\end{align}
and we abbreviate $f\equiv f_L$, and $\bar f = 1-f$.

\subsection{\label{SM:gaussian_fermionic_state}Gaussian fermionic state}
In general, a state defined on the $N$-fermion Hilbertspace requires a $2^N$ dimensional complex representation.
A particular class of states, \textit{Gaussian fermionic states}, however, allows a simplified description.
In our case, we use a particular special case of Gaussian fermionic states that are of Slater determinant form.
Such a state has well defined excitation number $0\leq M \leq N$ and can be written as~\cite{Surace2022},
\begin{align}
\label{eq:psi_slater}
    \ket{\psi} &= \prod_{i=1}^M \left(\sum_{j=1}^N U_{ji} c_j^\dagger\right)\ket{0},
\end{align}
where $\ket{0}$ is the vacuum state with respect to the operator algebra $\{c_i\}_i$.
Moreover, $U \in \mathbb{C}^{N\times M}$ is an isometry, i.e., it's column vectors are orthonormal with respect to the standard complex scalar product,
\begin{align}
    \sum_{j=1}^M U_{jk}^*U_{j\ell} = \delta_{k\ell}.
\end{align}
In fact, the transformation $U$ defines a new set of fermionic modes $e_1^\dagger,\dots,e_M^\dagger$ (satisfying $\{e_i^\dagger,e_j\}=\delta_{ij}$ and $\{e_i,e_j\}=0$), where $e_i^\dagger = \sum_{j=1}^N U_{ji}c_j^\dagger$, which allow us to write $\ket{\psi}=e_M^\dagger\cdots e_1^\dagger \ket{0}$.

For stochastic evolution following Eq.~\eqref{eq:dPsi_SSE} from Appendix~\ref{SM:Stochastic master equation and current}, we may assume that we initially start in the vacuum state and as time progresses the chain is gradually filled with excitations.
The initial state is thus (trivially) of Slater determinant form.
Evolution with the following operations:
\begin{enumerate}
    \item Evolution with a quadratic (even non-Hermitian) Hamiltonian,
\begin{align}
    H_{\rm eff} &= H - \frac{i}{2}\left( f c_1 c_1^\dagger + \bar f c_1^\dagger c_1 + f c_N^\dagger c_N + \bar f c_N c_N^\dagger\right) \nonumber \\
    &=H - \frac{i}{2}\left((\bar f - f)c_1^\dagger c_1 + (f-\bar f)c_N^\dagger c_N\right) - i\mathds 1,
    \label{eq:H_eff_fermionic}
\end{align}
    \item Mode creation with an operator $c_j^\dagger$,
    \item Mode annihilation with an operator $c_j$,
\end{enumerate}
preserves the Slater determinant state structure~\cite{Surace2022,Bravyi2005,Bettmann2024}, and thus, if the initial state is of the the type in Eq.~\eqref{eq:psi_slater} also the state at any later time is.

Any state $\ket{\psi}$ of Slater determinant form is also a fermionic Gaussian state.
Therefore, it satisfies the \textit{Wick-theorem}~\cite{Surace2022}.
That is, for any linear combinations $A,B,C,D$ of annihilation operators, which means $A=\sum_{i=1}^N A_i c_i$, we have
\begin{align}
    \label{eq:fermionic_wick}
    \langle A^\dagger B^\dagger C D \rangle = \langle A^\dagger D\rangle\langle B^\dagger C\rangle - \langle A^\dagger C\rangle \langle B^\dagger D\rangle.
\end{align}
Note that here we anticipated the vanishing of the \textit{anomalous terms} $\langle A^\dagger B^\dagger\rangle \langle C D \rangle = 0$ and implicitly used $\langle\cdot\rangle \equiv \langle \psi |\cdot|\psi\rangle$ to denote the average with respect to the Gaussian state $\ket{\psi}$.
Unless there is any ambiguity, from now on, the implicit average will be used.

\section{\label{SM:cov_matrix_evolution}Covariance matrix evolution}
In this section, we introduce the fermionic covariance matrix and derive its equations of evolution.
In Appendix~\ref{SM:stochastic_cov_evolution} we derive the conditioned dynamics or the jump-unraveled evolution. 
This is the framework in which the numerics are analyzed. 
In Appendix~\ref{SM:unconditional_cvm_evolution} we derive the Lyapunov equation which corresponds to the Lindblad Eq.~\eqref{eq:dotRho} in terms of the fermionic covariance matrices. 
This is crucial for the analytic derivation of the main result in Appendix~\ref{SM:analytical_derivation_variance}.

\subsection{\label{SM:stochastic_cov_evolution}Stochastic covariance evolution}
The main reason for considering the dynamics in terms of fermion covariance matrices is that it avoids simulations in the exponentially large spin Hilbert space ($2^N$-dimensional) in favor of $N^2$-dimensional state space of $N\times N$ dimensional covariance matrices.
For a state of Slater determinant form, or more generally for fermionic Gaussian states, the dynamics conditioned on the ticks is fully characterized in terms of a stochastic evolution of the covariance matrix ~\cite{Cao19,Alberton21,Bettmann2024}.

Since our underlying model is a spin-$\frac{1}{2}$-system the jump operator on site $N$ is decorated with the nonlocal Jordan--Wigner string operator [cf.~Eq.~\eqref{eq:J_R}].
Note that due to our choice of anchoring the Jordan--Wigner transformation in left of the chain the Jordan--Wigner string operators is not present for the jump operators in site $i=0$.
Following \cite{Prosen_2008,Yamanaka21} we explicitly show below that the effect of this non-local jump operator is greatly simplified. 

The in general the covariance matrix $C\in \mathbb C^{N\times N}$ of a state $\ket{\psi}$ can be defined as
\begin{align}
    C_{ij} = \langle \psi |c_i^\dagger c_j |\psi\rangle.
\end{align}
Note the convention of the indices $C_{ij}$ which is chosen differently from~\cite{Bettmann2024}.
The fact that $\ket{\psi}$ is a Slater determinant (or fermionic Gaussian state) plays a critical role in allowing us to write down an evolution equation for $C$, due to Wick's theorem~\eqref{eq:fermionic_wick}.

Some properties of the covariance matrix that will prove to be useful are:
\begin{itemize}
    \item Hermiticity, that is $C^\dagger = C$ or in index form $C_{ji}^* = C_{ij}$.
    \item Projector, i.e., $C^2 = C$, due to the Wick theorem 
    \item Singular value decomposition (SVD), $C = V \Sigma V^\dagger$, where the first $M$ columns $V^T = [U, \tilde U]$ equal the isometry $U$ that defines $\ket{\psi}$, and the singular values $\Sigma_{ii} = 1$ for $i=1,\dots,M$ and $\Sigma_{ii}=0$ for $i\geq M+1$.
\end{itemize}

The first property follows immediately from the definition.
For the second property, we note that $ C_{ik}C_{kj} = \langle e_i^\dagger e_j\rangle \langle e_k^\dagger e_k\rangle - \langle e_i^\dagger e_k^\dagger e_k e_j\rangle =C_{ij} M - (M-1)C_{ij}$ using implicit summation over $k=1,\dots,N$.
As for the last item, we note that we can extend the isometry $U\in \mathbb C^{N\times M}$ to a unitary $\mathcal U \in {\rm U}(N)$, with $\mathcal U = [U , \tilde U]$ using for example the Gram-Schmidt algorithm.
We can then extend the modes $e_1,\dots,e_M$ that define $\ket{\psi}=e_M^\dagger \cdots e_1^\dagger \ket{0}$ to modes $e_1,\dots,e_N$ by setting $e_i = \sum_{j=1}^N \mathcal U_{ji} c_j$.
Performing the inverse transformation yields the original modes, $c_i = \sum_{j=1}^N \mathcal U^*_{ij} e_j$, which we can insert into the expression for the covariance matrix to give
\begin{align}
    C_{ij} &= \sum_{k,\ell = 1}^N \mathcal U_{ki} \mathcal U^*_{\ell j}\langle 0| e_1 \cdots e_M  e_k^\dagger e_\ell  e_M^\dagger \cdots e_1 | 0 \rangle \\
    &=\left( \mathcal U^T \Sigma \mathcal U^*\right)_{ij},
\end{align}
where the singular values $\Sigma_{ii} = 1$ for $i=1,\dots,M$ and $\Sigma_{ii}=0$ for $i\geq M+1$.
When we set $V = \mathcal U^T$, we recover $C = V\Sigma V^\dagger$.
Finally, it shall be noted that the first $M$ columns of $V^T$ are unique, but not the remaining ones.
The numerical integration does not exactly preserve all those properties exactly, hence, we will later employ a projection technique to ensure that the numerical integration yields a consistent output.

\paragraph*{State updates.}
We work with a stochastic unravelling of the master equation as outlined in Appendix~\ref{SM:Stochastic master equation and current}.
For finite temperatures, we have four possible jumps: left in, left out, right in and right out; moreover there is the possibility of no jump.

Suppose at a given time $t$ we have state $\ket{\psi_t}$ of Slater-determinant form and with $M$ excitations.
The first possibility is that there is no jump, in which case, stochastic state update is ($t$-dependence implicit):
\begin{align}
\label{eq:psi_nojump}
    \dd |\psi^{(0)}\rangle_c = \dd t \left(-iH_{\rm eff} +\frac{1}{2} \sum_{\mu} \langle J_\mu^\dagger J_\mu \rangle \right) |\psi^{(0)}\rangle_c.
\end{align}
The sum $\sum_\mu$ goes over all four jumps $\mu = (L,{\rm in}), (L,{\rm out}),(R,{\rm in})$ and $(R,{\rm out})$.
The probability that no jump occurs within $\dd t$ can be calculated as,
\begin{align}
    P_0 &= 1 - \dd t \sum_\mu\langle J_\mu^\dagger J_\mu \rangle_c   \\
    &= 1 - \dd t\left((1-2f)C_{11} - (1-2f) C_{NN} + 1\right). \nonumber 
\end{align}

The other possibility is that one of the four jumps occurs with a state update
\begin{align}
\label{eq:psi_jump_general}
    \dd | \psi^{(\mu)}\rangle_c = \dd N_{\mu} \left(\frac{J_\mu}{\langle J_\mu^\dagger J_\mu\rangle_c} - 1 \right) \ket{\psi}_c.
\end{align}
The jump probability is given by $P_\mu = \dd t \langle J_\mu^\dagger J_\mu\rangle_c$, for the four jump operators $J_\mu$ as defined in Eqs.~\eqref{eq:J_L} and~\eqref{eq:J_R}.
Finally, we note the consistency equation $P_0 = 1 - P_{L,{\rm in}}-P_{L,{\rm out}}- P_{R,{\rm in}}-P_{R,{\rm out}}$, guaranteeing that one of the five cases occurs.

\paragraph*{Covariance matrix no jump update.}
In what follows, we write down the evolution equations for the covariance matrix, translating the prescriptions from Eqs.~\eqref{eq:psi_nojump} and~\eqref{eq:psi_jump_general} on the level of quantum states to the covariance matrix, akin to~\cite{Bettmann2024}.
That is, for any conditional evolution step on the state, we can write an equivalent evolution step in terms of the covariance matrix.
We start by considering the no-jump evolution step.
Element wise, it reads
\begin{align}
    \dd C_{ij}^{(0)} &= \big(\dd _c\langle\psi^{(0)} |\big) c_i^\dagger c_j |\psi\rangle_c + _c\langle\psi|c_i^\dagger c_j \big(\dd | \psi^{(0)} \rangle_c \big)  \\
    &= C_{ij}(1-P_0) + i\,\dd t \left \langle H_{\rm eff}^\dagger c_i^\dagger c_j  -  c_i^\dagger c_j H_{\rm eff} \right\rangle_c  \nonumber 
\end{align}
This expression can be further expanded to give:
\begin{widetext}
\begin{align}
    \dd C_{ij}^{(0)} &= C_{ij} \,\dd t\sum_\mu \langle J_\mu^\dagger J_\mu\rangle_c +  i\, \dd t \sum_{k\ell}h_{k\ell} \Big\langle \big[c_k^\dagger c_\ell, c_i^\dagger c_j\big] \Big\rangle_c -\frac{\dd t}{2}\Big\langle \big\{f c_1 c_1^\dagger + (1-f)c_1^\dagger c_1 + f c_N^\dagger c_N + (1-f) c_N c_N^\dagger, c_i^\dagger c_j\big\}\Big\rangle_c
    \label{eq:Cij_nojump_t+eps_UGLY_AF}
\end{align}
\end{widetext}
Here, we have used $H_{\rm eff}$ as in Eq.~\eqref{eq:H_eff_fermionic}, and defined $h_{k\ell}$ as the hermitian matrix elements of $H_{\rm eff}$ with respect to the operator algebra $c_k^\dagger c_\ell$.
Note that $h^T=h$ is real symmetric in our case.

The goal is now to express the increment $\dd C$ solely as a function of the covariance matrix $C$ itself.
We start by looking at the hermitian term in the commutator.
Generally, we can write,
\begin{align}
    \left\langle \left[c_k^\dagger c_\ell,c_i^\dagger c_j\right] \right\rangle_c &= \delta_{i\ell}\langle c_k^\dagger c_j\rangle_c -\delta_{kj}\langle c_i^\dagger c_\ell\rangle_c \\
    &=\delta_{i\ell} C_{kj}(t) - \delta_{kj}C_{i\ell}(t),
\end{align}
which implies (using $h^T=h$ here) that
\begin{align}
    \sum_{k\ell}h_{k\ell} \left\langle \left[c_k^\dagger c_\ell, c_i^\dagger c_j\right] \right\rangle_c &= \left(hC(t) - C(t)h\right)_{ij}.
\end{align}
This shows that the hermitian update to the covariance matrix can be written as a function of the initial covariance matrix.
What comes next is to show that also the anti-commutator terms can be expressed as functions of the covariance matrix.
To this end, we look at the generic expression
\begin{align}
    \left\langle \left\{c_k^\dagger c_k, c_i^\dagger c_j\right\} \right\rangle_c &= \langle c_k^\dagger c_k  c_i^\dagger c_j \rangle_c + \langle c_i^\dagger c_j c_k^\dagger c_k \rangle_c \nonumber \\
    &\hspace{-1.7cm}= \delta_{ik} C_{kj}(t) - \langle c_k^\dagger  c_i^\dagger c_k c_j \rangle_c + \delta_{jk} C_{ik}(t) - \langle c_i^\dagger c_k^\dagger c_j  c_k \rangle_c \nonumber \\
    &\hspace{-1.7cm}= \delta_{ik}C_{kj} + C_{ik}\delta_{kj} - 2(C_{ik}C_{kj} - C_{ij}C_{kk}), \label{eq:Ckk_Cij_anticomm}
\end{align}
where in the final line we did not explicitly write out the $t$-dependence of the covariance matrix.
The key step in expressing Eq.~\eqref{eq:Ckk_Cij_anticomm} in terms of the covariance matrix is to apply the fermionic Wick theorem discussed in Eq.~\eqref{eq:fermionic_wick}.
Without this property, the quartic terms could not have been simplified---highlighting once more the indispensable fact that our system is always in a Slater determinant state (or more generally, fermionic Gaussian state).
In the update equation~\eqref{eq:Cij_nojump_t+eps_UGLY_AF} we also have non-hermitian terms of the form $c_k c_k^\dagger$, however, since $c_k c_k^\dagger = 1-c_k^\dagger c_k$, we can reduce those terms to the expression in Eq.~\eqref{eq:Ckk_Cij_anticomm}.
In summary, we can express the covariance matrix evolution $\dot C = \dd C / \dd t$ conditioned on no jump in terms of a \textit{Riccati equation}~\cite{Brunton_19}
\begin{widetext}
\begin{align}
\label{eq:dotC}
    \dot C &= i(hC-Ch) - \frac{\bar f - f}{2}\big(\Pi_1 C +  C \Pi_1 - 2C\Pi_1 C\big) -  \frac{f - \bar f}{2}\big(\Pi_N C +  C \Pi_N - 2C\Pi_N C\big),
\end{align}
\end{widetext}
where we introduced $\Pi_k$, the projector onto the $k$th standard basis vector in $\mathbb C^N$ as follows: $(\Pi_k)_{ij} = \delta_{ik}\delta_{jk}$.
Note that we kept the \textit{no jump} superscript implicit.
We correctly recover the form of the equations of motion from Ref.~\cite{Bettmann2024}.

\paragraph{Covariance matrix left jump update.}
We now look at how the covariance matrix transforms when one of the four jumps occurs, starting with the transitions on the left end of the chain.
For a charge entering the left [Eq.~\eqref{eq:psi_jump_general} with $\mu=(L,{\rm in})$], we need to calculate
\begin{align}
\label{eq:dotC_jump1}
    \langle c_1 c_i^\dagger c_j c_1^\dagger \rangle_c &= \delta_{ij}\langle c_1 c_1^\dagger \rangle_c - \langle c_1  c_j c_i^\dagger c_1^\dagger \rangle_c \\
    &=(\delta_{i1} - C_{i1})(\delta_{1j} - C_{1j}) +  (1-C_{11})C_{ij}. \nonumber
\end{align}
Once more, the fermionic Wick theorem has been essential for going from the first to the second line.
In index free notation and including normalization by the probability that a charge enters on the left, the covariance matrix can be written as
\begin{align}
    C^{(L,{\rm in})} &= \frac{(\mathds 1 -C)\Pi_1(\mathds 1 -C) + (1-C_{11})C}{1-C_{11}}
\end{align}
Using $C^2=C$ and $\tr C = M$ we can show that the covariance matrix after the left jump is normalized, that is, $\tr C^{(L,{\rm in})}=M+1$.
Without further complications the Wick theorem can also be applied to the expression we encounter when a charge leaves on the left side [Eq.~\eqref{eq:psi_jump_general} with $\mu=(L,{\rm out})$],
\begin{align}
\label{eq:C1dagCidagCjC1}
    \langle c_1^\dagger c_i^\dagger c_j c_1\rangle_c &= C_{11}C_{ij} - C_{i1}C_{1j}.
\end{align}
As a normalized, index-free expression it reads, 
\begin{align}
    C^{(L,{\rm out})} &= \frac{C_{11} C - C\Pi_1 C}{C_{11}}.
\end{align}

\paragraph{Covariance matrix right jump update.}
For jumps on the right, the two correlators that appear are in principle of the same form as the jumps on the left terminal.
However, now, there is an additional corrections due to the Wigner string $\Omega_N$.
We first treat the case where a charge leaves the chain on the right-most site [Eq.~\eqref{eq:psi_jump_general} with $\mu=(R,{\rm out})$].
The relevant correlator is of the form
\begin{align}
    \langle c_N^\dagger \Omega_N c_i^\dagger c_j \Omega_N^\dagger c_N\rangle_c &= (-1)^{\delta_{iN}+\delta_{jN}}\langle c_N^\dagger c_i^\dagger  \Omega_N  \Omega_N^\dagger c_j c_N\rangle_c \\
     &= (-1)^{\delta_{iN}+\delta_{jN}}\left(C_{NN}C_{ij} - C_{iN}C_{Nj}\right),\label{eq:dotC_jumpN}
\end{align}
where we used that $c_i^\dagger\Omega_N = (-1)^{1-\delta_{iN}}\Omega_N c_i^\dagger$, meaning that for all $i<N$, $c_i^\dagger$ and $\Omega_N$ anticommute, whereas for $i=N$ the two operators commute.
We can introduce the matrix
\begin{align}
\label{eq:Omega_def}
    \Omega = \begin{bmatrix}
        1 & 0 & \cdots & 0 \\
        0 & \ddots & & \vdots \\
        \vdots &  & 1 & 0 \\
        0 & \cdots & 0 & -1
    \end{bmatrix},
\end{align}
such that in index-free notation, we have
\begin{align}
    C^{(R,{\rm out})} = \frac{\Omega (C_{NN}C - C\Pi_N C) \Omega}{C_{NN}}.
\end{align}
We similarly find for a charge entering on the right most site [Eq.~\eqref{eq:psi_jump_general} with $\mu=(R,{\rm in})$],
\begin{align}
    C^{(R,{\rm in})} &= \frac{\Omega\big((\mathds 1 -C)\Pi_N(\mathds 1 -C) + (1-C_{NN})C\big)\Omega}{1-C_{NN}},
\end{align}
again with the sign-correction coming from the Jordan--Wigner string.

\paragraph{Numerical method of integration.}
The prescription in the preceding paragraphs can be used for a stochastic evolution of the covariance matrix to obtain samples for the waiting times of the clock ticks.
We note that numerical integration of $C$ following the differential equation~\eqref{eq:dotC} for the no-jump evolution does not exactly preserve the singular value structure of $C$.
Hence, in the actual algorithm when $C$ is integrated, a singular value decomposition is performed after several integration steps, and the numerically attained singular values $\tilde\Sigma \mapsto \Sigma = [1_1,\dots,1_M,0_{M+1},\dots]$ are rounded to the physically ones consistent with the Slater determinant state.
This ensures the stability of the numerical integration over long times, as is used for this work where up to $2\times 10^4$ ticks were simulated per trajectory.

\subsection{\label{SM:unconditional_cvm_evolution}Unconditional covariance matrix evolution}
Here we show how the dynamics given by the master equation~\eqref{eq:dotRho} of the main text is represented in terms of an equation of motion for the covariance matrix.
The covariance matrix at time $t$ can thus be written as,
\begin{align}
    C_{ij} &= \langle c_i ^\dagger c_j \rangle =\tr\left[c_i^\dagger c_j \rho(t) \right],
\end{align}
where $\rho(t)=e^{\mathcal L t}\rho(0)$ is the solution of the Lindblad equation.
We stress that the expectation values $\langle O \rangle = \tr[O \rho] = \E[\tr(O \vert \psi\rangle_c \langle \psi\vert)]$ are with respect to the master equation or the averaged stochastic evolution. 

In the limit of fully biased transport, i.e., $\Sigma=\infty$, the Lindblad operator is given by
\begin{align}
\label{eq:Lind_fermionic}
    \mathcal L \rho &= -i[H,\rho] + c_1^\dagger \rho c_1 + \Omega_N^\dagger c_N \rho c_N^\dagger \Omega_N \nonumber  \\
    &\qquad - \frac{1}{2}\left\{c_1c_1^\dagger + c_N^\dagger c_N,\rho\right\}, 
\end{align}
where we remind ourselves of the operator $\Omega_N$ from the Jordan--Wigner transformation.
The time derivative of $C$ is
\begin{align}
    \dot C_{ij} &= \tr\left[c_i^\dagger c_j \mathcal L \rho (t)\right],
\end{align}
and can be simplified using cyclicity of the trace,
\begin{align}
    \dot C_{ij} &= \bigg\langle+i[H,c_i^\dagger c_j] + c_1 c_i^\dagger c_j c_1^\dagger + c_N^\dagger \Omega_N c_i^\dagger c_j \Omega_N^\dagger c_N \nonumber \\
    &\hspace{2cm} -\frac{1}{2}\left\{c_1c_1^\dagger + c_N^\dagger c_N, c_i^\dagger c_j\right\}\bigg\rangle_{\rho(t)}.
\end{align}
For a real symmetric Hamiltonian as in our case, we find, that the first term gives
\begin{align}
\label{eq:dotC_Hamiltonian}
    i\langle[H,c_i^\dagger c_j] \rangle &= i[hC - Ch]_{ij},
\end{align}
where $h$ are the matrix elements as already used in Eq.~\eqref{eq:Cij_nojump_t+eps_UGLY_AF}.
The jump terms are as in Appendix~\ref{SM:stochastic_cov_evolution}: for an excitation entering on the left, $\langle c_1 c_i^\dagger c_j c_1^\dagger \rangle$ is given in Eq.~\eqref{eq:dotC_jump1}, for one leaving on the right, $\langle c_N^\dagger \Omega_N c_i^\dagger c_j \Omega_N c_N \rangle$ is given by Eq.~\eqref{eq:dotC_jumpN}.
As for the anticommutators with site $1$:
\begin{align}
    \langle \{c_1 c_1^\dagger, c_i^\dagger c_j\} \rangle 
    &\!=\!\left[2(C+C\Pi_1 C - C_{11} C)-\Pi_1 C - C\Pi_1\right]_{ij}\!.
\label{eq:dotC_backaction1}
\end{align}
For site $N$, we get:
    \begin{align}
        \langle \{c_N^\dagger c_N, c_i^\dagger c_j\} \rangle  
        &\!=\!\left[2(C_{NN} C - C\Pi_N C) + \Pi_N C + C \Pi_N\right]_{ij}\!.
    \label{eq:dotC_backactionN}
\end{align}
Once all terms from Eqs.~\eqref{eq:dotC_Hamiltonian} to~\eqref{eq:dotC_backactionN} are added up, we find
\begin{align}
\label{eq:dotC_nonconditional_long}
    \dot C &= i[h,C] + \Pi_1 - \frac{1}{2}\left((\Pi_1 +\Pi_N)C + C(\Pi_1+\Pi_N)\right).
\end{align}

Notably, all contributions from the Jordan--Wigner string cancel out when we obtain Eq.~\eqref{eq:dotC_nonconditional_long}, as we argue in the following:
Using that $\Omega = \mathds 1 - 2\Pi_N$ and abbreviating $X:=C_{NN}C - C\Pi_N C$ we can write the contributions from Eq.~\eqref{eq:dotC_jumpN} and Eq.~\eqref{eq:dotC_backactionN} in index free notation as
\begin{align}
    \eqref{eq:dotC_jumpN}-\frac{1}{2}\eqref{eq:dotC_backactionN} &= (\mathds 1 - 2\Pi_N) X (\mathds 1 - 2\Pi_N) - X \nonumber \\
    &\hspace{1cm}- \frac{1}{2}(\Pi_NC + C\Pi_N) \\
    &=4\Pi_N X \Pi_N - 2(\Pi_N X- X\Pi_N) \nonumber \\
    &\hspace{1cm}- \frac{1}{2}(\Pi_NC + C\Pi_N).
\end{align}
Looking solely at the terms containing $X$, we see that both
\begin{align}
    \Pi_N X \Pi_N &=0,
\end{align}
as well as
\begin{align}
    \Pi_NX = X\Pi_N = 0.
\end{align}
Therefore, all Jordan--Wigner string contributions cancel out in the final evolution equation.

What remains is the expression from Eq.~\eqref{eq:dotC_nonconditional_long} which we can write in shorthand notation using $K=ih - \frac{1}{2}(\Pi_1 + \Pi_N)$ and in agreement with the expression from~\cite{Bettmann2024},
\begin{align}
\label{eq:dotC_nonconditional_short}
    \dot C &= KC + CK^\dagger + \Pi_1.
\end{align}
For finite bias, as one may readily verify, the expression is given by~\cite{Bettmann2024},
\begin{align}
\label{eq:dotC_nonconditional_finite_temp}
    \dot C &= KC + CK^\dagger + P,
\end{align}
where $P = f \Pi_1 + \overline f \Pi_N$.
This linear matrix equation is known as the \textit{Lyapunov equation}.

\section{\label{SM:noise_current_LB}Noise and current in the Landauer--B\"uttiker formalism}
In the Landauer--B\"uttiker (LB) framework for two-terminal coherent transport, the stationary particle current is derived using a scattering approach~\cite{Datta1995,Landauer1957,Buettiker1986}
\begin{equation}
  J_{\mathrm{LB}} \;=\; \int_{-\infty}^{\infty}\frac{\dd E}{2\pi}\;
  T(E)\,\big[f_L(E)-f_R(E)\big],
  \label{eq:LB-current}
\end{equation}
with Fermi functions \(f_{\alpha}(E)=\big(e^{\beta(E-\mu_\alpha)}+1\big)^{-1}\) for \(\alpha=L,R\), and energy-resolved transmission probability \(T(E)\).

Similarly, the zero-frequency current noise, corresponding to the diffusion constant $D\equiv S(\omega=0)$, is
\begin{align}
  \hspace{-0.1cm}D =& \int_{-\infty}^{\infty}\frac{\dd E}{2\pi}\,
  \Big\{\,T(E)\big[1-T(E)\big]\,[f_L(E)-f_R(E)]^2 \label{eq:LB-noise}\\
  &+T(E)\big[f_L(E)\big(1-f_L(E)\big)+f_R(E)\big(1-f_R(E)\big)\big]\Big\}.
  \nonumber
\end{align}
At zero temperature \(f_\alpha(E)=\Theta(\mu_\alpha-E)\), and therefore the expression for the current simplifies to
\begin{align}
\label{eq:J_zero}
  J_{\mathrm{LB}} \;=\; \frac{1}{2\pi}\int_{\mu_R}^{\mu_L}\dd E\, T(E)\,,
\end{align}
as well as the one for the noise
\begin{align}
    D \;=\; \frac{1}{2\pi}\int_{\mu_R}^{\mu_L}\dd E\, T(E)\,\big[1-T(E)\big]\,.
  \label{eq:D_zero}
\end{align}

As is well known from the Meir--Wingreen formulation~\cite{Meir1992}, the transmission function that enters the Landauer--B\"uttiker current and the zero-frequency noise reads
\begin{align}
\label{eq:T(E)_green}
    T(E)=\mathrm{Tr}\!\big[\Gamma_L\,G^{r}(E)\,\Gamma_R\,G^{a}(E)\big].
\end{align}
Here \(G^{r}(E)=[\,E\,\mathds{1}-h_{\mathrm{eff}}\,]^{-1}\) and \(G^{a}(E)=(G^{r}(E))^\dagger\) are the single-particle retarded (advanced) Green’s functions of the open system, encoding coherent propagation and contact-induced broadening. In the wide-band limit (WBL), the effective one-particle Hamiltonian is
\begin{equation}
\label{heff}
h_{\mathrm{eff}}=h-\tfrac{i}{2}\left(\Gamma_L\Pi_1 + \Gamma_R\Pi_N\right) 
\end{equation}
with \(\Pi_1=|1\rangle\!\langle1|\) and \(\Pi_R=|N\rangle\!\langle N|\) the projectors on the first and last site in the single-excitation subspace, and the couplings \(\Gamma_{L},\Gamma_{R}\) are energy-independent (flat) and in the main text assumed to be equal $\Gamma$.
For the chain described by Eq.~\eqref{eq:H}, \(h\) is the single-excitation tight-binding matrix obtained from \(H\) (after a Jordan--Wigner mapping for number-conserving spin models).
The explicit matrix form of \(h_{\mathrm{eff}}\) used here is given in Appendix~\ref{SM:calculating current and noise}, Eq.~\eqref{eq:h_eff_matrix}, where the numerical implementation is detailed.

\paragraph{Comparison with the master equation.\label{sec.MEvsLB_current}}
In this paragraph, we explain why LB theory coincides with predictions from the master equation (ME) by working out the case of the \emph{current}.
The argument for the zero-frequency noise proceeds analogously, and $D_{\mathrm{ME}}$ is given by~\eqref{eq:LB-noise} but with Fermi functions evaluated at the local transition energy~$\omega$.
See Appendix~\ref{ME_vs_LB_noise} for the detailed derivation in the limit of maximal bias with $f_L=1$ and $f_R=0$.

Using a Jordan--Wigner transformation, we write the Lindblad master equation for the spin chain in terms of the fermionic operators $c_i,c_i^\dagger$ as in Eq.~\eqref{eq:c_j_trafo}. 
In this representation, the average right-lead current can be written as
\begin{align}
  J_{\mathrm{ME}} &\;=\; \tr\!\big[(\mathcal L^{+}_N-\mathcal L^{-}_N)\,\rho_{\mathrm{ss}}\big] \nonumber \\
  &=\; \Gamma_R \big( \tr\big[C_{\rm ss} \Pi_N\big] - f_R \big),
  \label{eq:ME-current-both}
\end{align}
where $\mathcal L^{+}_N\rho=\Gamma_R(1-f_R)\,c_N\rho c_N^\dagger$ and $\mathcal L^{-}_N\rho=\Gamma_R f_R\,c_N^\dagger\rho c_N$ are the “out” and “in” jump superoperators at the right contact (see~\cite{Blasi2024}).
The operator $C_{\mathrm{ss}}$ is the single-excitation covariance matrix with entries $C_{ij}=\langle c_i^\dagger c_j\rangle$.

For quadratic systems as considered here, the one-body covariances obey a closed linear Lyapunov/Sylvester equation~\cite{Landi2022} (derivation see Eq.~\eqref{eq:dotC_nonconditional_short} in Appendix~\ref{SM:unconditional_cvm_evolution}):
\begin{equation}
  \dot C \;=\; K C + C K^\dagger + P,
  \qquad
  K \equiv +i h_{\mathrm{eff}}^\dagger,
  \label{eq:Lyap-eom}
\end{equation}
with $h_{\mathrm{eff}}$ the effective one-body Hamiltonian introduced in Eq.~\eqref{heff} and $P \equiv \Gamma_L f_L \Pi_1+\Gamma_R f_R \Pi_N$ with $f_{L},f_{R}$ being the Fermi functions evaluated at the local transition energy 
$\omega$ of the coupled site (hence, in the WBL these are energy-independent constants).
The steady value $C_{\mathrm{ss}}$ is obtained by imposing $\dot C=0$ in \eqref{eq:Lyap-eom}, which has the formal solution
\begin{equation}
  C_{\mathrm{ss}}
  \;=\; \int_0^\infty \dd t\, e^{K t}\,P\,e^{K^\dagger t}.
  \label{eq:Css-time}
\end{equation}
By inserting $\int \frac{\dd E}{2\pi}e^{i E t}=\delta(t)$ and using
$\int_0^\infty \dd t\,e^{(K+iE)t}=i\,[\,E\mathds 1-(-iK)\,]^{-1}=iG^r(E)$, we can express the above equation in terms of the retarded and advanced Green's functions as
\begin{equation}
  C_{\mathrm{ss}} \;=\; \int \frac{\dd E}{2\pi}\; G^r(E)\,P\,G^a(E).
  \label{eq:Css-Gr}
\end{equation}
Substituting \eqref{eq:Css-Gr} into \eqref{eq:ME-current-both} and using the Caroli identity~\cite{Caroli1971}, $G^r-G^a=-i\,G^r(\Gamma_L \Pi_1 +\Gamma_R \Pi_N) G^a$, together with trace cyclicity gives
\begin{equation}
  J_{\mathrm{ME}}
  \;=\; \int\!\frac{\dd E}{2\pi}\,
  \tr\!\big[\Gamma_L\,G^r(E)\,\Gamma_R\,G^a(E)\big]\;\big[\,f_L - f_R\,\big],
  \label{eq:IME-final}
\end{equation}
which has the same structure as the Landauer--B\"uttiker current \eqref{eq:LB-current}, except that here $[f_L-f_R]$ are \emph{energy-independent constants}, while in $J_{\mathrm{LB}}$ they are the \emph{energy-dependent} Fermi functions $[f_L(E)-f_R(E)]$.

It is straightforward that for large forward bias, i.e., $f_L=1,f_R=0$ (the regime considered in this work), one has $J_{\mathrm{ME}}=J_{\mathrm{LB}}$.
At finite bias and temperature, Refs.~\cite{Blasi2024,Blasi2025} proves analytically that the two coincide for any bias in the weak-coupling regime $\Gamma\ll k_B T$.

At absolute zero temperature but finite bias, the difference can be estimated explicitly,
\begin{equation}
  J_{\mathrm{ME}}-J_{\mathrm{LB}}
  \;=\; \!\int_{[\mu_R,\mu_L]^c} \frac{\dd E}{2\pi}\, T(E)\,,
  \label{eq:IME-ILB-diff}
\end{equation}
i.e., it is the integral of the tails of the transmission outside the transport window, the complement of $[\mu_R,\mu_L]$.
Since the spectral weight of $T(E)$ is broadened on a scale set by $\Gamma$, those tails decay as one moves a distance $\Delta_{\rm edge}$ away from the support; a crude bound is
\begin{equation}
  \big|J_{\mathrm{ME}}-J_{\mathrm{LB}}\big|
  \;\lesssim\; \frac{\|T\|_\infty}{2\pi}\,\frac{\Gamma}{\Delta_{\rm edge}},
  \label{eq:IME-ILB-bound}
\end{equation}
where $\Delta_{\rm edge}\equiv \min\{\mu_L-\max S,\;\min S-\mu_R\}$, and $S=\mathrm{supp}\,T(E)$ (the broadened spectral support).
For a single Lorentzian resonance one finds more explicitly
$J_{\mathrm{ME}}-J_{\mathrm{LB}}
\simeq \frac{\Gamma_L\Gamma_R}{\pi\Gamma}\big[\frac{\Gamma}{2\Delta_L}+\frac{\Gamma}{2\Delta_R}\big]$ with $\Delta_{L,R}$ the distances from $\mu_{L,R}$~\cite{Blasi2024}; in general chains the same scaling controls the band edges.
Thus, the mismatch vanishes when the entire support of $T(E)$ lies well inside the transport window, specifically when
\begin{equation}
  \mathrm{dist}\!\big(S,\{\mu_L,\mu_R\}\big)\ \gg\ \Gamma ,
  \label{eq:support-window-condition}
\end{equation}
which guarantees that the Fermi edges are far beyond the contact-induced broadening and the ME flat-$f$ approximation reproduces the LB current.

\section{\label{SM:calculating current and noise}Efficient current noise calculation with residue calculus}
In this section, we show how the \textit{Fano-Factor}, $D/J$, mentioned in the main text is computed efficiently.
We demonstrate how the current average and noise can be efficiently computed using the Landauer--Büttiker (LB) formalism with residue calculus, providing the basis for optimizing model parameters to minimize current fluctuations. 
At absolute zero temperature and infinite bias, both the current $J$ and noise $D$ can be evaluated numerically, enabling efficient minimization of the Fano factor $D/J$ as discussed in the main text.

For convenience, we explicitly write out the single-excitation subspace representation of the effective Hamiltonian governing the chain's dynamics:
\begin{align}
\label{eq:h_eff_matrix}
    h_{\rm eff} = 
     \begin{bmatrix}
-i\Gamma/2 & g_1 & 0 & \cdots & 0 \\
g_1 & 0 & g_2 & & \vdots \\
0 & g_2 & \ddots & \ddots  \\
\vdots &  & \ddots & 0 & g_{N-1} \\
0 & \cdots & & g_{N-1} & -i\Gamma/2 \\
\end{bmatrix}\!.
\end{align}
The effective Hamiltonian arises from the Lyapunov-equation~\eqref{eq:dotC_nonconditional_short}, and is related to the operator $K$ through
\begin{align}
    K = (-ih_{\rm eff})^\dagger.
\end{align}

The transmission function can then be calculated explicitly by using the Green's function expression in Eq.~\eqref{eq:T(E)_green} from~\cite{Sumetsky2003,Chak2006,Blasi2024}.
In matrix form, it reduces to
\begin{align}
\label{eq:T(E)_SM}
    T(E) = \left|\left(\left(h_{\rm eff} - E\mathds 1\right)^{-1}\right)_{1N}\right|^2,
\end{align}
where $(h_{\rm eff}-e\mathds 1)^{-1}$ is the matrix inverse of $h_{\rm eff}-E\mathds 1$.
Direct numerical integration is computationally expensive due to the matrix inversion.
A numerically more effective method is presented in the following:

Let $\Lambda$ be the diagonal matrix of $h_{\rm eff}$, with eigenvalues $\lambda \in \Lambda$ all distinct (we simplify the notation by using $\Lambda$ as both the symbol for the diagonal matrix and the spectrum).
Suppose the invertible matrix $Q$ diagonalizes $h_{\rm eff}$ such that we can write
\begin{align}
    h_{\rm eff} = Q \Lambda Q^{-1},
\end{align}
with $\Lambda_{ij}=\delta_{ij}\sigma_i$ diagonal with the eigenvalues of $h_{\rm eff}$.
The matrix element in Eq.~\eqref{eq:T(E)_SM} can be written as
\begin{align}
    \left((h_{\rm eff} - E\mathds 1)^{-1}\right)_{1N} &= \left(Q(\Lambda - E\mathds 1)^{-1}Q^{-1}\right)_{1N} \\
    &= \sum_{\lambda\in\Lambda} \frac{Q_{1\lambda} ({Q}^{-1})_{\lambda N}}{\lambda - E},
\end{align}
where we have implicitly written the sum over all eigenvalues $\lambda\in\Lambda$.
Taking the absolute value yields the $T(E)$.
For later purposes we replace the symbol $E\rightarrow z$, to highlight that the argument of the transmission function may formally be complex.
The transmission function is
\begin{align}
\label{eq:Tz_poles}
    T(z) = \sum_{\lambda,\mu\in\Lambda} \frac{C_{\lambda\mu}}{(\lambda-z)(\mu^*-z)},
\end{align}
where the matrix elements $C_{\lambda\mu}$ are a shorthand for
\begin{align}
    C_{\lambda\mu} = Q_{1\lambda}Q^{-1}_{\lambda N} Q^*_{1\mu} (Q^{-1})^*_{\mu N},
\end{align}
and need to be calculated only once for a given Hamiltonian, as they are independent of $z$.

To calculate average current $J$ using Eq.~\eqref{eq:J_zero} or diffusion constant $D$ according to Eq.~\eqref{eq:D_zero}, we can employ the residue theorem.
To provide a self-contained account of our techniques, we detail these calculations in the following; for additional background, we refer the reader to one of the standard textbooks on the matter like Ref.~\cite{Lang1999}.

For a general meromorphic function $f(z)$ that drops sufficiently quickly at the boundaries $|z|\rightarrow\infty$, we can write
\begin{align}
    \int_{-\infty}^\infty \dd z\, f(z) = 2\pi i\sum_{a\in \mathfrak{Jm}_+} {\rm Res}(f,a),
\end{align}
where the sum goes over all residues $a\in \mathfrak{Jm}_+$ of $T(z)$ in the positive complex plane.
Around a pole $a$ of $f$, the function $f(z)$ can be expanded as
\begin{align}
    f(z) = \frac{a_{-1}}{z-a} + a_0 + a_1(z-a) + \dots,
\end{align}
where ${\rm Res}(f,a) = a_{-1}$.
Using the explicit form~\eqref{eq:Tz_poles}, the poles can be identified as the complex conjugate of the eigenvalues of $h_{\rm eff}$.
To be explicit, $T(z)$ around the complex conjugate of an eigenvalue $\mu^*$ can be written as
\begin{align}
    T(z) = \frac{1}{z-\mu^*} \sum_{\lambda\in\Lambda} \frac{-C_{\lambda\mu}}{\lambda-z} + \sum_{\lambda,\nu\in\Lambda}\frac{C_{\lambda\nu}}{(\lambda-z)(\nu^*-z)}.\nonumber
\end{align}
More generally, $T(z)$ is of the form
\begin{align}
\label{eq:f(z)_general}
    f(z) = \frac{g(z)}{z-a} + h(z),
\end{align}
where both $g(z)$ and $h(z)$ are whole in the neighborhood of the pole $a$.
The residue of such a function is of the form ${\rm Res}(f,a)=g(a)$, and we can thus directly write the average current as
\begin{align}
    J &= i\sum_{\mu\in \Lambda} {\rm Res}(T,\mu^*) \\
    &=\sum_{\lambda,\mu\in\Lambda} \frac{iC_{\lambda\mu}}{\lambda - \mu^*}.
\end{align}
To calculate $D$, the integral in Eq.~\eqref{eq:D_zero} has a contribution linear in $T(z)$ for which the above expression can be used.
However, there is a second quadratic term where we integrate $T(z)^2$ which has to be treated separately.
Generally, for a meromorphic function $f(z)$ of the form~\eqref{eq:f(z)_general}, we can expand
\begin{align}
    f(z)^2 = \frac{a_{-1}^2}{(z-\mu^*)^2} + \frac{2a_0 a_{-1}}{z-\mu^*} + \dots,
\end{align}
which allows us to determine ${\rm Res}(f^2,\mu^*) = 2a_0 a_{-1}$.
As before $a_{-1}$ is the residue of $f$ and $a_0=h(a)$ for a function of the general expression~\eqref{eq:f(z)_general}.
For the transmission function, all terms again simplfy and we can sum over all poles,
\begin{align}
    \int_{-\infty}^\infty \dd z\,T(z)^2 &= 4\pi i \sum_{\mu \in \Lambda} a_0(\mu)a_{-1}(\mu),
\end{align}
and use the fact that we have
\begin{align}
    a_{-1}(\mu) = \sum_{\lambda\in\Lambda}\frac{C_{\lambda\mu}}{\lambda-\mu^*},
\end{align}
as before.
The other term is given by
\begin{align}
    a_{0}(\mu) = \sum_{\nu \in \Lambda \setminus \{\mu\}}\sum_{\lambda\in \Lambda}\frac{C_{\lambda \nu}}{(\lambda-\mu^*)(\nu^* - \mu^*)}.
\end{align}

These expressions combined with Eqs.~\eqref{eq:J_zero} and~\eqref{eq:D_zero} allow us to efficiently determine the relative fluctuations $D/J$ and minimize them as a function of the couplings $g_i$ and $\Gamma$, since the integrals need not be numerically approximated.
Instead, the effective Hamiltonian is numerically exactly inverted, and $D/J$ is calculated exactly using the residue theorem:
\begin{align}
    D/J &= \frac{i}{J} \left( \sum_{\mu\in\Lambda}ia_{-1}(\mu)\left(1 - 2 a_0(\mu)\right)\right).
\end{align}

\section{\label{SM:analytical_derivation_variance}Analytical number variance derivation}
In this section, we show the analytical derivation of the main result of this paper.
We characterize the connection between the tick current $I(t)$ at the boundary of the chain, and the coherent tick current in the chain's bulk.
Using an analytical expression for the bulk current fluctuations, we derive the $\Var[N_t]\sim \log t$ scaling.
To that end, we first derive the expressions for current and fluctuations at the boundary (Appendix~\ref{SM:fluctuations_boundary}) and in the bulk (Appendix~\ref{SM:fluctuations_bulk}).
Then, we determine a closed-form expression for the asymptotic state in the stationary regime (Appendix~\ref{SM:determining_stationary_state}).
With this state we can analytically derive the bulk number variance (Appendix~\ref{SM:calculating_two_point_fluctuations} and~\ref{SM:calculating_number_variance}). 
With the previous Appendix~\ref{SM:calculating_waiting_time_variance}, connecting $\Var[T_n]$ to the number variance $\Var[N_t]$, we obtain the main result.

\subsection{\label{SM:fluctuations_boundary}Boundary current fluctuations}
With $I(t)$ the tick current (cf.~Eq.~\eqref{eq:I_def_Stochastic}), we recall the the expression~\eqref{eq:VarNt_classical_main} for the tick number variance from the main text,
\begin{align}
\label{eq:VarNt_classical_SM}
    \Var[N_t] = \int_0^t \! \dd t' \! \int_0^t \! \dd t''\, \E[[I(t') I(t'')]].
\end{align}
Here, we provide an expression for $\Var[N_t]$ in terms of the covariance matrix for fermionic Gaussian states.
With the covariance matrix formulation, calculating $\Var[N_t]$ is numerically tractable for large system sizes and large times.

The connected, non-equal time current-current correlation in Eq.~\eqref{eq:VarNt_classical_SM} only depends on the time difference $\tau = t'-t''$ in the stationary regime.
It can be expressed as~\cite{Wiseman2009,Schaller2014,Landi2024,Blasi2024},
\begin{align}
    \E[[I(\tau)I(0)]] &= A \delta(\tau) + \tr\left[ \mathcal J e^{\mathcal L|\tau|} \mathcal J \rho_{\rm ss} \right] - J^2,
\end{align}
where: $J=\tr[\mathcal J \rho_{\rm ss}]$ is the stationary current value (also equal $J=\lim_{t\rightarrow\infty}\E[I(t)]$), with $\mathcal J = \mathcal J_+ - \mathcal J_-$ the net current.
The forward and backward current are given by,
\begin{align}
\label{eq:J+J-_deff}
    \mathcal J_+ \rho = \Gamma(1-f_R) \, c_N \rho c_N^\dagger,\quad \mathcal J_- \rho &=\Gamma f_R \, c_N^\dagger \rho c_N,
\end{align}
where we recall that we set $\Gamma\equiv 1$. 
Finally, the total rate of jumps, $A= \tr[(\mathcal J_+ + \mathcal J_-) \rho_{\rm ss}]$, is the dynamical activity.

The time derivative of the number variance then defines a time-dependent diffusion constant $D(t) = \dd \Var[N_t]/\dd t$.
To obtain a concise expression we first rewrite the double-integral that defines the number variance.
The singular dynamical activity term must be treated separately:
\begin{align}
    \int_0^t \! \dd t' \! \int_0^t \! \dd t'' \, A \,\delta(t'-t'') = At.
\end{align}
For the non-equal time integral, we can use the $\tau \leftrightarrow -\tau$ symmetry of the integrand to write
\begin{align}
    &\int_0^t \! \dd t' \! \int_0^t \! \dd t'' \left(\tr\left[ \mathcal J e^{\mathcal L|\tau|} \mathcal J \rho_{\rm ss} \right] - J^2\right)  \nonumber \\
    &\qquad =2{\rm Re}\int_0^t \! \dd t' \int_0^{t'} \! \dd \tau \left(\tr\left[ \mathcal J e^{\mathcal L \tau} \mathcal J \rho_{\rm ss} \right] - J^2\right).
\end{align}
Taking the time-derivative yields the expression~\cite{Landi2024,Blasi2024}
\begin{align}
\label{eq:D_ME}
    D(t) = A + 2\,{\rm Re}\int_0^t \dd \tau \, \Big(\tr\left[\mathcal J e^{\mathcal L \tau }\mathcal J \rho_{\rm ss}\right] - J^2\Big).
\end{align}

We now proceed with expressing Eq.~\eqref{eq:D_ME} in terms of the $N\times N$ covariance matrix, as opposed to the $2^N$-dimensional state space expression.
To start with, the dynamical activity can be expressed as follows:
\begin{align}
    A &=  (1-f_R) \tr[\Pi_N C_{\rm ss}] + f_R \tr[\Pi_N (\mathds 1 - C_{\rm ss})]  \nonumber \\
    &=(1-2f_R) \tr[\Pi_N C_{\rm ss}] + f_R,
    \label{eq:K_dynamical_activity}
\end{align}
where we recall $C_{ij}^{\rm ss} = {\rm tr}[c_i^\dagger c_j \rho_{\rm ss}]$ are the covariance matrix elements.
Also the integral part on the right-hand side of Eq.~\eqref{eq:D_ME} can be expressed in terms of the covariance matrix.
We define $\sigma = \mathcal J \rho_{\rm ss} / \tr[\mathcal J\rho_{\rm ss}]$ as the jump steady state, and denote it's covariance matrix using $C^\sigma$.
The matrix elements can be calculated explicitly,
\begin{align}
\label{eq:Csigma_ij_def}
    &\tr\left[c_i^\dagger c_j \sigma\right] \\
    &\quad =\frac{1}{J}\left((1-f_R)\langle c_N^\dagger c_i^\dagger c_j c_N \rangle_{\rm ss} + f_R \langle c_N c_i^\dagger c_j c_N^\dagger \rangle_{\rm ss}\right). \nonumber
\end{align}
The correlation function can be calculated using the Wick theorem. The first term as in Eq.~\eqref{eq:C1dagCidagCjC1},
\begin{align}
    \langle c_N^\dagger c_i c_j c_N \rangle_{\rm ss} &= C_{ij}^{\rm ss} C_{NN}^{\rm ss} - C_{iN}^{\rm ss}C_{Nj}^{\rm ss},
\end{align}
and the second as in Eq.~\eqref{eq:dotC_jump1},
\begin{align}
    &\langle c_N c_i^\dagger c_j c_N^\dagger \rangle_{\rm ss}  \\
    &\quad = C_{ij}^{\rm ss}(1-C_{NN}^{\rm ss}) + (\delta_{iN}-C_{iN}^{\rm ss})(\delta_{Nj}-C_{Nj}^{\rm ss}). \nonumber
\end{align}
In index-free notation, we thus find that
\begin{align}
\label{eq:Csigma}
    C^{\sigma} &= \frac{1}{J}\Big((1-f_R)(C_{\rm ss} \tr[\Pi_N C_{\rm ss}] - C_{\rm ss}\Pi_N C_{\rm ss})  \\
    & + f_R(C_{\rm ss}(1-\tr[\Pi_N C_{\rm ss}]) + (\mathds 1 - C_{\rm ss})\Pi_N (\mathds 1 - C_{\rm ss}))\Big). \nonumber
\end{align}
Next, we consider the state $\sigma_t = e^{\mathcal L t} \sigma$.
Its time evolution on the level of the covariance matrix $C^{\sigma_t}$ is given by Eq.~\eqref{eq:dotC_nonconditional_finite_temp} (also see~\cite{Bettmann2024}), written out for completeness
\begin{align}
    \frac{\dd }{\dd t} C^{\sigma_t} &= KC^{\sigma_t} + C^{\sigma_t}K^\dagger + P,
\end{align}
where $P=f_L \Pi_1 + f_R \Pi_N$.
This equation is formally solved by,
\begin{align}
\label{eq:C_sigmat}
    C^{\sigma_t} &= e^{Kt} C^\sigma e^{K^\dagger t} + \int_0^t \dd s\, e^{K(t-s)} P e^{K^\dagger (t-s)},
\end{align}
as one can readily verify. 
Performing a change of variables $t-s\mapsto s$, the integral on the right-hand side can be simplified as
\begin{align}
\label{eq:C_sigmat_integral_manipulation}
    \int_0^t \dd s\, e^{K(t-s)} P e^{K^\dagger (t-s)} &= \int_0^t \dd s\, e^{Ks} P e^{K^\dagger s}  \\
    &= C_{\rm ss} -  \int_t^\infty \dd s\, e^{Ks} P e^{K^\dagger s} \nonumber \\
    &= C_{\rm ss} + \mathcal K^{-1}  e^{Kt} P e^{K^\dagger t}. \nonumber
\end{align}
Note that $C_{\rm ss}$ is the steady-state covariance matrix obtained from the $t\rightarrow\infty$ limit.
Further, the superoperator $\mathcal K$ is defined by,
\begin{align}
    \mathcal K X = KX + XK^\dagger,
\end{align}
that is, $\mathcal K^{-1}Y$ is the solution of the Lyapunov equation $\mathcal KX=Y$.
By using that $\mathcal K^{-1}$ commutes with the superoperator exponential $e^{\mathcal K\tau}$, we can further simplify using the expression $C_{\rm ss}=-\mathcal K ^{-1}P$,
\begin{align}
\label{eq:C_sigmat_integral_manipulation_2}
    \int_0^t \dd s\, e^{K(t-s)} P e^{K^\dagger (t-s)} &= C_{\rm ss} - e^{Kt}C_{\rm ss}e^{K^\dagger t}.
\end{align}
This expression can be used to simplify the integral term in Eq.~\eqref{eq:D_ME} by using:
\begin{align}
    \tr\left[\mathcal J e^{\mathcal L \tau }\mathcal J \rho_{\rm ss}\right] &= J \tr\left[\mathcal J \sigma_t \right].
\end{align}
The right-hand side can be explicitly calculated,
\begin{align}
    \tr[\mathcal J \sigma_t] &= \left(C^{\sigma_t}_{NN} - f_R\right) \\
    &=\tr\left[ \Pi_N e^{Kt} \left( C^\sigma - C_{\rm ss}\right)e^{K^\dagger t}\right] + J, \nonumber
\end{align}
using the formal solution from Eq.~\eqref{eq:C_sigmat} together with~\eqref{eq:C_sigmat_integral_manipulation_2}.
Moreover, we again identified the steady-state current $J=\tr[\Pi_N (C_{\rm ss}-f_R\mathds 1)]$.

\begin{figure*}
    \centering
    \includegraphics[width=\linewidth]{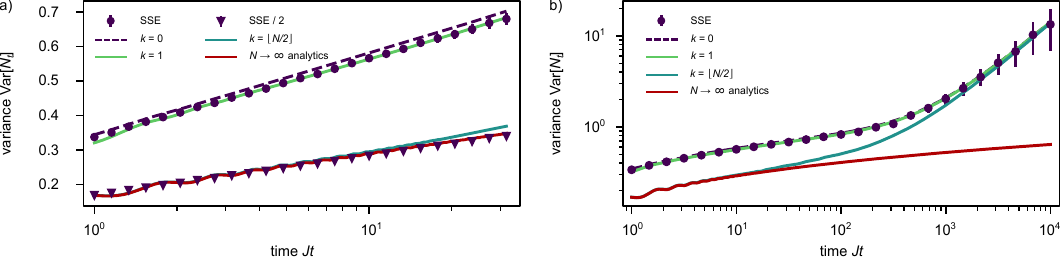}
    \caption{Number variance bulk vs.\ boundary for $N=40$ sites. 
    a) Short time fluctuations are shown in a log-linear scale.
    The dots show results from the Monte-Carlo stochastic Schr\"odinger equation (SSE) evolution, obtained through histogram binning of the tick times.
    Error bars show the sample error.
    In addition, the numerically exact results for the classical current fluctuations from Eq.~\eqref{eq:VarNt_COV_matrix} are shown in dashed lines ($k=0$).
    The fluctuations of the coherent current at the boundary, (Eq.~\eqref{eq:VarNtcoh_k} with $k=1$, green line), closely match the fluctuations of the classical current.
    In contrast, the fluctuations in the bulk ($k=\lfloor N/2\rfloor$) are shown to be approximately half as large as the fluctuations at the boundary.
    For comparison $0.5$ times the SSE result is shown (diamond symbol).
    The red solid line is the analytical prediction for the bulk fluctuations in the limit of $N\rightarrow\infty$ sites (Eq.~\eqref{eq:VarNT_FINALFULL}).
    b) The number variance is shown over a longer time scale.
    In addition to part a), it can be seen here that while the bulk fluctuations ($k=\lfloor N/2\rfloor$) at short times are approximately half of those at the boundary; after very long times, they increase and match the boundary fluctuations.
    The $N\rightarrow\infty$ analytics maintain the logarithmic scaling indefinitely because of the unbounded chain length.
    }
    \label{fig:Fig4_VarNt}
\end{figure*}

In the last step, all terms are inserted into the integral from Eq.~\eqref{eq:D_ME}.
Notably, the constant $J^2$-term exactly cancel with the steady-state contribution from $C^{\sigma_\tau}$:
\begin{align}
    &\tr\left[\mathcal J e^{\mathcal L \tau} \mathcal J \rho_{\rm ss}\right] - J^2 \\
    &\quad = J\tr\left[\Pi_N \left( e^{K\tau}\left( C^\sigma -C_{\rm ss}) \right) e^{K^\dagger \tau} \right) \right]
\end{align}
Inserting into the time integral yields,
\begin{align}
    &\int_0^t \dd \tau\, \left( \tr\left[\mathcal J e^{\mathcal L \tau} \mathcal J \rho_{\rm ss}\right] -J^2\right) \\
    &=  -J \tr\left[\Pi_N \mathcal K^{-1} \left(C^\sigma - C_{\rm ss}-e^{Kt} \left(C^\sigma  -C_{\rm ss} \right)e^{K^\dagger t}\right) \right]. \nonumber
\end{align}
This defines the time-dependent diffusion constant together with Eqs.~\eqref{eq:D_ME} and~\eqref{eq:K_dynamical_activity}:
\begin{align}
\label{eq:D(t)_full}
    &D(t) = +A\\
    &-2\,{\rm Re}\, J \tr\left[\Pi_N \mathcal K^{-1} \left(C^\sigma - C_{\rm ss}-e^{Kt} \left(C^\sigma  -C_{\rm ss} \right)e^{K^\dagger t}\right) \right]\!.\nonumber
\end{align}
Integration yields $\Var[N_t]=\int_0^t \dd \tau\,D(\tau)$, the number variance,
\begin{align}
    \int_0^t \!\! \dd \tau \,D(\tau) &= t\left(A-2 \,{\rm Re}\,J \tr\left[\Pi_N \mathcal K^{-1} \left(C^\sigma - C_{\rm ss}\right) \right] \right) \nonumber  \\
    &\hspace{-1cm} +2 \,{\rm Re}\, J \int_0^t \! \dd \tau\tr\left[\Pi_N \mathcal K^{-1}\left(e^{K\tau} \left(C^\sigma  -C_{\rm ss} \right)e^{K^\dagger \tau}\right) \right] \nonumber  \\
    &= t\left(A-2 \,{\rm Re}\,J \tr\left[\Pi_N \mathcal K^{-1} \left(C^\sigma - C_{\rm ss}\right) \right] \right)  \nonumber  \\
    &\hspace{-1cm} -2 \,{\rm Re}\, J \tr\left[\Pi_N \mathcal K^{-2}\left(e^{K\tau} \left(C^\sigma  -C_{\rm ss} \right)e^{K^\dagger \tau}\right) \right] \bigg|_0^t\!.
\end{align}
Abbreviating $D=A-2\,{\rm Re}\,J\tr[\Pi_N \mathcal K^{-1} (C^\sigma-C_{\rm ss})]$, the diffusion constant in the $t\rightarrow\infty$ limit, the number variance can be written in the following concise form:
\begin{align}
\label{eq:VarNt_COV_matrix}
    \Var[N_t] &= Dt - 2\,{\rm Re}\, J \tr\left[\Pi_N e^{Kt} \mathcal K^{-2}\left(C^\sigma\!-C_{\rm ss}\right)e^{K^\dagger t}\right]\!\bigg|_0^t \!\!,
\end{align}
In Fig.~\ref{fig:Fig4_VarNt}, the number variance calculated according to Eq.~\eqref{eq:VarNt_COV_matrix} is shown and compared to its coherent counterpart in the bulk; which we discuss in the following section.

\subsection{\label{SM:fluctuations_bulk}Bulk current fluctuations}
Excitations travel coherently through the bulk.
The current operator describing this transport is defined by the continuity equation for the time derivative of the excitation number $n_{N'}=c_{N'}^\dagger c_{N'}$ at some site $N'$:
\begin{align}
\label{eq:dnN'dt_continuity}
    \frac{\dd n_{N'}}{\dd t}&=j_{{N'-1}\rightarrow {N'}} - j_{{N'}\rightarrow {N'+1}}.
\end{align}
Here, $j_{k\rightarrow k+1}$ is the current from site $k$ to $k+1$ ($k=N'-1$ and $k=N'$). 
Away from the boundary we obtain the continuity equation~\eqref{eq:dnN'dt_continuity} from the Heisenberg equation of motion 
\begin{align}
\label{eq:dnN'dt_adjoint}
    \frac{\dd n_{N'}}{\dd t} 
    &= +i\left[H,n_{N'}\right]  \nonumber \\
    & = -(j_{N^\prime \rightarrow N^\prime+1} - j_{N^\prime -1 \rightarrow N^\prime}).
\end{align}
from which we can read off the current
\begin{align}
    j_{N'\rightarrow N'+1} = -ig_{N'}\left(c_{N'}^\dagger c_{N'+1} - c_{N'+1}^\dagger c_{N'}\right).
\end{align}
For $N'=N$, the dissipative contribution from the tick current $\mathcal J$ which is leaving the chain (cf.~Eq.~\eqref{eq:J+J-_deff} in Appendix~\ref{SM:fluctuations_boundary}) must be considered:
\begin{align}
    \frac{\dd \langle n_N\rangle }{\dd t} &= \langle j_{N-1\rightarrow N}\rangle - (\langle n_N \rangle - f_R).
\end{align}

In the stationary regime where $\langle n_{N'}\rangle$ is constant in time, Eq.~\eqref{eq:dnN'dt_continuity} implies $\langle j_{N'\rightarrow N'+1} \rangle_{\rm ss} = \langle j_{N'-1\rightarrow N'} \rangle_{\rm ss}$, for all $2\leq N' \leq N-1$.
Further, at the boundary $N'=N$, we have $\langle j_{N-1\rightarrow N}\rangle_{\rm ss} = \tr[\mathcal J \rho_{\rm ss}]$.
The current is therefore everywhere exactly the same, equal to the steady-state current $J$,
\begin{align}
\label{eq:equal_currents}
    \langle j_{N'\rightarrow N'+1} \rangle_{\rm ss} = J,
\end{align}
where we recall that the expectation values are given by $\langle j\rangle_{\rm ss}=\tr[j\rho_{\rm ss}]$, with $\rho_{\rm ss}$ the steady state.
We may therefore introduce the time-integrated bulk current as,
\begin{align}
    N_t^{{\rm bulk}, k} &= \int_0^t \dd \tau \, j_{N-k\rightarrow N-k+1}(\tau).
\end{align}
That is, $N_t^{{\rm bulk},k}$ can be understood as the integrated number of excitations that have passed the bond $N-k \rightarrow N-k+1$ in the time-interval $[0,t]$.
In the stationary state, we thus have $\langle N_t^{{\rm bulk},k}\rangle_{\rm ss} = Jt$, equal to the integrated tick number $\E[N_t]=Jt$.

Looking at the fluctuations, coherent bulk current and classical tick current need not give exactly the same predictions.
To quantify the relationship of the coherent bulk vs.\ classical boundary current, we first derive an expression for the current fluctuations in the bulk: using the expression as in Eq.~\eqref{eq:VarNt_classical_SM}, but with classical expectation values replaced with quantum expectation values~\cite{Wiseman2009}, we obtain
\begin{align}
\label{eq:VarNtcoh_k}
    \Var[N_t^{{\rm bulk}, k}] &= \int_0^t \! \dd t' \! \int_0^t \! \dd t'' \langle\langle j_{k}(t') j_{k}(t'') \rangle\rangle_{\rm ss}.
\end{align}
Note the abbreviation $j_{k}:=j_{N-k\rightarrow N-k+1}$ for the coherent current $k$ sites into the chain from the right.
Further, the two-time connected correlator is,
\begin{align}
\label{eq:jk_twotimecorrelator}
    \langle\langle j_{k}(t')j_{k}(t'') \rangle\rangle_{\rm ss} & \equiv \tr[j_{k} e^{\mathcal L\tau} j_{k} \rho_{\rm ss}]-J^2,
\end{align}
which depends only on the time difference $\tau=t''-t'$ due to stationarity.

To express Eq.~\eqref{eq:jk_twotimecorrelator} in terms of the steady-state covariance matrix like Eq.~\eqref{eq:VarNt_COV_matrix} for the boundary current, we first determine the covariance matrix of $j_{k}\rho_{\rm ss}$.
To that end, we introduce the current operator matrix elements $(J_k)_{m,n}$ as
\begin{align}
    j_{k} &= \sum_{m,n=1}^N (J_k)_{m,n} c_n^\dagger c_m. 
\end{align}
The average current $\langle j_k\rangle_{\rm ss}=\tr[j_{k}\rho_{\rm ss}]$ can be written as
\begin{align}
    \tr[j_{k}\rho_{\rm ss}] &= \sum_{m,n=1}^N (J_k)_{m,n} \tr[c_n^\dagger c_m \rho_{\rm ss}] \\
    &= \tr[J_k C_{\rm ss}],
\end{align}
with $C_{\rm ss}$ the steady-state covariance matrix and with the first trace in Hilbert space and the second trace in the $N\times N$ covariance-matrix space.
Current conservation~\eqref{eq:equal_currents} implies that $\tr[J_k C_{\rm ss}]=J$.
Further, we set $\sigma_k = j_{k}\rho_{\rm ss} / J$, but note that $\sigma_k$ is not Hermitian, i.e., not a state.
The covariance matrix elements of $\sigma_k$ can nonetheless be calculated formally like in Eq.~\eqref{eq:Csigma_ij_def}, $C^{\sigma,k}_{ij}=\tr[c_i^\dagger c_j \sigma_k]$.
To be explicit, we find,
\begin{align}
    \tr[c_i^\dagger c_j \sigma_k] &= \frac{1}{J} \sum_{m,n=1}^N (J_k)_{m,n} \tr[c_i^\dagger c_j c_n^\dagger c_m \rho_{\rm ss}] \\
    &\hspace{-0.5cm}= \frac{1}{J} \sum_{m,n=1}^N (J_k)_{m,n} (\delta_{jn}C_{im} - C_{nj}C_{im} + C_{ij}C_{nm}). \nonumber
\end{align}
In index-free notation, that is,
\begin{align}
    C^{\sigma,k} = \frac{1}{J} \left( C J_k - C J_k C \right) + C
\end{align}
Time evolution $e^{\mathcal L\tau}\sigma$ can be calculated as in Eq.~\eqref{eq:C_sigmat},
implying that:
\begin{align}
\label{eq:<<jkjk>>_res}
    \langle\langle j_{k}(\tau)j_{k}(0) \rangle\rangle_{\rm ss} &= J \tr\left[J_k \left( e^{K\tau} \left( C^{\sigma,k} - C_{\rm ss}\right) e^{K^\dagger \tau} \right) \right].
\end{align}

In Fig.~\ref{fig:Fig4_VarNt}, $\Var[N_t^{\rm bulk,k}]$ is numerically evaluated for a chain of $N=40$ sites, using the expression in Eq.~\eqref{eq:<<jkjk>>_res}.
We see that for $k=1$, the number variance closely matches the the variance of the classical tick current determined in the previous section, Eq.~\eqref{eq:VarNt_COV_matrix}, for all times.
For $k=\lfloor N/2\rfloor$, and short times, the fluctuations are however only approximately half as large as those of the classical tick current at the boundary.
At longer times in the diffusive regime, however, the fluctuations again coincide.

\subsection{\label{SM:determining_stationary_state} Analytical approximation of steady state}

For the analytical analysis in the bulk, we first determine a closed-form expression for the steady state.
To this end, we split the full Hamiltonian into two parts
\begin{align}
    H = -\underbrace{g\sum_{i=1}^N \left(c_i^\dagger c_{i+1} + c_{i+1}^\dagger c_i\right)}_{=:H_{\rm bulk}} + H_{\rm boundary},
\end{align}
where $H_{\rm bulk}$ is the bulk Hamiltonian with periodic boundaries ($c_{N+1}\equiv c_1$) and constant coupling nearest-neighbor hopping interaction.
Far away from the boundary, $H_{\rm bulk}$ is a good approximation for the true dynamics.

To determine the stationary state, the position operators $c_j$ can be expanded in the momentum basis $e_\kappa$ given by the unitary operator transformation
\begin{align}
\label{eq:c_j_expand_e_kappa}
    c_j = \frac{1}{\sqrt{N}}\sum_{k=0}^{N-1} e^{i\kappa_k j}e_{\kappa_k},
\end{align}
where $\kappa_k = 2\pi k / N - \pi/2$.
This transformation diagonalizes the constant coupling Hamiltonian,
\begin{align}
    H_{\rm bulk} &= -g\sum_{j=0}^{N-1} c_j^\dagger c_{j+1} + {\rm h.c.} \\
    &=-\frac{g}{N}\sum_{j,k,\ell=0}^{N-1} e^{i(\kappa_\ell - \kappa_k)j + i\kappa_\ell} e_{\kappa_k}^\dagger e_{\kappa_\ell} + {\rm h.c.} \\
    &=-2g\sum_{k=0}^{N-1} \cos(\kappa_k) e_{\kappa_k}^\dagger e_{\kappa_k},
\end{align}
with energy-momentum dispersion relation
\begin{align}
    E(\kappa) = -2g\cos(\kappa),
\end{align}
and lattice momentum $\kappa \in [-\pi,\pi)$.

\begin{figure}
    \centering
    \includegraphics[width=\linewidth]{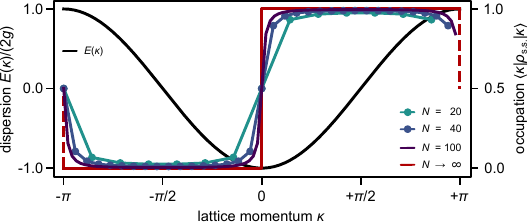}
    \caption{Dispersion relation and occupation probabilities.
    Left $y$-axis: we show the dispersion relation as a function of the bulk lattice momentum.
    Right $y$-axis: numerical results for the occupation probabilities of the momentum modes in the stationary regime are shown for $N=20,40$ and $100$ together with the shifted Fermi sea ansatz for the limit $N\rightarrow\infty$.}
    \label{fig:Fig_Dispersion}
\end{figure}

To numerically determine the momentum occupation in the stationary regime as shown in Fig.~\ref{fig:Fig_Dispersion}, we solve the stationary state equation for the covariance matrix given by Eq.~\eqref{eq:dotC_nonconditional_long}, by setting
\begin{align}
    KC + CK^\dagger + \Pi_1 &= 0.
\end{align}
Here, we work in the regime of full bias, i.e., $f_L=1$ and $f_R=0$.
The equation is of Lyapunov form, as commonly encountered in the context of master equations and control theory~\cite{Nicacio2016,Rouchon2013,Bartels1972,Brunton_19}.

For our setup we have numerically calculated the stationary value of $C$ for the Hamiltonian which minimizes the current noise for $N$ up to 100, i.e., the Hamiltonian with apodized couplings at the boundaries.
Since we are interested in the momentum space representation of the stationary state, we can transform the covariance matrix using the transformation in Eq.~\eqref{eq:c_j_expand_e_kappa}.
To this end, let us define the matrix
\begin{align}
    E_{\kappa_k,\kappa_\ell} &= \langle e_{\kappa_k}^\dagger e_{\kappa_\ell} \rangle.
\end{align}
With the discrete Fourier transform defined by the matrix elements $\mathcal F_{kj} = e^{i\kappa_k j}/\sqrt{N}$, the transformation of the covariance matrix $C_{mn}=\langle c_m^\dagger c_n\rangle$ can be written as
\begin{align}
\label{eq:cmn_to_ekl_trafo}
    \langle c_m^\dagger c_n \rangle &= \frac{1}{N}\sum_{k,\ell=0}^{N-1} e^{-i(\kappa_km-\kappa_\ell n)}\langle e_{\kappa_k}^\dagger e_{\kappa_\ell}\rangle \\
    &= [\mathcal F^\dagger E \mathcal F]_{mn}.
\end{align}
In index-free notation, this is $C=\mathcal F^\dagger E \mathcal F$.
While the momentum space transformation could in principle be restricted to the sub-matrix of the constant bulk couplings, the boundary contributions are negligible in the limit of long chains.
Thus, we transform the entire covariance matrix according to Eq.~\eqref{eq:cmn_to_ekl_trafo}.
In the momentum space representation, the diagonal entries $p(\kappa_k) = E_{\kappa_k,\kappa_k}$ are the occupation probability of the mode with momentum $\kappa_k$.
As the number of sites $N$ grows, the numerically obtained occupation probabilities tend toward $p(\kappa)\rightarrow 0$ for $\kappa \in (-\pi,0)$ and $p(\kappa)\rightarrow 1$ for $\kappa \in (0,\pi)$ (see Fig.~\ref{fig:Fig_Dispersion}).

We can also obtain this momentum-space occupation analytically, using symmetry arguments in the $N\rightarrow\infty$ limit as follows.
Because the equations of evolution are particle-hole symmetric under switching left and right, the momentum-space occupation probability of the stationary state must also exhibit this symmetry.
Switching left and right (parity) flips the momentum sign $\kappa\mapsto -\kappa$, together with exchanging particles and holes gives:
\begin{align}
\label{eq:p_momentum_symmetry}
    p(\kappa) = 1-p(-\kappa).
\end{align}
Since as $N\rightarrow \infty$ the transmission function approaches a perfect box-car shape for appropriately chosen couplings, all modes are being transmitted and none are being reflected.
As a consequence, only right-traveling modes where $\dd E / \dd \kappa \geq 0$ in the chain are being occupied in that limit: $p(\kappa< 0)=0$.
Together with the symmetry~\eqref{eq:p_momentum_symmetry}, this implies $p(\kappa>0)=1$.
The state in the bulk of the chain, for large $N$, can thus be well approximated by a shifted Fermi sea state, given by
\begin{align}
\label{eq:FS+}
    \ket{{\rm FS}^+} &= \prod_{\kappa \in [0,\pi)} e_\kappa^\dagger \ket{0},
\end{align}
with $\ket{0}$ the Fermionic vacuum state.
Note that in contrast to a standard low-energy limit, here, both highest and lowest energy states are occupied.

\subsection{\label{SM:calculating_two_point_fluctuations}Calculating the two-point fluctuations}
We now determine the first two moments of $N_t^{{\rm bulk},\lfloor N/2\rfloor}$, with the Hamiltonian $H_{\rm bulk}$.
Further, we work with the shifted Fermi-sea ansatz from Eq.~\eqref{eq:FS+}, which, in the limit of large $N$, becomes exact. 
In that limit, the site at which the current is considered is irrelevant.
Instead of the current from site $N'=\lfloor N/2\rfloor$ to $N'+1$ we can thus look at the current from site $0$ to site $1$ for the ease of notation,
\begin{align}
\label{eq:j}
    j = -ig\left(c_0^\dagger c_1 - c_1^\dagger c_0\right).
\end{align}
Further, we simplify notation and write,
\begin{align}
\label{eq:N_bulk_simpl}
N_t^{\rm bulk} &= \int_0^t \dd t\, j(t).
\end{align}

Let us start with the current expectation value.
The time dependency drops because $\ket{\rm FS^+}$ is stationary,
\begin{align}
    \langle j(t) \rangle_{\rm FS^+} &= -ig \langle  c_0^\dagger c_1  \rangle_{\rm FS^+} + {\rm h.c.} \\
    &=-\frac{ig}{N}\sum_{\kappa_k,\kappa_\ell} e^{+i\kappa_\ell} \langle e_{\kappa_k}^\dagger e_{\kappa_\ell} \rangle_{\rm FS^+} + {\rm h.c.}.
\end{align}
The expectation-value $\langle e_{\kappa_k}^\dagger e_{\kappa_\ell}\rangle_{\rm FS^+ }$ can be calculated explicitly using Eq.~\eqref{eq:FS+}.
It gives
\begin{align}
    \langle e_{\kappa_k}^\dagger e_{\kappa_\ell}\rangle_{\rm FS^+ } &= \delta_{k,\ell} \chi_{[0,\pi)}({\kappa_k}),
\end{align}
where $\chi_{[0,\pi)}$ is the indicator function on the momentum interval $[0,\pi)$.
To simplify the following derivation, we assume that $N$ is even to avoid notational clutter with the floor function.
The generalization to odd $N$ is straightforward.
Continuing the calculation of the average current,
\begin{align}
    \langle j(t) \rangle_{\rm FS^+} 
    &=-\frac{ig}{N} \sum_{\kappa_k \in [0,\pi)} \left(e^{+i\kappa_k} - e^{-i\kappa_k}\right)  \\
    &\rightarrow \frac{g}{\pi} \int_0^\pi \dd \kappa \sin(\kappa) 
    =\frac{2g}{\pi}, \label{eq:j_t_stationary}
\end{align}
using that in the limit $N\rightarrow\infty$ we have the sum over the discrete momenta $\kappa_k$ converges to the Riemann integral $\frac{2\pi}{N}\sum_{\kappa_k\in[0,\pi)}\rightarrow \int_0^\pi \dd \kappa$.
Note that the average current in the stationary state regime from Eq.~\eqref{eq:j_t_stationary} is in agreement with the average current prediction from the Landauer-Büttiker formalism.
There, the average current $J$ at the boundary is given by the expression from Eq.~\eqref{eq:J_zero} in the limit of $\Sigma\rightarrow\infty$.
Explicit integration using the asymptotic boxcar transmission function with $T(E)=1$ for $E\in[-2g,2g]$ and $T(E)=0$ otherwise results in $J = \int_{-2g}^{2g} \frac{\dd E}{2\pi} = 2g/\pi$ as the calculation in Eq.~\eqref{eq:j_t_stationary}.

Going towards the second moment, we have to derive the two-time correlation function $\langle j(t) j(s) \rangle_{\rm FS^+}$.
Since $\ket{\rm FS^+}$ is a stationary state, we can instead look at $\langle j(\tau) j(0) \rangle_{\rm FS^+}$ with $\tau = t-s$.
\begin{widetext}
    \begin{align}
        \langle j(\tau) j(0)\rangle_{\rm FS^+} &= -g^2\left\langle \left(c_0^\dagger(\tau) c_1(\tau) - c_1^\dagger(\tau) c_0(\tau)\right)\left(c_0^\dagger(0)c_1(0) - c_1^\dagger(0)c_0(0)\right) \right\rangle_{\rm FS^+} \\
        &=-\frac{g^2}{N^2}\sum_{\kappa_p\kappa_q\kappa_k\kappa_\ell} \left(e^{+i\omega_{\kappa_p} \tau + i\kappa_q -i\omega_{\kappa_q}\tau} - e^{+i\omega_{\kappa_p}\tau - i\kappa_p - i\omega_{\kappa_q}\tau}\right) \left(e^{i\kappa_\ell} - e^{-i\kappa_k}\right) \left\langle e_{\kappa_p}^\dagger e_{\kappa_q} e_{\kappa_k}^\dagger e_{\kappa_\ell} \right\rangle_{\rm FS^+}\label{eq:jtau_j0_expand}
    \end{align}
\end{widetext}
where we used that $\dot e_\kappa =+i[H_{\rm bulk},e_\kappa]$ and thus $e_\kappa(\tau) = e^{-i\omega_\kappa\tau} e_\kappa$, with $\omega(\kappa)=-2g\cos(\kappa)$.
The four-point correlator in Eq.~\eqref{eq:jtau_j0_expand} can be expanded using the fact that $\ket{{\rm FS}^+}$ is a Slater-determinant state,
\begin{align}
    \langle e_{\kappa_p}^\dagger e_{\kappa_q} e_{\kappa_k}^\dagger e_{\kappa_\ell} \rangle_{\rm FS^+}  
    &=\delta_{kq}\langle e_{\kappa_p}^\dagger e_{\kappa_\ell} \rangle_{\rm FS^+} \label{eq:delta_kq_pl} \\
    &\quad - \langle  e_{\kappa_k}^\dagger e_{\kappa_q} \rangle_{\rm FS^+}  \langle e_{\kappa_p}^\dagger e_{\kappa_\ell} \rangle_{\rm FS^+} \label{eq:kq_pl}  \\
    &\quad +  \langle  e_{\kappa_k}^\dagger e_{\kappa_\ell}  \rangle_{\rm FS^+}  \langle e_{\kappa_p}^\dagger e_{\kappa_q}  \rangle_{\rm FS^+}. \label{eq:kl_pq}
\end{align}
In the limit of $N\rightarrow\infty$ the terms in lines~\eqref{eq:delta_kq_pl},~\eqref{eq:kq_pl} and~\eqref{eq:kl_pq} together with the sum from Eq.~\eqref{eq:jtau_j0_expand} can be written as integrals.
For some general function $f$ we can write line~\eqref{eq:delta_kq_pl} as
\begin{align}
    &-\frac{g^2}{N^2}\sum_{\kappa_p\kappa_q\kappa_k\kappa_\ell} \delta_{kq} \langle e_{\kappa_p}^\dagger e_{\kappa_\ell} \rangle_{\rm FS^+} f(\kappa_k,\kappa_q,\kappa_p, \kappa_\ell) \nonumber \\
    &\qquad \rightarrow -\frac{g^2}{4\pi^2} \int_{-\pi}^\pi \dd \kappa \int_0^\pi \dd \varsigma \, f(\kappa,\kappa,\varsigma,\varsigma), \label{eq:kq_pl_term1}
\end{align}
and line~\eqref{eq:kq_pl} as
\begin{align}
    &+\frac{g^2}{N^2}\sum_{\kappa_p\kappa_q\kappa_k\kappa_\ell} \langle  e_{\kappa_k}^\dagger e_{\kappa_q} \rangle_{\rm FS^+}  \langle e_{\kappa_p}^\dagger e_{\kappa_\ell} \rangle_{\rm FS^+} f(\kappa_k,\kappa_q,\kappa_p, \kappa_\ell) \nonumber \\
    &\qquad \rightarrow +\frac{g^2}{4\pi^2} \int_{0}^\pi \dd \kappa \, \dd \varsigma \, f(\kappa,\kappa,\varsigma,\varsigma). \label{eq:kq_pl_term2}
\end{align}
Both terms together yield
\begin{align}
    \eqref{eq:kq_pl_term1}+\eqref{eq:kq_pl_term2} &= -\frac{g^2}{4\pi^2}\int_0^\pi \dd \kappa\,\dd \varsigma f(-\kappa,-\kappa,\varsigma,\varsigma). \label{eq:integral_A}
\end{align}
The third term in line~\eqref{eq:kl_pq} yields
\begin{align}
    &-\frac{g^2}{N^2}\sum_{\kappa_p\kappa_q\kappa_k\kappa_\ell}\langle  e_{\kappa_k}^\dagger e_{\kappa_\ell}  \rangle_{\rm FS^+}  \langle e_{\kappa_p}^\dagger e_{\kappa_q}  \rangle_{\rm FS^+} f(\kappa_k,\kappa_q,\kappa_p,\kappa_\ell) \nonumber\\
    &\qquad\rightarrow -\frac{g^2}{4\pi^2} \int_0^\pi \dd\kappa\,\dd\varsigma f(\kappa,\varsigma,\varsigma,\kappa). \label{eq:integral_B}
\end{align}
\begin{widetext}
\noindent Inserting the functional form of $f$ from Eq.~\eqref{eq:jtau_j0_expand} into the integrals yields,
\begin{align}
    \eqref{eq:integral_A} &= -\frac{g^2}{4\pi^2}\int_0^\pi \dd\kappa\,\dd\zeta \left( e^{i\omega_\varsigma \tau - i \kappa -i\omega_\kappa\tau} - e^{i\omega_\varsigma \tau -i\varsigma - i\omega_\kappa\tau} \right) \left(e^{i\varsigma} - e^{i\kappa}\right) \\
    &=-\frac{g^2}{4\pi^2}\int_0^\pi \dd\kappa\,\dd\zeta 
    \left(
          e^{i(\omega_\varsigma \tau + \varsigma) - i(\omega_\kappa\tau + \kappa)}
        + e^{i(\omega_\varsigma\tau - \varsigma) - i(\omega_\kappa - \kappa)} 
        - e^{i\omega_\varsigma\tau -i\omega_\kappa\tau}  
        - e^{i\omega_\varsigma\tau - i\omega_\kappa\tau} 
    \right) \\
    &=-\frac{g^2}{4\pi^2}
    \left(
        \left|\int_0^\pi \dd \kappa \, e^{i(\omega_\kappa \tau + \kappa)} \right|^2
        +\left|\int_0^\pi \dd \kappa \, e^{i(\omega_\kappa \tau - \kappa)} \right|^2
        -2\left|\int_0^\pi \dd \kappa \, e^{i\omega_\kappa \tau} \right|^2
    \right).\label{eq:ugly_integral_ABC}
\end{align}
\end{widetext}
Recalling that $\omega_\kappa=-2g\cos(\kappa)$ we can express the integrals from Eq.~\eqref{eq:ugly_integral_ABC} in closed form using Bessel functions~\cite{NIST:DLMF10.9i}.
Denoting by $J_n(z)$ the Bessel function of kind zero, order $n$ (following the convention of~\cite{NIST:DLMF10.9i}), we have
\begin{align}
    \int_0^\pi \dd\kappa\, e^{-2i g\tau \cos(\kappa) + i\kappa} &= i\left(\frac{\sin(2g\tau)}{g\tau} - \pi J_1(2g\tau)\right),
\end{align}
as well as
\begin{align}
    \int_0^\pi \dd \kappa\, e^{-2ig\tau \cos(\kappa)} &= \pi J_0(2g\tau).
\end{align}
Consequently, we find
\begin{align}
    \eqref{eq:integral_A} 
    &=\frac{g^2}{2}\left(J_0(2g\tau)^2 - J_1(2g\tau)^2\right) - \frac{2g^2}{\pi^2}\frac{\sin(2g\tau)^2}{(2g\tau)^2}.
    \label{eq:ugly_integral_finalA}
\end{align}
When considering the remaining terms from Eq.~\eqref{eq:jtau_j0_expand} given by the integral from Eq.~\eqref{eq:integral_B}, we find
\begin{align}
    \eqref{eq:integral_B} 
        &= -\frac{g^2}{4\pi^2} \int_0^\pi \dd \kappa \dd \varsigma \, \left( e^{i\kappa} - e^{-i\kappa} \right)\left( e^{i\varsigma} - e^{-i\varsigma} \right) \nonumber \\
        &= -\left(\frac{ig}{\pi} \int_0^\pi \dd\kappa\, \sin(\kappa)\right)^2 = +\frac{4g^2}{\pi^2},\label{eq:ugly_integral_finalB}
\end{align}
which equals the average current squared already obtained in Eq.~\eqref{eq:j_t_stationary}.
Combining the two resulting expressions from Eq.~\eqref{eq:ugly_integral_finalA} and Eq.~\eqref{eq:ugly_integral_finalB} gives us
\begin{align}
    \langle\langle j(\tau)j(0) \rangle\rangle_{\rm FS^+} &= \langle j(\tau)j(0) \rangle_{\rm FS^+} - \langle j(\tau) \rangle_{\rm FS^+} \langle j(0)\rangle_{\rm FS^+} \nonumber \\
    &\hspace{-2cm}= \frac{g^2}{2}\left(J_0(2g\tau)^2 - J_1(2g\tau)^2\right) - \frac{2g^2}{\pi^2}\frac{\sin(2g\tau)^2}{(2g\tau)^2}.
\end{align}
This concludes the calculation of the two-time fluctuations of the current $j$ in the chain's bulk.
In the limit of $N\rightarrow\infty$ with stationary state equal to the shifted Fermi sea $\ket{\rm FS^+}$, this is the analytically exact result.

\subsection{\label{SM:calculating_number_variance}Calculating the number variance}
The variance $\Var[N^{\rm bulk}_t]$ can be calculated using Eq.~\eqref{eq:VarNtcoh_k}.
With the two-time correlation function from Appendix~\ref{SM:calculating_two_point_fluctuations}, it is given by
\begin{align}
    \Var[N_t^{\rm bulk}] 
    &= \int_0^t \dd t'\dd t''\, \langle\langle j(t''-t')j(0) \rangle\rangle_{\rm FS^+}\label{eq:Var_NT_line1}\\
    &=2\int_0^t \dd \tau (t-\tau) \langle\langle j(\tau)j(0) \rangle\rangle_{\rm FS^+}. \label{eq:Var_NT_line2}
\end{align}
Going from line~\eqref{eq:Var_NT_line1} to line~\eqref{eq:Var_NT_line2} relies on a change of variables and the integrand only depending on the time difference $\tau = t''-t'$.
Further abbreviating $a:=2gt$ for the time being, we can write
\begin{align}
    \text{\eqref{eq:Var_NT_line2}} &= \! \int_0^a \! \dd x \frac{a-x}{4}\left(J_0(x)^2 - J_1(x)^2 - \frac{4\,{\rm sinc}(x)^2}{\pi^2}\right),
\end{align}
with ${\rm sinc}(x)=\sin(x)/x$.
To determine the number variance we split the integration into three parts.
The first term can be determined using the integral identity for Bessel functions (e.g.\ Ref.~\cite{Rosenheinrich2016}) stating that
\begin{align}
    \frac{a}{4}\int_0^a \dd x \left(J_0(x)^2 - J_1(x)^2\right) &= \frac{ax}{4}\left(J_0(x)^2 + J_1(x)^2\right)\bigg|_0^a \nonumber \\
    &=\frac{a^2}{4}\left(J_0(a)^2 + J_1(a)^2\right). \label{eq:VarNT_partA}
\end{align}
By using the asymptotic expansions for $x\rightarrow\infty$ of the first two Bessel functions,
\begin{align}
    J_0(x) &= \sqrt{\frac{2}{\pi x}} \sin\left(x + \frac{\pi}{4}\right) -\frac{1}{4\sqrt{2\pi x^3}}\cos\left(x + \frac{\pi}{4}\right) \nonumber  \\
    &\qquad + \mathcal O(x^{-5/2}),
\end{align}
as well as,
\begin{align}
    J_1(x) &= -\sqrt{\frac{2}{\pi x}} \cos\left(x + \frac{\pi}{4}\right) + \frac{3}{4\sqrt{2\pi x^3}}\sin\left(x+\frac{\pi}{4}\right) \nonumber  \\
    &\qquad +  \mathcal O(x^{-5/2}),
\end{align}
we can expand Eq.~\eqref{eq:VarNT_partA} to leading order,
\begin{align}
    \eqref{eq:VarNT_partA} &= +\frac{a}{2\pi} - \frac{1}{4\pi}\cos(2a) + \mathcal O(a^{-1}), \label{eq:VarNT_partA_NICE}
\end{align}
as $a=2gt \rightarrow\infty$.
The next term in the integral for $\Var[N_t]$ can be written in closed form as well using Bessel functions and integral identities from Ref.~\cite{Rosenheinrich2016},
\begin{align}
    -\frac{1}{4}\int_0^a \dd x\, x\left(J_0(x)^2 - J_1(x)^2\right) 
    &= -\frac{x}{4}J_1(x)J_0(x)\bigg|_0^a \nonumber \\
    &= -\frac{a}{4}J_1(a)J_0(a). \label{eq:VarNT_partB}
\end{align}
Asymptotically expanding the result for large values of $a$ yields,
\begin{align}
    \eqref{eq:VarNT_partB} &= \frac{1}{4\pi} \cos(2a) + \mathcal O(a^{-2}),\label{eq:VarNT_partB_NICE}
\end{align}
precisely cancelling the constant order ferm from Eq.~\eqref{eq:VarNT_partA_NICE}.
Finally, we wish to calculate the integral
\begin{align}
    -\frac{1}{\pi^2} \int_0^a \dd x (a-x)\, {\rm sinc}(x)^2.\label{eq:sinc_integral}
\end{align}
Using partial integration and trigonometric identities such as $\sin(x)^2 = (1-\cos(2x))/2$, we can write
\begin{align}
    \int_0^a \dd x\,{\rm sinc}(x)^2 &= {\rm Si} (2a) - \frac{\sin(a)^2}{a},
\end{align}
where ${\rm Si}(x)=\int_0^x \dd x \,{\rm sinc}(x)$ is the Sine integral.
The other term can also be expressed in closed form using the Cosine integral~\cite{NIST:DLMFChapter6},
\begin{align}
    \int_0^a \dd x\,x\,{\rm sinc}(x)^2 &=\frac{1}{2}(-{\rm Ci}(2a) + \log(2a) + \gamma),
\end{align}
where ${\rm Ci}(x) = \int_x^\infty \dd x\cos(x)/x$, and $\gamma$ is the Euler-Mascheroni constant.
All in all, Eq.~\eqref{eq:sinc_integral} can thus be written in closed form using the Sine and Cosine integrals,
\begin{align}
    \eqref{eq:sinc_integral}
        &= \frac{1}{\pi^2}\Big(\sin(a)^2 - a\,{\rm Si}(2a)\nonumber \\
        &\hspace{1cm} +\frac{1}{2}\left(\log(2a) +\gamma - {\rm Ci}(2a)\right)\Big).\label{eq:sinc_integral_next}
\end{align}
Asymptotically as $x\rightarrow\infty$ the above expression can be further expanded using Ref.~\cite{NIST:DLMFChapter6}, with
\begin{align}
    {\rm Si}(x) &= \frac{\pi}{2} - \frac{\cos(x)}{x} + \mathcal O(x^{-1}),
\end{align}
and
\begin{align}
    {\rm Ci}(x) &= \frac{\sin(x)}{x} + \mathcal O(x^{-2}).
\end{align}
Inserting for large values of $a\rightarrow\infty$ yields
\begin{align}
    \eqref{eq:sinc_integral_next} &= -\frac{a}{2\pi} + \frac{1}{2\pi^2}\left(\log 2a + \gamma + 1\right) + \mathcal O(a^{-1}).\label{eq:VarNT_partC_NICE}
\end{align}
Combining all terms gives the number variance in full functional form, expressed in terms of Bessel functions, Sine and Cosine integrals,
\begin{align}
    \Var[N_t^{\rm bulk}] &= \frac{a^2}{4}\Big(\left(J_0(a)^2 + J_1(a)^2\right) - \frac{J_1(a)J_0(a)}{a}\Big) \nonumber \\
    & \hspace{-1.25cm}+ \frac{1}{\pi^2}\Big(\sin(a)^2 - a\,{\rm Si}(2a) +\frac{1}{2}\left(\log(2a) +\gamma - {\rm Ci}(2a)\right)\Big) .\label{eq:VarNT_FINALFULL}
\end{align}
Taking the large-$t$ expansions by combining Eq.~\eqref{eq:VarNT_partA_NICE} with~\eqref{eq:VarNT_partB_NICE} and~\eqref{eq:VarNT_partC_NICE}, we obtain the asymptotic form:
\begin{align}
    \Var[N_t^{\rm bulk}] &= \frac{1}{2\pi^2}\left(\log(4gt) + \gamma + 1\right) + \mathcal O\left((gt)^{-1}\right).
    \label{eq:VarNT_FINALderivation}
\end{align}

In Fig.~\ref{fig:Fig4_VarNt}, the analytical expression is compared to the numerical simulation for finite $N$ and the stochastic Schr\"odinger evolution data.

\section{\label{SM:Connection_RMT}Connection random matrix theory}
The scaling behavior observed for a discrete fermionic chain in the limit of $N\rightarrow\infty$ can be recovered in a one-dimensional continuum scattering problem.
There, the connection to random matrix universality can be made fully exact and analytical as we outline in this section.
We start this appendix with the Levitov--Lesovik expression for full counting statistics (Appendix~\ref{SM:levitov-lesovik}), and then explicitly solve the one-dimensional scattering problem (Appendix~\ref{SM:1D_scattering_setup}).
We conclude by showing how the identical result arises in random matrix theory (Appendix~\ref{SM:random_matrix_universality}).

\subsection{\label{SM:levitov-lesovik}Levitov--Lesovik formula}
The bulk number statistics can be more formally expressed using the moment generating function $M(\lambda)=\langle e^{\lambda N_t}\rangle$, where the average in our case is taken with respect to the shifted Fermi sea state $\ket{\rm FS^+}$.
The moments can thus conveniently be obtained using
\begin{align}
    \langle N_t^m \rangle &=\partial_\lambda^m M(\lambda) \Big|_{\lambda=0}.
\end{align}
In the seminal works by Levitov, Lesovik and Lee~\cite{Levitov1993,Levitov1996,Avron2008}, a concise formal expression for the moment generating function has been derived for Gaussian fermionic systems.
The identity reads,
\begin{align}
\label{eq:e^lambdaN}
    \left\langle e^{\lambda N_t} \right\rangle_{\rm FS^+} = \det[\mathds 1 - C(e^{\lambda A}-\mathds 1)],
\end{align}
where $C_{ij}=\langle c_i^\dagger c_j\rangle$ is the usual covariance matrix, and the matrix $A$ in the determinant are the matrix elements from expanding
\begin{align}
    N_t = \sum_{ij=1}^N A_{ij} c_i^\dagger c_j.
\end{align}
The Levitov--Lesovik identity~\eqref{eq:e^lambdaN} holds for fully generally for all finite Gaussian fermionic systems.
In the case of our clock,
using the state $\ket{\rm FS^+}$, the covariance matrix $C$ is a projector onto all positive momenta $\kappa\in[0,\pi)$.
The counting operator $N$ takes a particular form given by
\begin{align}
    N_t &= \int_0^t \dd t' \, j(t') \\
    &= -i g\int_0^t \dd t'  \left( c_0^\dagger(t') c_1(t') - c_1^\dagger(t') c_0(t')\right),
\end{align}
where the current operator can be further expanded in the momentum basis using
\begin{align}
    &c_0^\dagger(t) c_1(t) - c_1^\dagger(t) c_0(t) =\\
    &\quad \frac{1}{N}\sum_{\kappa_p\kappa_q}\left(e^{+i\omega_{\kappa_p} t - i\omega_{\kappa_q} t + i\kappa_q} - e^{+i\omega_{\kappa_p}t - i\kappa_p - i\omega_{\kappa_q} t}\right) e_{\kappa_p}^\dagger e_{\kappa_q}. \nonumber 
\end{align}
To simplify the notation, we can write $\kappa_p \to \kappa$ and $\kappa_q\to \kappa'$, keeping the discreteness of the $N$ momenta in $[-\pi,\pi)$ implicit.
The time-integration can be done in closed form to yield the matrix elements of $N_t$ with respect to the momentum basis,
\begin{align}
\label{eq:N_kappakappa'_discrete}
    A_{\kappa\kappa'} &=\frac{g}{N}\frac{e^{i(\omega_\kappa -\omega_{\kappa'})t}-1}{\omega_\kappa-\omega_{\kappa'}}\left(e^{i\kappa} - e^{-i\kappa'}\right) \\
    &\propto \frac{gt}{N}\,{\rm sinc}\left(\frac{(\omega_\kappa-\omega_{\kappa'})t}{2}\right)\sin\left(\frac{\kappa+\kappa'}{2}\right),
\end{align}
where in the second line we factored out phases.

This can be compared to the expression obtained in the context of a continuum 1D scattering problem with unit transmission as we derive in the following Appendix~\ref{SM:1D_scattering_setup}.
There, the covariance matrix in momentum space is also a projector onto the scattering momenta, and the number operator can be expressed as
\begin{align}
\label{eq:N_kappakappa'_continuous}
    A_{\kappa\kappa'} &\propto \frac{vt}{L}\,{\rm sinc}\left(\frac{(\kappa-\kappa')vt}{2}\right).
\end{align}
While the sinc-kernel in~\eqref{eq:N_kappakappa'_continuous} is equidistantly sampled by the momenta $\kappa,\kappa'$, we see that in contrast, the discrete chain (due to nonlinear dispersion) samples the kernel~\eqref{eq:N_kappakappa'_discrete} non-equidistantly with $\omega_\kappa-\omega_{\kappa'}$.
Remarkably, in the late time limit $t\rightarrow\infty$, we recover the universal log-scaling of the variance in both cases.

\subsection{\label{SM:1D_scattering_setup}One-dimensional scattering setup}
Similar number statistics are obtained when considering a periodic, 1D scattering region of length $L$, occupied by plane waves of the form
\begin{align}
    \psi_k(x) = \frac{1}{\sqrt{L}} e^{i\kappa x},
\end{align}
where $\kappa$ is the momentum number, and where the excitations obeying fermi statistics.
Due to periodicity, the momenta are discretized to take values $\kappa_n=2\pi n / L$ where $n\in \mathbb Z$ is an integer.
If the transmission band is filled with $N$ excitations between momenta $k=0$ and $\kappa=2\pi N / L$, the many-body particle wave function can be expressed as a Slater determinant
\begin{align}
\label{eq:psi(x1,...,xN)}
    \psi(x_1,\dots,x_N)= \frac{1}{\sqrt{N!}}\det\left[\psi_{\kappa_n}(x_m)\right]_{m,n=1}^N.
\end{align}
For a linear dispersion relation $E(\kappa)=v\kappa$, time evolution rigidly translates all the excitations with velocity $v$.
Ticks can be identified by excitations passing the boundary $x=0$.
To determine the number of ticks $N_t$ that occurred in an interval $[0,t]$, we can---since all excitations travel with the same velocity $v$---simply look at the number of excitations in the spatial interval $[0,vt]$, so long as $vt<L$ is less than a round trip time.
In second-quantized form, the number operator is then given by
\begin{align}
\label{eq:N_realspace}
    N_t = \int_0^{vt}\dd x \, \Psi(x)^\dagger \Psi(x),
\end{align}
where $\Psi(x)$ is the position field operator, satisfying canonical commutation relations $\{\Psi^\dagger(x),\Psi(y)\}=\delta(x-y)$,  and $\{\Psi(x),\Psi(y)\}=0$.
In analogy to the discrete chain, the moment generating function can be simplified using the Levitov--Lesovik identity,
\begin{align}
\label{eq:e^lambdaN_cont}
    \left\langle e^{\lambda N_t}\right\rangle_\psi &= \det\left[\mathds 1 - C\left(e^{\lambda A} - 1\right)\right].
\end{align}
In this case, the covariance matrix is simply given by occupation of the momenta within the interval $[0,2\pi N/L]$,
\begin{align}
    C_{\kappa_m\kappa_n} &= \delta_{mn} \chi_{[0,2\pi N / L]}(k_n).
\end{align}
As for the number operator~\eqref{eq:N_realspace}, the momentum space representation can be written as
\begin{align}
    N_t &= \sum_{m,n\in\mathbb Z} A_{\kappa_m,\kappa_n} c_{\kappa_m}^\dagger c_{\kappa_n},
\end{align}
where the transformation into momentum space is defined by the Fourier transformation
\begin{align}
    \Psi(x) &= \frac{1}{\sqrt{L}}\sum_{n\in \mathbb Z} e^{ik_n x}c_{k_n}.
\end{align}
The fermionic mode operators in momentum space $c_\kappa$ also satisfying canonical (anti-)commutation relations (Poisson summation formula for the proof).
As a result, the matrix elements for $N_t$ can be written as
\begin{align}
    A_{\kappa \kappa'} &= \frac{e^{-i(\kappa-\kappa')vt}-1}{-iL(\kappa-\kappa')} \propto \frac{vt}{L}\,{\rm sinc}\left(\frac{(\kappa-\kappa')vt}{2}\right),
\end{align}
like in Eq.~\eqref{eq:N_kappakappa'_continuous}.

Such a setup can, for example, be physically obtained by considering electrons, spinless and non-interacting for simplicity, that travel through a scattering potential on an interval $[-L,0]$ as for example in~\cite{Albert2012,Dasenbrook2015}.
By applying a chemical potential bias $\Delta$ across the scatterer by having a chemical potential $\mu_L$ on the left and $\mu_R$ on the right with $\mu_L-\mu_R=\Delta$, we can achieve that electrons travel from $x<-L$ through the scatterer to $x>0$ between the energy window $[\mu_R,\mu_L]$.
For a sufficiently large chemical potential in comparison to the bias, i.e., $\Delta\ll \min\{\mu_l,\mu_R\}$, the free Hamiltonian $H=P^2/2m$, where $m$ is the particle mass and $P$ the momentum operator, can be linearized as $H \approx v_F P$.
Where we identified $v_F=\sqrt{2\mu_R/m}$ as the \textit{fermi-velocity}, and ignored constant shifts of the Hamiltonian.
In the stationary regime, for unit transmission probability, all states in the transmission window $[\mu_R,\mu_L]$ are occupied, and up to a a shift in momenta, the state takes of the form~\eqref{eq:psi(x1,...,xN)}.
The number of particles in the band can then be obtained using momentum discretization, $\kappa_n = 2\pi n / L$ and the linear dispersion $E(\kappa)=v_F\kappa$ by fixing $E(\kappa_N)=\Delta$.

\subsection{\label{SM:random_matrix_universality}Random matrix universality}

Taking the absolute square of~\eqref{eq:psi(x1,...,xN)} gives the probability distribution of the particle's positions (and by identifying $x=vt$ also tick times),
\begin{align}
\label{eq:p(x1,...,xN)}
    p(x_1,\dots,x_N) &= |\psi(x_1,\dots,x_N)|^2 \\
    &= \frac{1}{N!}\det\left[K_N(x_i,x_j)\right]_{i,j=1}^N.
\end{align}
The determinant kernel $K_N$ is obtained using the product identity $\det[AB]=\det[A]\det[B]$ for two square matrices $A,B$ with matching dimensions, and given by,
\begin{align}
    K_N(x,y) = \sum_{m=0}^{N-1} \psi_{\kappa_m}^* (x) \psi_{\kappa_m}(y).
\end{align}
Up to normalization of the positions in the interval $[0,L)$, the particles following the distribution in Eq.~\eqref{eq:p(x1,...,xN)} have the identical distribution as the phases in the interval $[0,2\pi)$ of the circular unitary ensemble~\cite{Mehta1967,Tracy1993,Tracy1998}.
In fact, the Kernel $K_N(x,y)$ can be expressed in simplified form
\begin{align}
\label{eq:K_N(x,y)_SINE}
    K_N(x,y) &= \frac{1}{L}\sum_{m=0}^{N-1} e^{2\pi i m (y-x) / L} \\
    &\propto \frac{\sin\left(\pi N  (x-y) / L\right)}{L\sin(\pi(x-y)/L)},
\end{align}
factoring out constant phases irrelevant to $p(x_1,\dots,x_N)$.
By choosing units such that $L=N$, and taking the limit of $N\rightarrow\infty $ the kernel converges to the well-known sine kernel for all $x$ and $y$ such that $|x-y|$ does not grow with $N$,
\begin{align}
    K_N(x,y) &\rightarrow K(x,y) = \frac{\sin(\pi(x-y))}{\pi(x-y)},
\end{align}
using that $N\sin(x/N) \to x$ as $N\rightarrow\infty$.
In fact, all local correlations between positions (and thus tick times) can be obtained from the kernel with
\begin{align}
    \rho^{(n)}_N(x_1,\dots,x_n) = \det\left[K_N(x_i,x_j)\right]_{i,j=1}^n,
\end{align}
where $\rho^{(n)}_N$ is the correlation function obtained from marginalizing over last the $N-n$ variables of the full $N$-particle distribution $p(x_1,\dots,x_N)$,
\begin{align}
    \rho^{(n)}_N(\dots) := \frac{N!}{(N-n)!}\int \dd x_{n+1}\cdots \dd x_N p(x_1,\dots,x_N).
\end{align}
For any fixed $n$ and $x_1,\dots,x_n$ in a bounded interval, the correlation function $\rho_N^{(n)}$ converges to a universal expression as $N\rightarrow\infty$ depending only on the sine-kernel~\eqref{eq:K_N(x,y)_SINE},
\begin{align}
    \rho_N^{(n)}(\dots)\rightarrow  \rho^{(n)}(x_1,\dots,x_n)=\det\left[K(x_i,x_j)\right]_{i,j=1}^n.
\end{align}

In the context of random matrix theory, the sine kernel is the universal kernel that describes the bulk statistics of the unitary ensembles with order parameter $\beta=2$.
That includes in particular the Gaussian unitary ensemble as well as the circular unitary ensemble~\cite{Mehta1967}.
Both tick times and eigenvalues of the unitary ensemble (as well as their respective correlations) are thus fully determined by sine-kernel universality.
Ultimately, this is also the rigorous explanation for why in~\cite{Albert2012} the Wigner-Dyson distribution is obtained for the waiting time between two electrons in a unit transmission scatterer: the probability that no tick occurs in a time interval $[0,t]$ is given by the probability that there is no particle in the interval $[0,x]$ with $x=t/v_F$.
In the picture of random matrices, this maps to the probability that there is no eigenvalue in $[0,\lambda]$, i.e., the gap probability distribution which is famously known to be well-approximated by a Wigner-Dyson distribution~\cite{Mehta1967}.
Further, the correlations between neighboring ticks as found in~\cite{Dasenbrook2015}, are also a direct consequence of this.

Finally, this also allows us to look at the correlations between ticks over longer intervals, captured for example by the variance of the number of ticks $N_t$ in an interval $[0,t]$.
In the formulation of the PDF~\eqref{eq:p(x1,...,xN)}, this translates into the number of particles on an interval of length $x=v_Ft$, given by the linear estimator
\begin{align}
    N_x(x_1,\dots,x_N) = \sum_{i=1}^N \chi_{[0,x]}(x_i),
\end{align}
with $\chi_{[0,x]}(y)$ the indicator function on the interval $[0,x]$.
When comparing to the operator form in Eq.~\eqref{eq:N_realspace}, this is in fact simply the position representation,
\begin{align}
    N_x(x_1,\dots,x_N) = \langle x_1,\dots,x_N | N_t | x_1,\dots,x_N \rangle.
\end{align}
The variance of $N_x$, in the context of random matrix theory, is known as the \textit{level number variance} which captures the spectral rigidity of the eigenvalues.
It can be directly calculated using
\begin{align}
    \E[N_x^m] &= \int \dd x_1\cdots\dd x_N p(x_1,\dots,x_N) N_x(x_1,\dots,x_N)^m,
\end{align}
which simplifies neatly for the lower moments.
In particular,
\begin{align}
    \E[N_x] &= \int \dd x_1 \chi_{[0,x]}(x_1) \rho_N^{(1)}(x_1)
\end{align}
because $\rho_N^{(1)}(x)=1$ (unit particle density).
For the second moment,
\begin{align}
    \E[N_x^2] &= \int \dd x_1 \dd x_2 \chi_{[0,x]}(x_1)\chi_{[0,x]}(x_2) \rho_N^{(2)}(x_1,x_2)  \nonumber \\
    & \qquad+ \E[N_x].
\end{align}
The integrals can be expanded for large values of $x$ in the limit of $N\rightarrow\infty$ which gives~\cite{French1978,Brody1981},
\begin{align}
    \Var[N_x] = \frac{1}{\pi^2} (\log(2\pi x) + \gamma + 1) + \mathcal{O}(x^{-1}),
\end{align}
asymptotically resulting in the same variance as for the case of the finite spin chain, despite the differences in the matrix elements Eq.~\eqref{eq:N_kappakappa'_discrete} and Eq.~\eqref{eq:N_kappakappa'_continuous}.
It thus appears that even in the discretized spin chain, the excitation current is distributed like the eigenvalues of random matrices from the unitary ensemble, up to finite-size corrections for short times.

\section{\label{SM:robustness_realistic_implementation}Robustness of realistic implementation}
In this section, we provide additional details for how the log-linear cross-over times are calculated (Appendix~\ref{SM:cross_over_times}) in the presence of the following imperfections: Finite temperature (Appendix~\ref{SM:finite_temperature_perturbation}), disorder in the spin frequencies (Appendix~\ref{SM:energy_disorder}) and the coupling between neighboring spins (Appendix~\ref{SM:coupling_disorder}).

\subsection{\label{SM:cross_over_times}Cross-over times}
Here, the cross-over time $t^*$ between logarithmic and linear variance scaling is derived.
We define the cross-over time as the time $t^*$ where the asymptotic logarithmic scaling of $\Var[N_t]$ as in Eq.~\eqref{eq:VarNT_FINALderivation} crosses with the linear diffusive scaling $\sim Dt$ at very large times $t\rightarrow\infty$.
Since the cross-over $t^*$ is not a sharp point and we are only interested in the asymptotic scaling with $D\rightarrow 0$, we ignore additive constants and look for the solution of,
\begin{align}
    \frac{1}{2\pi} \log (Jt) &= Dt.
\end{align}

Generally, two solutions exist to that equation, but we are interested in the large-$t$ solution.
This equation can be recast in the more general form,
\begin{align}
    ax &= \log(x),
\end{align}
where we abbreviate $x=Jt$ (recall $J=2g/\pi$), such that $a=2\pi D / J$.
This equation is solved by the Lambert-$W$ function~\cite{Kalugin2011}, 
\begin{align}
    x &=- \frac{W_{k}(-a)}{a},
\end{align}
with branches indexed by the integer $k$.
For real values only the branches $k=0$ and $k=-1$ are relevant.
The principal branch expands as $W_0(z)\sim z$ around $z=0$, and thus gives the smaller solution for $t^*$ which we are not interested in.
The correct cross-over time $t^*$ is thus obtained from the branch $W_{-1}(z)$.
Around $z=0$, the this branch admits an expansion in log-functions to leading order~\cite{Kalugin2011}
\begin{align}
    W_{-1}(z) = \log(-z) - \log(-\log(-z)) + o(1),
\end{align}
as $z\rightarrow 0$.
Inserting physical units yields a leading order scaling proportional to
\begin{align}
\label{eq:t^*_res_derivation}
    t^* \sim \frac{\log(JD^{-1})}{D}.
\end{align}
In some cases, however, the scaling exponent of $D$ is not exactly known and only numericaly approximated.
In such cases, the log-correction in~\eqref{eq:t^*_res_derivation} is asymptotically dominated by any arbitrarily small correction of the scaling exponent of $D$, which is why in most cases, the asymptotic scaling 
\begin{align}
\label{eq:t^*_linear_approx}
    t^* \sim \frac{1}{ D},
\end{align}
is sufficient.

\begin{figure}
    \centering
    \includegraphics[width=\linewidth]{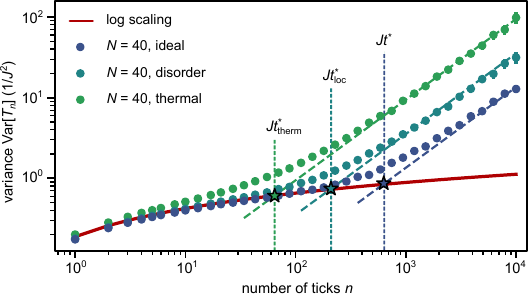}
    \caption{Scaling cross-over with imperfections.
    Here we show the variance of the waiting time for the $n$th tick as a function of $n$, for a chain of length $N=40$.
    The dots show results from a stochastic Monte Carlo simulation with $10^5$ ticks.
    In order of increasing variance: first, the case where the numerically optimized couplings are chosen like in the main text; second, in the presence of coupling disorder of $0.5\%$; and third, in the presence of non-zero temperature environments where the entropy has been chosen as $\Sigma = 5.5 k_B$.
    Dashed diagonal lines indicate the theoretical prediction for the asymptotic diffusive scaling $Dn/J^3$, the vertical lines with the star symbol indicate the cross-over time $t^*$ between the logarithmic and linear scaling as predicted by the theory.}
    \label{fig:Fig3_EndMatter}
\end{figure}

For finite $N$, we have numerically determined the scaling $D/J\sim N^{1.86}$, indicating that the diffusion constant almost quadratically vanishes with system size.
This implies that the cross-over time scales as
\begin{align}
    t^* \sim \frac{1}{J N^{1.86}},
\end{align}
where we dropped the $\log(N)$ correction because any arbitrarily small numerical uncertainty in the exponent $1.86$ asymptotically dominates the logarithmic factor.
Figure~\ref{fig:Fig3_EndMatter} shows the cross-over behavior for a chain of $N=40$ sites using a stochastic Monte Carlo simulation of over $10^5$ ticks, together with the cross-over due to thermal and disorder effects.

In the following appendices, we derive analogous cross-over time scalings, for finite temperatures $t_{\rm therm}^*$ in Appendix~\ref{SM:finite_temperature_perturbation}, and disorder $t_{\rm loc}^*$ in Appendices~\ref{SM:energy_disorder} and~\ref{SM:coupling_disorder} (also shown in Fig.~\ref{fig:Fig3_EndMatter}).

\subsection{\label{SM:finite_temperature_perturbation}Finite temperature perturbations}
\begin{figure}
    \centering
    \includegraphics[width=\linewidth]{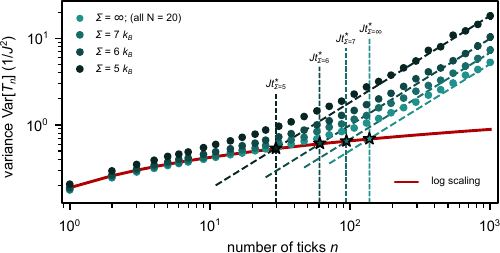}
    \caption{Waiting time variance with finite temperature. Here, the coarse-grained variance of $T_n$ is shown for $N=20$ sites, and finite temperature.
    Temperature is parametrized by entropy dissipation $\Sigma = \beta(\mu_L-\mu_R)$, where $\beta$ is the inverse temperature of the environment and $\mu_{L(R)}$ the chemical potential of the leads.
    The case of infinite dissipation arises from absolute zero temperature or infinite bias.
    The smaller the dissipation, the more noise affects the ticks and thus the log-scaling of the variance breaks down earlier.
    The figure shows, however, already moderate dissipation is sufficient to recover the log-scaling from the ideal setting.}
    \label{fig:Fig5_Thermal}
\end{figure}

The effect of finite-temperature imperfections can be analytically treated in the wide-band limit (WBL) of the master equation (see Methods).
In WBL it is assumed that the Fermi-functions,
\begin{align}
\label{eq:f_Fermi}
    f_{L(R)}(E) = \frac{1}{1+e^{\beta(E-\mu_{L(R)})}},
\end{align}
do not change appreciably within the support of the transmission function $T(E)$ defined by the chain.
Thus, they can be approximated as constants, evaluated at the on-site energy $\omega$.
Recall that $\mu_{L(R)}$ denotes the left (right) chemical potential and $\beta$ is the inverse temperature.
For simplicity of the analysis, we assume that the chemical potentials are symmetric and parametrized as $\mu_{L(R)} = \omega \pm V$.
The WBL is valid, for example, for large potential bias $V\gg g$ quantified by large entropy $\Sigma:= \beta V\gg 1$, in which case we can approximate
\begin{align}
    f_{L(R)} &= \frac{1}{1+e^{\pm \Sigma}}.
\end{align}
The diffusion constant $D_\Sigma$ in the WBL can be explicitly calculated using Eq.~\eqref{eq:LB-noise} from the Methods:
\begin{align}
    D_\Sigma &= D(f_L-f_R)^2 + J\big(f_L(1-f_L) + f_R(1-f_R)\big) \nonumber \\
    &=D \tanh\left(\Sigma/2\right)^2 + \frac{J}{2\cosh(\Sigma/2)^2}.
\end{align}
Here, $D$ and $J$ are the diffusion constant and the current in case of absolute zero temperature and infinite bias as considered in the main text.
Furthermore, the current can also be explicitly determined,
\begin{align}
    J_\Sigma = J(f_L-f_R) = J\tanh(\Sigma/2).
\end{align}
When we consider the limit of $D\rightarrow 0$ for very long chains $N\rightarrow\infty$, only the thermal contribution to the noise remains and is given by $J/(2\cosh(\Sigma/2)^2)$.
Using the asymptotic expansion of $\cosh(x)=\exp(x)/2 + \dots$ for large $x$, we can approximate the scaling of the diffusion constant in the limit of $\Sigma\rightarrow\infty$ (ignoring constant factors)
\begin{align}
    D_\Sigma 
    &\sim Je^{-\Sigma}. \label{eq:D_Sigma_JexpSigma}
\end{align}
Further note that $J_\Sigma$ and $J$ only differ by an exponentially small correction and thus $D_\Sigma/J_\Sigma$ and $D_\Sigma/J$ are asymptotically equal.
Using the relationship~\eqref{eq:t^*_res_derivation}, the cross-over time $t^*$ due to thermal noise can thus be inferred to be
\begin{align}
\label{eq:t*_expSigma}
    t^* \sim \frac{1}{J}\Sigma e^\Sigma.
\end{align}
In Fig.~\ref{fig:Fig5_Thermal}, we showcase a numerical example for $N=20$ sites where the cross-over from the logarithmic to linear scaling of the variance is shown for different values of $\Sigma/k_B = 5,6,7$ and the limit $\Sigma/k_B=\infty$.

\paragraph*{Thermal noise beyond the wide-band limit.}
In the limit of $N\rightarrow\infty$ where we can isolate thermal from finite-size effects, the thermal noise can in fact be derived in closed form beyond the WBL approximation.
In the that limit with perfect apodization, the transmission function approaches
\begin{align}
\label{eq:T(E)_HEaviside}
    T(E) = \Theta\big((E-\omega)-2g\big) \Theta\big((E-\omega)+2g\big),
\end{align}
where we remind the reader that $\Theta$ is the Heaviside step function.
The shot-noise term vanishes in that limit and the diffusion constant can thus be explicitly expressed as,
\begin{align}
    D &= \int_{\omega- 2g}^{\omega+2g} \frac{\dd E}{2\pi} \, \sum_{\alpha=L,R} f_\alpha(E)(1-f_\alpha(E)) \\
    &= \frac{1}{2\pi} \int_{-2g}^{2g} \dd E\, \frac{1}{1+\cosh\big(\beta(E-V)\big)},
    \label{eq:D_WBL_temp}
\end{align}
where we used symmetry between left and right for going to the second line.
The integral admits a closed solution thanks to the identity $\tanh(x)'=(1+\cosh(x))^{-1}$ and is given by
\begin{align}
\label{eq:D_tanh}
    D &= \frac{1}{2\pi\beta} \Big(\tanh\big(2\beta g-\Sigma\big) + \tanh\big(2\beta g+\Sigma\big) \Big).
\end{align}

In the regime of large potential bias $\Sigma \gg \beta g$ we can use the asymptotic expansion $\tanh(x)=1-2e^{-2x}$ for $x\rightarrow\infty$ and analogously for $\tanh(-x)=-\tanh(x)$.
Thus, the diffusion constant scales as
\begin{align}
    D & \sim \frac{2\sinh(4\beta g)}{\beta\pi} e^{-2\Sigma},
\end{align}
and is exponentially suppressed in the parameter $\Sigma$.
If we further work in the WBL where $\beta g \ll 1$, the $\sinh$-term can be expanded using $\sinh(x)=x+\dots$ for small $|x|\ll 1$, giving
\begin{align}
    D \sim \frac{8g}{\pi} e^{-\Sigma},
\end{align}
in agreement with the scaling directly obtained using the WBL~\eqref{eq:D_Sigma_JexpSigma}.
To leading order, the cross-over time $t^*$ obtained from the relation~\eqref{eq:t^*_res_derivation} thus scales exponentially in the entropy $\Sigma$, even beyond the WBL approximation.

In contrast, the cross-over time-scale $t^*\sim \beta$ obtained by Levitov, Lee and Lesovik~\cite{Levitov1996} is proportional to the Planckian time-scale $\beta$.
Whereas here, for large bias, the time-scale is exponentially enhanced by the entropy $\Sigma$.
We can recover the linear scaling in the case where the potential bias defined by $V$ is set within the transport window, i.e., $0<V<2g$.
Then, the $\tanh(x)$ terms on the right-hand side of Eq.~\eqref{eq:D_tanh} are of order unity giving $D \sim \frac{1}{\beta}$. 
Expressing in terms of the cross-over time (and ignoring the almost constant $\log(J\beta)$ term) we get, 
\begin{align}
    t^* \sim \beta,
\end{align}
the scaling from~\cite{Levitov1996}.

\paragraph*{Unidirectional current noise from the Landauer--Büttiker formalism via FCS.}
The diffusion constant $D$ as derived in Eq.~\eqref{eq:D_Sigma_JexpSigma} describes the asymptotic fluctuations of the net current $N_t = N_t^+ - N_t^-$.
The net current is given by $N_t^+$, the total number of excitations that have left the chain on the right, minus $N_t^-$, the total number of excitations that have entered the chain on the right.
If, as opposed to the net current, one only considers the forward current $N_t^+$ and discards all backwards jumps, the diffusion constant $D_+$ in general differs from $D$ as defined by the net current $N_t$.
Asymptotically, in the regime of large entropy $\Sigma$ and thus very few reverse jumps, the statistical distribution of $N_t$ differs only little from that of $N_t^+$.
As $\Sigma\rightarrow\infty$, both diffusion constants $D$ as well as $D_+$ approach a fixed value, exponentially quickly in $\Sigma$, showing that the cross-over time scaling in Eq.~\eqref{eq:t*_expSigma} remains.

Starting from the Landauer--Büttiker (LB) formalism, one can isolate the unidirectional 
transport contributions by employing the framework of Full Counting Statistics 
(FCS). 
In this approach, the full counting statistics of coherent two-terminal transport can be 
constructed from the contribution of each independent energy channel. For a fixed energy $E$, 
a single scattering attempt results in a net transferred charge $\Delta n \in \{+1,0,-1\}$ 
with probabilities
\begin{align}
P_E(+1) &= T(E)\, f_L(1-f_R),\\
P_E(-1) &= T(E)\, f_R(1-f_L),\\
P_E(0) &= 1 - P_E(+1) - P_E(-1),
\end{align}
corresponding respectively to forward transfer, backward transfer, or no net transfer, 
with $T(E)$ the transmission function and $f_L(E)$, $f_R(E)$ the Fermi distributions of 
the left and right reservoirs.
For this single channel, the cumulant generating function (CGF) is
\begin{equation}
\chi_E(\lambda)
= \big\langle e^{i\lambda \Delta n} \big\rangle
= P_E(0) + P_E(+1)e^{i\lambda} + P_E(-1)e^{-i\lambda},
\end{equation}
which can be written compactly as
\begin{equation}
\chi_E(\lambda)
= 1 + P_E(+1)(e^{i\lambda}-1) + P_E(-1)(e^{-i\lambda}-1).
\end{equation}
To keep forward and backward transfer events distinct, it is convenient to introduce 
two counting fields $\lambda_+$ and $\lambda_-$, assigning the phases $e^{+i\lambda_+}$(for forward), $e^{-i\lambda_-}$ (for backward).
The single-energy CGF then becomes
\begin{equation}
\chi_E(\lambda_+,\lambda_-)
= 1 + P_E(+1)\,(e^{i\lambda_+}-1)
      + P_E(-1)\,(e^{-i\lambda_-}-1).
\end{equation}
Different energies contribute independently, and in a measurement time $t$ the 
number of scattering attempts in an energy interval $[E,E+\dd E]$ is $t\, \dd E/(2\pi)$ (in units of $\hbar$).
The total CGF is therefore the product
\begin{equation}
\chi_t(\lambda_+,\lambda_-)
= \prod_E \big[\chi_E(\lambda_+,\lambda_-)\big]^{t\, \dd E/2\pi}.
\end{equation}
Taking the logarithm and passing to the continuum,
\begin{equation}
\ln \chi_t(\lambda_+,\lambda_-)
= t\!\int\!\frac{\dd E}{2\pi}\,
\ln\chi_E(\lambda_+,\lambda_-).
\end{equation}
Inserting the explicit form of $\chi_E$ yields the Levitov--Lesovik CGF~\cite{Levitov1996,Segal2018},
\begin{align}
\ln\chi(\lambda_+,\lambda_-)
= t\!\int\!\frac{\dd E}{2\pi}
\ln&\!\Big[
1+T(E)\big(f_L(1-f_R)(e^{i\lambda_+}-1)\nonumber\\&
+f_R(1-f_L)(e^{-i\lambda_-}-1)\big)
\Big],
\end{align}

The statistics of the \emph{net} transferred charge,
$N_{\mathrm{net}} = N_{+}-N_{-}$,
are obtained by setting the two counting fields equal, i.e. $\lambda_+ = \lambda_- = \lambda$,
so that forward and backward events acquire phases $e^{+i\lambda}$ and $e^{-i\lambda}$.

The first and second derivatives of $\ln\chi$ with respect to $i\lambda$ at $\lambda=0$ yield, respectively, the average current
\begin{equation}
J=\frac{1}{t}\frac{\partial\ln\chi}{\partial(i\lambda)}\Big|_{\lambda=0},
\end{equation}
and the corresponding zero-frequency fluctuations
\begin{equation}
D=\frac{1}{t}\frac{\partial^2\ln\chi}{\partial(i\lambda)^2}\Big|_{\lambda=0},
\end{equation}
which coincide with the standard LB results.

To focus only on forward tunneling events (excitations transmitted from $L$ to $R$), one sets the backward counting field to zero, $\lambda_- = 0$, obtaining the unidirectional CGF
\begin{equation}
\ln\chi_+(\lambda)
= t\!\int\!\frac{\dd E}{2\pi}\,
\ln\!\Big[1+T(E)\,f_L(1-f_R)\,(e^{i\lambda}-1)\Big].
\end{equation} 
From this, one obtains the forward current and its fluctuations as
\begin{align}
J_+ &= \int\frac{\dd E}{2\pi}\,T(E)\,f_L(1-f_R), \\[3pt]
D_+ &= \int\frac{\dd E}{2\pi}\,
\Big[T(E)f_L(1-f_R)-T(E)^2(f_L(1-f_R))^2\Big].
\end{align}
Analogously, a backward CGF $\chi_-(\lambda)$ can be defined by setting $\lambda_+=0$ and keeping $\lambda_-\neq0$, yielding
\begin{equation}
J_- = \int\frac{\dd E}{2\pi}\,T(E)\,f_R(1-f_L).
\end{equation}
The physical current and total noise then follow from the combination of the two directions,
\begin{equation}
J = J_+ - J_-, \qquad
D = D_+ + D_- - 2\mathcal{C}_{+-},
\end{equation}
where the cross-correlation term,
\begin{equation}
\mathcal{C}_{+-} = \int\frac{\dd E}{2\pi}\,T(E)^2\,f_L(1-f_R)f_R(1-f_L),
\end{equation}
accounts for the mutual exclusion of forward and backward events in the same energy channel.

Therefore, $J_+$ and $D_+$ naturally emerge as the forward-only contributions that can be directly extracted from the LB framework through its FCS generating function, while the standard LB noise is recovered when both directions and their cross-correlations are included.
In the limit of $N\rightarrow\infty$ with transmission $T(E)$ given by Eq.~\eqref{eq:T(E)_HEaviside} and the WBL, we also find $D_+ \sim  Je^{-\Sigma}$, albeit with different prefactors than~\eqref{eq:D_Sigma_JexpSigma} where the net current is considered.

\subsection{\label{SM:energy_disorder}Disorder in the qubit frequencies}

\begin{figure}
    \centering
    \includegraphics[width=\linewidth]{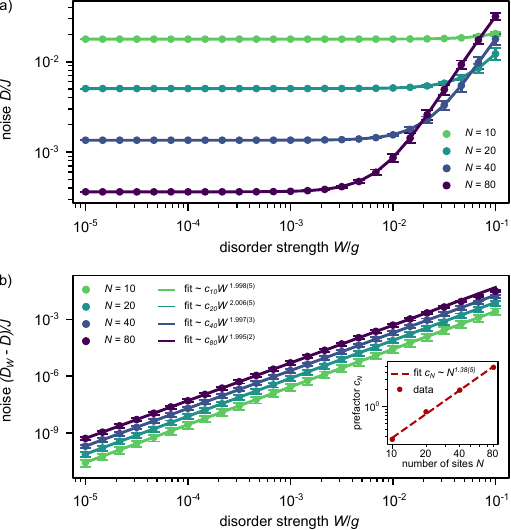}
    \caption{Impact of on-site energy disorder.
    Note the resemblence with Fig.~\ref{fig:Fig6b_CouplingLocalization} where a separate analysis is shown for disorder in the couplings.
    a) The Fano factor $D/J$ is shown for various chain lengths $N$ in the presence of disorder in the on-site energies.
    Each dot corresponds to the sample average over 1000 numerical samples of the disorder-affected frequencies for fixed disorder strength $W$.
    The error bars are the sample variance.
    b) The difference between the numerical expectation value of the diffusion constant $D_W$ with disorder [as in a)] and the constant $D$ in the absence of disorder is shown, together with a fit of the model~\eqref{eq:DW_D_model_anderson}.
    The inset shows how the prefactor scales with length $N$ of the chain.
    All fit exponents are obtained using least square in log-log scale, with the uncertainty given by the covariance matrix of the least square.
    Only the smallest third of disorder strengths was used for the least square fit.}
    \label{fig:Fig6a_EnergyDisorder}
\end{figure}

Here, we consider how disorder in the on-site energies of the chain affects the clock's behavior.
For simplicity, we model such perturbations by adding a site-dependent correction to the energy
\begin{align}
    \label{eq:omega_disorder}
    \hat\omega_i = \omega + \delta \hat\omega_i,
\end{align}
where $\delta\hat\omega_i$ independently and uniformly randomly sampled from the interval $-W/2\leq \delta \hat\omega_i \leq W/2$.
We note that the mean is zero, $\E[\delta\hat\omega_i]=0$ and the covariance is given by $\E[\delta\hat\omega_i \delta\hat\omega_j]=\delta_{ij} W^2/12$.
Such errors are known to lead to localization in one dimension, which negatively impact the clock's transport behavior and therefore also the precision scaling, because ticks may get ``stuck'' in the chain.
To quantify the impact this has on the log-linear cross-over time, we numerically estimate the expected diffusion constant $D_W$ of a model subject to the noise defined in Eq.~\eqref{eq:omega_disorder}, for chains of length $N=10,20,40$ and $80$, and different values of $W$.
In Fig.~\ref{fig:Fig6a_EnergyDisorder}a, the numerical expectation value of $D_W$ over 1000 samples is shown as a function of disorder strength $W$.
Then, in Fig.~\ref{fig:Fig6a_EnergyDisorder}b, we fit the the following error model,
\begin{align}
\label{eq:DW_D_model_anderson}
    (D_W- D)/J = c_N (W/g)^\alpha,
\end{align}
to the numerically sampled data, where $J$ is the current in the absence of disorder.
Numerically we find the exponent $\alpha = 1.999(4)$ (sample average and error from Fig.~\ref{fig:Fig6a_EnergyDisorder}b), in excellent agreement with $\alpha=2$ as one would theoretically expect, as detailed in the following.
Furthermore, the prefactor is determined numerically to scale with the system size as $c_N \sim N^{1.38(5)}$ (inset of Fig.~\ref{fig:Fig6a_EnergyDisorder}).

\paragraph*{Theoretical derivation.}
The central quantity from which transport properties and their fluctuations are inferred is the end-to-end disorder-averaged retarded Green's function $\E[G^r(E)]_{N1}$.
A fundamental result of Anderson~\cite{Anderson1958,Abrahams1979} is that in one dimension, this Green's function always decays exponentially as $G^r(E)\sim e^{-N/\xi(E)}$, where $\xi(E)$ is the \textit{localization length}.
In the following treatment, we determine how the $\xi(E)$ scales with the disorder $W$ to recover the scaling from Eq.~\eqref{eq:DW_D_model_anderson}.
The localization length is obtained from the lifetime of momentum mode $\psi_\kappa$, with $\kappa$ the lattice momentum.
This lifetime emerges because the perfect eigenmodes of the clean chain, labeled by momentum $\kappa$, are perturbed by the disorder, which causes them to decay exponentially in time with characteristic timescale $\tau_\kappa$. 
The distance over which the absolute square of the wave function decays as $|\langle N |\psi_\kappa\rangle|^2 \sim e^{-N/\ell_\kappa}$ defines the scattering length $\ell_\kappa$, which is related to the lifetime by 
\begin{align}
\ell_\kappa = v_\kappa \tau_\kappa,
\end{align}
where $v_\kappa = \partial_\kappa E(\kappa) = 2g\sin(\kappa)$ is the group velocity of momentum mode $\kappa$, and $g$ is the bulk coupling constant of the chain.
The localization length $\xi_\kappa$ is defined in terms of the wave function amplitude (rather than its square) \cite{LeeRamakrishnan1985,BruusFlensberg2004}, which upon taking the square root yields the relation
\begin{align}
\xi_k = 2 \ell_k = 2 v_k \tau_k.
\end{align}

Here, we reproduce a perturbative expression for $\xi_\kappa$ due to Thouless~\cite{Thouless1974,Thouless1979}, invoking only the \textit{Fermi Golden Rule}~\cite{Sakurai1967} as our technical tool. 
This approach is fully equivalent to a diagrammatic treatment of the self-energy in the first Born approximation (including only ``wigwam'' diagrams \cite{BruusFlensberg2004}). 
While it is well known that this expression has its shortcomings \cite{KappusWegner1981,DerridaGardner1984}, in the limit of weak disorder $W\ll g$ it correctly reproduces the scaling we observe in our numerical simulations.
We begin by expressing the perturbation $H_W$ arising from disordered qubit frequencies in fermionic momentum space:
\begin{align}
    \hat H_W &= \sum_j \delta \hat\omega_j c^\dagger_j c_j  \\
    &= \sum_{\kappa_n,\varsigma_m} e^\dagger_{\kappa_n + \varsigma_m} \hat V_{\varsigma_m} e_{\kappa_n}.
\end{align}
Note that here we already sitting in a rotating frame with frequency $\omega$.
The disorder potential expressed in momentum space is
\begin{equation}
    \hat V_{\varsigma} = \frac{1}{N}\sum_j \delta\hat\omega_j e^{{-}i \varsigma j}.
\end{equation}
The first and the second order of the disorder-averaged Fourier components $V_{\varsigma_n}$ are
\begin{align}
\label{eq:EV_EV2}
    \E[V_{\varsigma_n}] = 0,
    \quad
    \E[V_{\varsigma_n}V_{\varsigma_m}] = \frac{W^2}{12 N}\delta_{\varsigma_n + \varsigma_m,0}.
\end{align}
The lifetime $\tau_{\kappa_n}$ of momentum mode $\kappa_n$ due to the scattering with the disorder potential can now be computed by applying the Fermi Golden rule,
\begin{align}
    \tau_{\kappa_n} &=
    2\pi \sum_{\varsigma_m} \E\left[\vert\langle \kappa_n + \varsigma_m \vert \hat H_W \vert \kappa_n \rangle \vert^2\right] \\
    &\hspace{1.8cm} \times\delta(E(\kappa_n+\varsigma_m) - E(\kappa_n)). \nonumber
\end{align}
The disorder averaged matrix element can be directly computed using the relations in Eq.~\eqref{eq:EV_EV2},
\begin{align}
    \E\left[\vert \langle \kappa_n + \varsigma_m \vert \hat H_W \vert \kappa_n \rangle \vert^2 \right] &= \E \left[V_{-\varsigma_m}V_{\varsigma_m}\right] \\
    &= \frac{W^2}{12N}.
\end{align}
Using the identity $\sum_{\varsigma_n} \rightarrow N\int_{-\pi}^\pi\frac{\dd \varsigma}{2\pi}$ to convert the sum over the momenta to an integral, we obtain
\begin{align}
    \tau_\kappa  =  \frac{24 g \vert \sin(\kappa)\vert}{W^2}.
\end{align}
Substituting this into the relation for the localization length yields $\xi_\kappa = 96 \frac{g^2}{W^2} \sin^2(\kappa)$. 
By identifying the dispersion relation $E(k) = -2g\cos(k)$, the localization length can be rewritten as in the seminal result~\cite{Thouless1979},
\begin{align}
\label{eq:xiE_result}
 \xi(E) = \frac{96g^2-24E^2}{W^2}.
\end{align}
in the energy band $-2g \le E \le 2g$.
The correlation length thus scales quadratically in the disorder strength $g/W$, due to the vanishing first moments of the disorder distribution. 

Since the transmission function $T(E)$ is given by the absolute square of the matrix element $|G^r(E)_{N1}|^2$ (see Eq.~\eqref{eq:T(E)_green} in the Methods), the transmission function also decays as $T(E) \sim e^{-N/\xi(E)}$.
This deviation from the ideal box-car shape of the transmission function leads to a noisy current and thus an early cross-over into the diffusive regime of the clock.
In the limit of a small perturbation, the diffusion constant can be expanded to leading order in the disorder strength $W$ using the result~\eqref{eq:xiE_result},
\begin{align}
    D_W = D + \mathcal O\left(J(W/g)^2\right),
\end{align}
in agreement with the numerically determined exponent $\alpha=1.999(4)$ in Fig.~\ref{fig:Fig6a_EnergyDisorder}b.

\paragraph*{Cross-over time.}
The error model defined in Eq.~\eqref{eq:DW_D_model_anderson} also allows for expressing the correction to the cross-over time.
The numerically determined constant prefactor $c_N$ can be interpreted as a length-dependent error threshold $W_0 := g/\sqrt{c_N}$, such that we can write the error model as $(D_W-D)/J=(W/W_0)^2$.
In a next step, the relation between the diffusion constant and the cross-over time, $t^* \sim \log(J/D)/D$, from Eq.~\eqref{eq:t^*_res_derivation} can be used.
However, we assume the log-correction $\log(J/D)$ in Eq.~\eqref{eq:t^*_res_derivation} to be constant, since any arbitrarily deviation of the exponent of $(W/W_0)^\alpha$ from $\alpha=2$ will dominate the $\log$-contribution as $W\rightarrow 0$.
Thus dropping the logarithm does not further impact the quality of the approximation.
That being said, inserting the diffusion constant $D_W$ of the disorder-affected model into the cross-over time $t^*\sim 1/D_{W}$ yields
\begin{align}
    t^* &\sim \frac{1}{D + J(W/W_0)^2},
\end{align}
with $D$ the diffusion constant in the absence of disorder.
Two limits are interesting to consider:
\begin{enumerate}
    \item The limit of small errors $(W/W_0)^2 \ll D$. 
    To be explicit by inserting the $N$-dependence of $D\sim N^{1.86}$ as well as that of $W_0 \sim N^{-0.69(3)}$, we find that if $W$ satisfies
    \begin{align}
        W \ll \sqrt{D} W_0 \sim N^{0.32(3)},
    \end{align}
    then we can expand
    \begin{align}
    \label{eq:t^*_disorder_small}
        t^* \sim \frac{1}{D} - J\left(\frac{W}{W_0}\right)^2.
    \end{align}
    \item The other way around, we consider the limit where $D/J\ll (W/W_0)^2 \ll 1$. 
    That is, the error contribution $(W/W_0)^2$ is still small compared to $1$ but larger than $D$.
    Then, the cross-over time is dominated by disorder errors,
    \begin{align}
        t^* \sim J\left(\frac{W_0}{W}\right)^2 \sim \frac{1}{N^{1.38(5)} (W/g)^{2}}.\label{eq:t^*_disorder_eps0/eps}
    \end{align}
\end{enumerate}
To conclude the discussion of disorder effects, we see from the scaling in Eq.~\eqref{eq:t^*_disorder_eps0/eps} that the cross-over time scales inverse quadratically with the error: smaller error gives quadratically larger cross-over time, until one goes into the regime of case 1 where the scaling~\eqref{eq:t^*_disorder_small} holds.
The $N$-dependence of $W_0\sim N^{-0.69(3)}$, is sublinear, implying that the tolerable error shrinks with system size, but only moderately.

\subsection{\label{SM:coupling_disorder}Coupling disorder}

The impact of localization effects due to coupling disorder is analyzed as a perturbation effect for finite chain lengths $N<\infty$ akin to how disorder in the on-site energies is modeled in Appendix~\ref{SM:energy_disorder}.
To that end, consider site-dependent perturbations
\begin{align}
    \hat g_i = g_i + \delta \hat g_i
\end{align}
where the error $\delta \hat g_i$ are sampled independently and uniformly randomly from the interval $-\Delta/2 \leq \delta \hat g_i \leq \Delta/2$.
The constant $0<\Delta<g$ is the error strength and we consider the limit where $\Delta/g \ll1$ is very small.

\begin{figure}
    \centering
    \includegraphics[width=\linewidth]{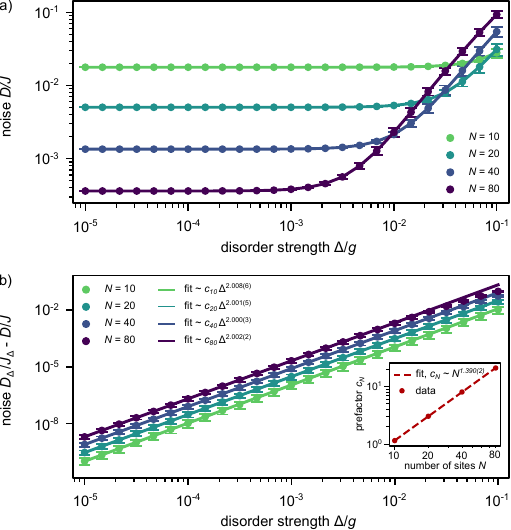}
    \caption{Impact of coupling disorder.
    Also note the resemblence with Fig.~\ref{fig:Fig6a_EnergyDisorder} for the on-site energy disorder analysis.
    a) We show the Fano factor $D/J$ for the perturbed couplings.
    The dots are sample averages over 1000 random realizations of the couplings for given disorder strenght $\Delta$, and the error bars indicate the sample error.
    The four colours indicate the chain lengths $N=10,20,40$ and $80$.
    b) Here, instead, the difference of the perturbed Fano factor and the ideal Fano factor without disorder is shown, together with a fit of model~\eqref{eq:DDelta_model_coupling}.
    The inset shows how the prefactor $c_N$ scales with $N$.
    All uncertainties in the scaling exponents are derived from the least-square fit covariance matrix.
    }
    \label{fig:Fig6b_CouplingLocalization}
\end{figure}

Like in Appendix~\ref{SM:energy_disorder}, we analyze this error model numerically by sampling many random realizations of noisy couplings.
In Fig.~\ref{fig:Fig6b_CouplingLocalization}a, the impact on the diffusion constant $D$ is shown for different chain lengths $N=10,20,40$ and $80$.
For every value of $\Delta$, $1000$ random samples of the couplings $\hat g_i$ are chosen to determine an average value for the diffusion constant.
It can be seen that as the error vanishes, $\Delta\rightarrow 0$, the value of $D$ approaches a constant.
To analyze this convergence behavior more quantitatively, we denote the average of the diffusion constant for a fixed error $\Delta$ by $D_{\Delta}$ (note that the chain length $N$ is implicit).
For the case of $\Delta=0$ as in the ideal setting of the main text we simply denote the diffusion constant by $D$.
In Fig.~\ref{fig:Fig6b_CouplingLocalization}b, the same data as in Fig.~\ref{fig:Fig6b_CouplingLocalization}a is shown, though in double-log scale where the following model is fitted to the data:
\begin{align}
\label{eq:DDelta_model_coupling}
    D_{\Delta}/J_\Delta - D/J \sim  c_N (\Delta / g)^\alpha,\quad c_N \sim N^\beta,
\end{align}
in the limit of small $\varepsilon$.
We numerically determine that the average exponent of $\varepsilon$ is given by $\alpha=2.003(3)$ (uncertainty from the sample error).
Moreover, the exponent for the $N$ scaling comes from the inset of Fig.~\ref{fig:Fig6b_CouplingLocalization}b and is found to be $\beta=1.390(2)$ (uncertainty from the least-square fit).
Remarkably, to a very approximation, disorder in the couplings affects the diffusion constant only to second order $\alpha= 2$, similar to the case of on-site energy disorder.

Therefore, the cross-over time scales quadratically with the error like discussed in Appendix~\ref{SM:energy_disorder}.
Moreover, the prefactor $c_N$ defines a length-dependent error threshold $\Delta_0 := g/\sqrt{c_N}\sim g N^{-0.695(1)}$ which vanishes sub-linearly in $N$.
To leading order, in the limit where $D/J\ll (\Delta/\Delta_0)^2\ll 1$, we thus also find
\begin{align}
    t^* \sim \frac{1}{N^{1.390(2)} (\Delta/g)^2},
\end{align}
like in~\eqref{eq:t^*_disorder_eps0/eps}, up to differences in constant factors and numerical uncertainty in the exponent of the $N$-scaling.

\section{Zero-frequency noise master equation vs.\ Landauer--B\"uttiker}
\label{ME_vs_LB_noise}

In this appendix we extend the correspondence between the master-equation (ME) and Landauer--B\"uttiker (LB) approaches to current fluctuations. In particular, we show that, in the wide-band and fully biased limit $f_L=1$, $f_R=0$, the ME expression for the zero-frequency noise reproduces the LB result
\begin{equation}
    D_{\rm LB}
    =\int_{-\infty}^{\infty}\frac{\dd E}{2\pi}\,T(E)\bigl[1-T(E)\bigr].
\end{equation}
For simplicity, but without loss of generality, we now set $\Gamma_L=\Gamma_R=1$,
which simplifies the notation without affecting the structure of the derivation.
In this case, the transmission function introduced in Eq.~\eqref{eq:T(E)_green} is given by $T(E) = \tr[\Pi_N G^r(E) \Pi_1 G^a(E)]$.

\paragraph*{Boundary current and noise definitions.}
For completeness, we recall the boundary current superoperator 
\begin{equation}
    \mathcal{J}=\mathcal{J}_+ - \mathcal{J}_-,
\end{equation}
with $\mathcal{J}_{+}\rho=\Gamma_R(1-f_R)\,c_N\rho c_N^\dagger$
and $\mathcal{J}_{-}\rho=\Gamma_R f_R\,c_N^\dagger\rho c_N$.
The associated ME current, identical to Eq.~\eqref{eq:ME-current-both}, is
\begin{equation}
    J_{\rm ME}
    =\mathrm{tr}[\mathcal{J}\rho_{\rm ss}]
    =\Gamma_R\bigl(\mathrm{tr}[C_{\rm ss}\Pi_N]-f_R\bigr),
\end{equation}
where $C_{\rm ss}$ is the stationary covariance matrix. The zero-frequency noise is defined as
\begin{equation}
    D_{\rm ME}
    =\int_{-\infty}^\infty \dd \tau\, S(\tau),
\end{equation}
where the stationary current--current correlator is given by~\cite{Blasi2024,Bourgeois2024}
\begin{equation}
\begin{aligned}
S(\tau)
&=\delta(\tau)\,A
+\Theta(\tau)\,\mathrm{tr}\!\big[\mathcal{J}e^{\mathcal{L}\tau}\mathcal{J}\rho_{\rm ss}\big] \\
&\qquad
+\Theta(-\tau)\,\mathrm{tr}\!\big[\mathcal{J}e^{\mathcal{L}|\tau|}\mathcal{J}\rho_{\rm ss}\big]
-J_{\rm ME}^2,
\end{aligned}
\end{equation}
where $A=\Gamma_R(1-2f_R)\,\tr[\Pi_N C_{\rm ss}] + \Gamma_R f_R$
is the dynamical activity [cf.~Eq.~\eqref{eq:K_dynamical_activity}]. Stationarity implies $S(-\tau)=S(\tau)^*$,
so that
\begin{equation}
    D_{\rm ME}
    =A+2\,\mathrm{Re}\!\int_0^\infty \dd\tau
    \left[
        \mathrm{tr}\bigl(\mathcal{J}e^{\mathcal{L}\tau}\mathcal{J}\rho_{\rm ss}\bigr)
        -J_{\rm ME}^2
    \right].
\label{eq:D-ME-pos-times}
\end{equation}

\paragraph*{Covariance-matrix formulation.}
In Appendix~\ref{SM:fluctuations_boundary}, we established that the integral above can be rewritten in terms of the effective generator $\mathcal{K}$ acting on covariance matrices. The resulting expression is
\begin{align}
\label{eq:D_ME_initial}
    D_{\rm ME}
    =A
    -2\,\mathrm{Re}\,J_{\rm ME}\,
    \mathrm{tr}\!\left[
        \Pi_N\,\mathcal{K}^{-1}(C^\sigma - C_{\rm ss})
    \right],
\end{align}
which corresponds to Eq.~\eqref{eq:D-ME-pos-times} and is equivalent to Eq.~\eqref{eq:D(t)_full} when expressed through the stationary covariance. For the case of full bias, the expression in Eq.~\eqref{eq:D_ME_initial} can be simplified by inserting the definition of $C^\sigma$ from Eq.~\eqref{eq:Csigma},
\begin{align}
    C^\sigma -C_{\rm ss} &= -\frac{1}{J_{\rm ME}}C_{\rm ss}\Pi_N C_{\rm ss},
\end{align}
and therefore:
\begin{equation}
    D_{\rm ME}
    =\mathrm{tr}[\Pi_N C_{\rm ss}]
    +2\,\mathrm{Re}\,
    \mathrm{tr}\!\left[
        \Pi_N\mathcal{K}^{-1}(C_{\rm ss}\Pi_N C_{\rm ss})
    \right].
    \label{eq:DME_Css}
\end{equation}

Using the Green's-function representation of the stationary covariance matrix (see Eq.~\eqref{eq:Css-Gr}),
\begin{equation}
    C_{\rm ss}
    =\int_{-\infty}^{\infty}\frac{\dd E}{2\pi}\,
    G^r(E)\,P\,G^a(E),
\end{equation}
with $P=\Pi_1$, the first term of Eq.~\eqref{eq:DME_Css} can be written as
\begin{align}
    \tr[\Pi_N C_{\rm ss}] &=\int_{-\infty}^\infty \frac{\dd E}{2\pi} T(E).
    \label{eq:trPiCss}
\end{align}
It therefore remains to show that the second part
\begin{align}
    I:=+2\,{\rm Re}\tr\left[\Pi_N \mathcal K^{-1} \left( C_{\rm ss} \Pi_N C_{\rm ss}\right)\right] &= -\int_{-\infty}^\infty\frac{\dd E}{2\pi} T(E)^2.
\end{align}

\paragraph*{Derivation.}
We recall that, formally, the inverse superoperator is given by
\begin{align}
    \mathcal K^{-1} &= -\int_0^\infty \dd t\, e^{\mathcal K t},
\end{align}
and furthermore $C_{\rm ss} = -{\mathcal K}^{-1} \Pi_1$.
The term $I$ can thus be rewritten as
\begin{align}
\label{eq:I_step1}
    I &= -2 \,{\rm Re}\int_{\mathbb R^3} \dd^3 t \,\Theta(\vec t) \times \\
    &\hspace{1cm}\tr\Big[\Pi_N \, e^{K(t_1+t_2)} \,\Pi_1\, e^{K^\dagger t_2}  \,\Pi_N\, e^{Kt_3} \,\Pi_1\, e^{K^\dagger(t_3+t_1)}\Big].  \nonumber
\end{align}
Here, we used that $\Theta(\vec t) = \Theta(t_1)\Theta(t_2)\Theta(t_3)$ is a product of Heaviside-$\Theta$ functions.
Furthermore, we may use the functional identity
\begin{align}
\label{eq:Theta(t)exp(Kt)}
    i\int_{\mathbb R}\frac{\dd E}{2\pi} e^{-iEt} G^r(E) &= \Theta(t) e^{Kt},
\end{align}
and analogously for the advanced Green's function.
By inserting Eq.~\eqref{eq:Theta(t)exp(Kt)} into Eq.~\eqref{eq:I_step1}, we find
\begin{align}
\label{eq:I_step2}
    I &= -2 \,{\rm Re}\int_{\mathbb R^3} \dd^3 t \,\Theta(t_1) \int_{\mathbb R^4} \frac{\dd^4 E}{(2\pi)^4} f(\vec t,\vec E)\times \\
    &\hspace{1cm} \tr\Big[\Pi_N  G^r(E_1) \Pi_1 G^a(E_2) \Pi_N G^r(E_3) \Pi_1 G^a(E_4)\Big].\nonumber
\end{align}
Note that $\Theta(t_2)$ and $\Theta(t_3)$ have been factored into the Fourier integral~\eqref{eq:Theta(t)exp(Kt)} when going from Eq.~\eqref{eq:I_step1} to Eq.~\eqref{eq:I_step2}.
Therefore, in Eq.~\eqref{eq:I_step2}, only the $\Theta(t_1)$ factor remains.
The term $f(\vec t,\vec E)$ comes from the Fourier transform and can be rearranged as
\begin{align}
\label{eq:f(t,E)_foufou}
f(\vec t,\vec E)
&=e^{-it_1(E_1-E_4) -it_2(E_1-E_2)-it_3 (E_3-E_4)}.
\end{align}

In the next step, the integrals $\dd^4 E$ and $\dd^3 t$ can be interchanged, and we find
\begin{align}
\label{eq:shokhotsky-identity}
    \int_{\mathbb R^3}\!\! \dd^3 t\,\Theta(t_1) f(\vec t,\vec E) &= 4\pi^2 \,\delta(E_1-E_2)\delta(E_3-E_4) \\
    &\hspace{-0.5cm}\times\left\{\pi \delta(E_1-E_4) - i{\rm PV}\frac{1}{E_1-E_4}\right\}. \nonumber
\end{align}
Here, ${\rm PV}(\,\cdot\,)$ is the Cauchy principal value, which arises due to the Sokhotski--Plemelj identity in the semi-infinite integral $\int_0^\infty \dd t\, e^{-itE}=\pi\delta(E) - i{\rm PV}(1/E)$.

Then, inserting Eq.~\eqref{eq:shokhotsky-identity} into the integral in Eq.~\eqref{eq:I_step2} gives
\begin{align}
\label{eq:I_step3}
    I&=-2\,{\rm Re}\! \int_{-\infty}^\infty\!\frac{\dd E_1\dd E_4}{(2\pi)^2}\left\{\pi \delta(E_1-E_4) - i{\rm PV}\frac{1}{E_1-E_4}\right\} \nonumber \\
    &\qquad\times\tr\Big[\Pi_N  G^r(E_1) \Pi_1 G^a(E_1) \Pi_N G^r(E_4) \Pi_1 G^a(E_4)\Big].
\end{align}
The trace can be factorized because $\tr[\Pi_N A\Pi_N B]=\tr[\Pi_NA]\tr[\Pi_N B]$, which reveals the transmission function,
\begin{align}
    &\tr\Big[\Pi_N  G^r(E_1) \Pi_1 G^a(E_1) \Pi_N G^r(E_4) \Pi_1 G^a(E_4)\Big] \\
    &=\tr\Big[\Pi_N  G^r(E_1) \Pi_1 G^a(E_1)\Big] \tr\Big[ \Pi_N G^r(E_4) \Pi_1 G^a(E_4)\Big]\nonumber\\
    &=T(E_1)T(E_4).
\end{align}
Since the transmission function is real, the principal-value integral is also real.
In Eq.~\eqref{eq:I_step3}, due to the imaginary factor $i$ in front of the principal-value term, only the Dirac-$\delta$ contribution from the Sokhotski--Plemelj factor survives under the real part.
The result is
\begin{align}
    I &= -\int_{-\infty}^\infty \frac{\dd E}{2\pi} T(E)^2,
    \label{eq:final_I}
\end{align}
which, together with Eq.~\eqref{eq:trPiCss}, can be substituted into Eq.~\eqref{eq:DME_Css} to give
\begin{equation}
    D_{\rm ME}=D_{\rm LB}
    =\int_{-\infty}^{\infty}\frac{\dd E}{2\pi}\,T(E)\bigl[1-T(E)\bigr].
\end{equation}

\end{appendices}

\end{document}